\documentclass[preprint,12pt,authoryear]{elsarticle}

\usepackage{amssymb}
\usepackage{booktabs}
\usepackage{multirow}
\usepackage{bbm}
\usepackage{amsmath}
\usepackage{xcolor}
\usepackage{hyperref}

\usepackage{graphicx}
\usepackage{array}

\usepackage{caption}

\journal{Computer Methods in Applied Mechanics and Engineering}

\begin{document}

\begin{frontmatter}



\title{Stress-constrained Topology Optimization for Metamaterial Microstructure Design}


\author[1]{Yanda Chen}
\ead{yanda.chen@ensam.eu}
\author[1]{Sebastian Rodriguez}
\author[1]{Beatriz Moya}
\author[1,2]{Francisco Chinesta}

\affiliation[1]{organization={Laboratory PIMM, Arts et Metiers Institute of Technology, CNRS, CNAM},
            addressline={151 Bd de l'Hôpital}, 
            city={Paris},
            postcode={75013}, 
            country={France}}
\affiliation[2]{organization={CNRS @ CREATE},
            addressline={1 Create Way, 08-01 CREATE Tower}, 
            postcode={138602}, 
            country={Singapore}}

\begin{abstract}
Although stress-constrained topology optimization has been extensively studied in structural design, the development of optimization frameworks to enable the creation of metamaterials with optimal mechanical performance is still an open problem. This study incorporates local stress constraints into the topology optimization framework for metamaterial microstructure design, aiming to avoid the stress concentration in the optimized microstructure. For the efficient solution of multi-constraint topology optimization problems, the Augmented Lagrangian formulation is extended to address local minimization problems subjected to the combined action of local and global constraints. Additionally, as an extension of static load conditions, this study further investigates the design of metamaterial microstructures under cyclic loading. Finally, the effectiveness of the proposed approach is demonstrated through a series of two-dimensional and three-dimensional benchmark problems.
\end{abstract}

\begin{graphicalabstract}
\end{graphicalabstract}

\begin{highlights}
\item The augmented Lagrangian method is applied to efficiently solve the topology optimization problem of metamaterial microstructure with a large number of constraints
\item The fatigue problem of the metamaterial microstructure under a large number of constant amplitude cyclic loading is considered
\item Considers and compares the topology optimization results of metamaterial microstructure subjected to different stress-based high-cycle fatigue criteria constraints
\item The augmented Lagrange formula is extended to simultaneously integrate different types of constraints.
\end{highlights}


\end{frontmatter}


\section{Introduction}
\label{sec1}
Metamaterials are artificial materials whose exceptional properties do not arise from their chemical composition but from the deliberate design of their internal microstructures (\cite{jiao_mechanical_2023}). Unlike conventional materials, which are limited by intrinsic atomic or molecular structures, metamaterials enable unprecedented control over mechanical behavior by tailoring the geometry, topology, and spatial distribution of solid and void phases within a defined design domain. This structural programmability allows for the realization of unusual and improved mechanical characteristics, including negative Poisson’s ratios (\cite{zhang_stretchable_2024}), high stiffness-to-weight ratios (\cite{efa_computational_2025}), tunable anisotropy (\cite{zheng_data-driven_2021}), improved fracture resistance (\cite{wang_superior_nodate}), among many others. As a result, metamaterials have emerged as a promising platform for designing lightweight, multifunctional, and mechanically robust structures. Their versatility has enabled a broad range of applications, spanning aerospace, robotics, biomedical devices, and energy absorption systems (\cite{wu_superelastic_2022}, \cite{Brandenbourger_reciprocal_2019}, \cite{zhao_hydrogel-based_2022}, \cite{Cheng_energy_2023}).

Since the material distribution at the microscale determines the macroscopic behavior of metamaterials, the targeted performance at the macroscale can be achieved by precisely designing the microstructure and its spatial arrangement. Structural topology optimization has been widely recognized as an indispensable tool for systematically exploring and designing high-performance metamaterials. It provides a mathematical framework for determining the optimal material distribution within a prescribed design domain, with the aim of maximizing specific performance metrics and satisfying functional constraints (\cite{bendsoe_2003_topology}). Theoretically, the optimal topology is composed of periodic composite materials with infinitesimally small features (\cite{groen_homogenization-based_2018}). However, in practical applications, the admissible design space is appropriately relaxed in homogenization-based topology optimization to allow the modeling of microstructure composites through effective material properties, thereby achieving an approximate representation of composite materials (\cite{allaire_homogenization_2019}). In this context, homogenization theory plays a key role in addressing engineering problems characterized by multiscale features. The core idea is to separate the macroscopic and microscopic scales, allowing the complex and heterogeneous characteristics of materials or structures on the microscale to be translated into equivalent macroscopic responses (\cite{hassani_review_1998, hassani_review_1998_II}). The effective macroscopic physical properties such as elastic modulus can be extracted by numerically simulating the response of the representative unit cell (RUC) under specific loads or boundary conditions  (\cite{andreassen_how_2014, dong_149_2018}), which makes it possible to simplify modeling and effectively predict complex multiscale systems.

Bendsøe and Kikuchi (\cite{BendsoeKikuchi1988}) introduced the numerical homogenization method into the structural topology optimization framework for the first time, to compute the effective elastic properties of periodic porous microstructures. Sigmund (\cite{Sigmund1994}) proposed an inverse homogenization method to generate periodic mechanical metamaterials with tailored effective properties. Since then, the design of metamaterials based on numerical homogenization and topology optimization has been further extended. Xia and Breitkopf (\cite{Xia_design_2015}) developed an energy-based homogenization MATLAB code to facilitate the implementation of topology optimization for metamaterials. Zhang and Khandelwal (\cite{zhang_computational_2019}) combined density-based topology optimization with nonlinear homogenization to design metamaterials exhibiting negative Poisson’s ratio over large strain ranges. Chen and Huang (\cite{chen_topological_2019}) integrated couple-stress homogenization with topology optimization to successfully design 3D chiral metamaterials capable of exhibiting torsional responses under compressive loading. Huo et al. (\cite{huo_bi-directional_2025}) proposed a bidirectional homogenization method based on multiscale topology optimization principles, significantly improving the design efficiency of functionally graded auxetic metamaterials. For more details on homogenization theory and topology optimization of multiscale structures, please refer to Xia and Breitkopf (\cite{Xia_topology_2016}), Allaire et al. (\cite{allaire_homogenization_2019}) and Wu et al. (\cite{wu_topology_2021}).

Previous studies have primarily focused on stiffness optimization of microstructures, which typically leads to significant stress concentration around the boundaries of holes, thereby substantially increasing the risk of material fatigue and fracture. To ensure that metamaterial microstructures meet specific functional requirements, it is essential to prevent structural failure at any location within the component. This can be achieved by developing specialized topology optimization frameworks that enforce the design satisfies prescribed stress-based constraints and maintains sufficient fatigue life over a given number of loading cycles. Solving stress-constrained topology optimization problems involves two distinct challenges. The first is the singularity phenomenon(\cite{rozvany_difficulties_1996}). Several relaxation techniques have been proposed to address this issue, such as $\varepsilon $-relaxation (\cite{cheng_-relaxed_1997, paris_topology_2009}) or $qp$-method (\cite{bruggi_alternative_2008}) or damage approach (\cite{verbart_damage_2016}). The second challenge arises from the large-scale nature of the optimization problem caused by the local behavior of stress constraints (\cite{duysinx_topology_1998}). In terms of this issue, the constraint aggregation techniques can be used to transform a large number of local constraints into one or more aggregated global constraints by using a smoothed approximation of the maximum function, such as the Kreisselmeier–Steinhauser function (\cite{kreisselmeier_systematic_1979, paris_topology_2009}) and the P-norm function (\cite{le_stress-based_2010, lee_novel_2016}). Several studies have already applied the aggregation techniques in the topology optimization of metamaterial microstructures to reduce computational cost(\cite{Thillaithevan_stress-constrained_2020, alacoque_stress-based_2021, gupta_computational_2024}). However, they lost control over the local behavior of the constraints and the quality of this local control depends on both the number of aggregated constraints and the parameters used in the aggregation technique. Thus, the solution to the global problem may differ from that of the corresponding local problem. 

At present, only a small number of studies have considered the local properties of stress constraint in the topology optimization of metamaterial microstructures. Collet et al. (\cite{collet_topology_2018}) implemented an active set selection strategy to accelerate the process. Coelho et al. (\cite{Coelho_topology_2019}) employed parallel computing techniques to reduce the computational burden associated. Although the above strategies have demonstrated effective acceleration in two-dimensional applications, their scalability to large-scale three-dimensional problems is limited due to the need to solve a large number of adjoint equations. As a promising alternative technique, the augmented Lagrangian (AL) method has been shown to significantly reduce the costs associated with a large number of constraints while providing more consistent models than aggregation techniques (\cite{giraldo-londono_polystress_2021, chen_stress-constrained_2024}). To the best of our knowledge, the AL method has not yet been applied to the stress-constrained topology optimization problem of metamaterial microstructures. Furthermore, as an extension of stress-constrained problems under static loading, this study presents for the first time the topology optimization of metamaterial microstructures subjected to a large number of cyclic loads, where the constraints are based on the implementation of stress-based high-cycle fatigue criteria (\cite{papadopoulos_comparative_1997, carpinteri_multiaxial_2001, chen_fatigue-constrained_2024}).

This study aims to incorporate stress responses under different load conditions, i.e., static and cyclic loading, into the topology optimization framework for designing novel periodic metamaterial microstructures to prevent the occurrence of high stress concentration. This paper is organized as follows. Section \eqref{sec2} provides an overview of topology optimization framework. Section \eqref{sec3} presents the topology optimization problem with stress constraints for continuum structures, and several numerical examples are shown in Section \eqref{sec4}, followed by a discussion of the predicted results under different load conditions and different stress criteria in Section \eqref{sec5}. Finally, Section \eqref{sec6} gives concluding remarks to complete the paper. The procedure of sensitivity analysis is given in \eqref{secA}.

\section{Topology optimization framework} \label{sec2}
\begin{figure}[ht]%
\centering
\includegraphics[width=0.6\textwidth]{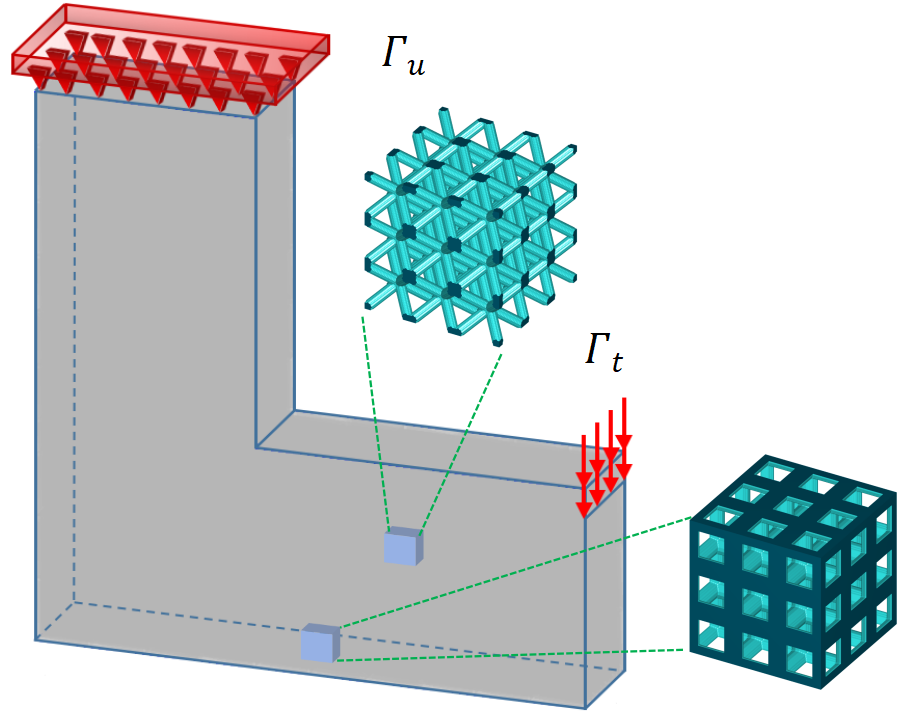}
\caption{Multiscale topology optimization}
\label{fig1}
\end{figure}

With the rapid advancement of additive manufacturing technologies, it has become feasible to fabricate structures with intricate hierarchicals. This capability enables the resurgent interest in optimal design of multiscale structures. Fig. \eqref{fig1} shows a schematic diagram of multiscale topology optimization. To reduce the computational cost, the homogenization method is usually applied to represent the complex behavior of periodic microstructures through equivalent material properties, thereby allowing for structural optimization at the macroscopic scale without the need to explicitly parse the microscopic scale at each iteration.
\subsection{Numerical homogenization}\label{subsec2.1}
Numerical homogenization is an efficient method to determine the effective elastic tensor of periodic microstructures. It is based on two fundamental assumptions: (1) the size of the RUC is much smaller than that of the overall material body; and (2) the RUCs are periodically distributed throughout the material. In this study, the constitutive behavior of the microstructure is assumed to be linear elastic, and the homogenized constitutive matrix of a periodic microstructure $\boldsymbol{C}^{H}$ is estimated using the asymptotic homogenization method (\cite{BendsoeKikuchi1988, andreassen_how_2014, dong_149_2018}): 
\begin{equation}\label{Eq. (1)}
\begin{aligned}
C_{i j}^{H}= & \frac{1}{V} \int_{\Omega}\left[\boldsymbol{\varepsilon}_{i}^{0}-\boldsymbol{\varepsilon}_{i}(\boldsymbol{u}_{i})\right]^{T} \boldsymbol{C}\left[\boldsymbol{\varepsilon}_{j}^{0}-\boldsymbol{\varepsilon}_{j}(\boldsymbol{u}_{j})\right] d\Omega
\end{aligned}
\end{equation}
where $i,j=1,\dots,d$ means the Voigt dimension ($d=6$ in 3D and $d=3$ in 2D), $V$ represents the area (2D) or the volume (3D) of the RUC, $\Omega$ denotes the solid part in the RUC, $\boldsymbol{C}$ is the constitutive matrix of the solid material, $\boldsymbol{u}_{i}$ signifies the nodal displacement field corresponding to the $i$-th disturbance strain field, $\boldsymbol{\varepsilon}^{0}_{i}$ and $\boldsymbol{\varepsilon}_{i}$ are the $i$-th prescribed macroscopic unit strain field and the locally varying strain field, respectively. In the 3D scenario, $\boldsymbol{\varepsilon}^{0}$ contains six linearly independent test strains $\boldsymbol{\varepsilon}^{0}=\boldsymbol{I}_{6 \times 6}$. In a two-dimensional plane stress problem, it can be simplified to the form of a $3\times3$ matrix $\boldsymbol{\varepsilon}^{0}=\boldsymbol{I}_{3 \times 3}$.

To satisfy the periodicity of the microstructures, the periodic boundary conditions can be implemented using the elimination method (\cite{andreassen_how_2014, collet_topology_2018}), the Lagrange multiplier method (\cite{miehe_computational_2002, van_dijk_formulation_2016}) or the penalty function method (\cite{temizer_computation_2008}) on the RUC. In structured periodic meshes, the elimination method is the most efficient where the periodic boundary conditions can be implemented by directly pairing corresponding nodes and eliminating redundant degrees of freedom. In contrast, for unstructured or non-periodic meshes where one-to-one correspondence between boundary nodes is not available, more general approaches such as the Lagrange multiplier method or the penalty method are commonly employed to impose boundary constraints and accommodate complex geometrical configurations. Subsequently, the unknown displacement field $\boldsymbol{u}^{i}$ is determined by solving the weak form of the linear elasticity equilibrium equations subject to periodic boundary conditions:
\begin{equation}\label{Eq. (2)}
\int_{\Omega} \boldsymbol{\varepsilon}_{i}(\boldsymbol{v})^{T} \boldsymbol{C}\boldsymbol{\varepsilon}_{i}\left(\boldsymbol{u}_{i}\right) d {\Omega}=\int_{\Omega} \boldsymbol{\varepsilon}_{i}(\boldsymbol{v})^{T} \boldsymbol{C} \boldsymbol{\varepsilon}_{i}^{0} d {\Omega}
\quad \forall \boldsymbol{v} \in \boldsymbol{H}^{1} (\Omega)
\end{equation}
where $\boldsymbol{v}$ is a virtual displacement field that belongs to the allowed displacement space (Sobolev space) $\boldsymbol{H}^{1} (\Omega)$.

To solve the weak form in Eq. \eqref{Eq. (2)} numerically, it is necessary to discretize the RUC into $N$ elements. Then it can be approximated through the sum of finite element integrals as follows:
\begin{equation}\label{Eq. (3)}
\begin{aligned}
\sum_{I=1}^{N} \sum_{J=1}^{N} \boldsymbol{v}_{I}^{T}\left(\int_{\Omega} \boldsymbol{B}_{I}^{T} \boldsymbol{C} \boldsymbol{B}_{J} d \Omega\right) \boldsymbol{u}_{iJ} = \sum_{J=1}^{N} \boldsymbol{v}_{J}^{T}\left(\int_{\Omega} \boldsymbol{B}_{J}^{T} \boldsymbol{C} \boldsymbol{\varepsilon}_{i}^{0} d \Omega\right)
\end{aligned}
\end{equation}
where $\boldsymbol{B}_{I}$ represents the strain–displacement matrix of element $I$, $\boldsymbol{u}_{iJ}$ is the displacement field of element $J$ under $i$-th unit strain field. Finally, Eq.\eqref{Eq. (3)} can be simplified to the following static equilibrium equation:
\begin{equation}\label{Eq. (4)}
\boldsymbol{K} \boldsymbol{u}_{i}=\boldsymbol{f}_{i}, \quad i=1,...,d
\end{equation}
where $\boldsymbol{K}$ represents the stiffness matrix, $\boldsymbol{f}_{i}$ is the force vector in correspondence with the predefined unit strain $\boldsymbol{\varepsilon}_{i}^{0}$.

\subsection{Topology optimization formulation}\label{subsec2.2}
The solid isotropic material with penalization (SIMP) method (\cite{sigmund_99_2001, andreassen_efficient_2011}) is used to solve the optimization problems and their general statement can be expressed mathematically as follows:
\begin{align}  \label{Eq. (5)}
\min& \  c(\boldsymbol{\bar{\rho}}) \notag \\
&\text{s.t.} \quad 
\begin{cases} 
g_{I}(\boldsymbol{\bar{\rho}}, \boldsymbol{u}) \leq 0, & I=1, \ldots, N_{t} \\
0 \leq \rho_{J} \leq 1,      & J=1, \ldots, N 
\end{cases} \\
&\text{with:} \qquad 
\begin{aligned}[t]   \notag
&\boldsymbol{K}(\boldsymbol{\bar{\rho}}) \boldsymbol{u} = \boldsymbol{f} \\
&\boldsymbol{\bar{\rho}} = \boldsymbol{\mathcal{H}} (\boldsymbol{\mathcal{F}} \boldsymbol{\rho})
\end{aligned}
\end{align}
where $c$ denotes the objective function, $\boldsymbol{\rho}$ stands for the vector of design variables, $g_{I}$ is the $I$-th constraint among a total of $N_{t}$ constraints. To avoid numerical instabilities, the polynomial filter developed by Zegard and Paulino \cite{zegard_bridging_2016} is adopted, resulting in the filtered design variable $\widetilde{\boldsymbol{\rho}} = \boldsymbol{\mathcal{F}} \boldsymbol{\rho}$:
\begin{equation}\label{Eq. (6)}
\mathcal{F}_{I J}=\frac{L_{I J} \rho_{J}}{\sum_{K=1}^{N} L_{I K} \rho_{K}}, \text { with } L_{I J}=\max \left[0,1-\frac{d\left(\boldsymbol{x}_{I}, \boldsymbol{x}_{J}\right)}{R}\right]^{s}
\end{equation}
where $R$ denotes the radius of the filter, $s\geq 1$ is the exponent, $d\left(\boldsymbol{x}_{I}, \boldsymbol{x}_{J}\right)$ represents the Euclidean distance between the centroid coordinates of elements $I$ and $J$ respectively.

The Heaviside projection function $\boldsymbol{\mathcal{H}}$ is used in Eq.\eqref{Eq. (5)} to obtain a black-and-white design, which can be expressed as follows (\cite{guest_achieving_2004, wang_projection_2011}):

\begin{equation}\label{Eq. (7)}
\bar{\rho}_{I}=\frac{\tanh (\beta \eta)+\tanh \left[\beta\left(\tilde{\rho}_{I}-\eta\right)\right]}{\tanh (\beta \eta)+\tanh [\beta(1-\eta)]}
\end{equation}

\noindent where $\eta$ is the projection threshold and $\beta$ controls the slope of the function near the threshold parameter $\eta$.

The stiffness matrix $\boldsymbol{K}$ in Eq. \eqref{Eq. (5)} can be calculated through a typical assembly process:

\begin{equation}\label{Eq. (9)}
\boldsymbol{K}(\boldsymbol{\bar{\rho}})=\sum_{I=1}^{N}\mathbbm{A}^{I} \boldsymbol{K}_{I}(\boldsymbol{\bar{\rho}}), \text { with } \boldsymbol{K}_{I}(\boldsymbol{\bar{\rho}})=\left[\epsilon+(1-\epsilon) \bar{\rho}_{I}^{p}\right] \boldsymbol{K}_{I}
\end{equation}

\noindent where $\boldsymbol{K}_{I}(\boldsymbol{\bar{\rho}})$ is the stiffness matrix of $I$-th element, $\mathbbm{A}$ is an assembly operator, $\epsilon$ is the Ersatz parameter to prevent singularity, $p$ is the penalty factor, $\boldsymbol{K}_{I}$ is the stiffness matrix of $I$-th solid element when $\bar{\rho}_{I}=1$.

Due to the local nature of stress-based constraints, each finite element is typically associated with an individual constraint. As a result, a large number of adjoint equations need to be solved when performing sensitivity analysis using the adjoint method. However, the AL approach enables the efficient solution of topology optimization problems using gradient-based algorithms, while preserving the local nature of stress constraints (\cite{senhora_topology_2020, giraldo-londono_polystress_2021, chen_fatigue-constrained_2024, chen_stress-constrained_2024}). This efficiency is attributed to its sensitivity analysis scheme, which necessitates the computation of only one adjoint vector per independent load vector at each optimization step. This work presents an extension of the AL formulation that concurrently handles local and global constraints. Taking into account all types of constraints required in this work, a generalized AL formulation is expressed as follows:

\begin{equation}\label{Eq. (10)}
\begin{aligned}
\min _{\boldsymbol{\rho} \in[\boldsymbol{0},\boldsymbol{1}]} \mathcal{J}^{(k)}(\overline{\boldsymbol{\rho}}, \boldsymbol{\lambda}^{(k)}, \mu^{(k)}) & =c(\overline{\boldsymbol{\rho}})+\frac{1}{N_s} P_{s}^{(k)}(\overline{\boldsymbol{\rho}}, \boldsymbol{u})+P_{v}^{(k)}(\overline{\boldsymbol{\rho}}, \boldsymbol{u})+P_{iso}^{(k)}(\overline{\boldsymbol{\rho}}, \boldsymbol{u}) \\
\text { with: } \quad P_s^{(k)}(\overline{\boldsymbol{\rho}}, \boldsymbol{u}) &  =\sum_{J=1}^{N_s}\left[\lambda_{sJ}^{(k)} h_{J}^{s}(\overline{\boldsymbol{\rho}}, \boldsymbol{u})+\frac{\mu^{(k)}}{2} h_{J}^{s}(\overline{\boldsymbol{\rho}}, \boldsymbol{u})^{2}\right] \\
\quad P_v^{(k)}(\overline{\boldsymbol{\rho}}, \boldsymbol{u}) &  =\lambda_{v}^{(k)} h^{v}(\overline{\boldsymbol{\rho}}, \boldsymbol{u})+\frac{\mu^{(k)}}{2} h^{v}(\overline{\boldsymbol{\rho}}, \boldsymbol{u})^{2} \\
P_{iso}^{(k)}(\overline{\boldsymbol{\rho}}, \boldsymbol{u}) & =\lambda_{iso}^{(k)} h^{iso}(\overline{\boldsymbol{\rho}}, \boldsymbol{u})+\frac{\mu^{(k)}}{2} h^{iso}(\overline{\boldsymbol{\rho}}, \boldsymbol{u})^{2}
\end{aligned}
\end{equation}
where

\begin{equation}\label{Eq. (11)}
\begin{aligned}
h_{J}^{s}(\overline{\boldsymbol{\rho}}, \boldsymbol{u})& =\max \left\{\left[\epsilon+(1-\epsilon) \bar{p}_{J}^{p}\right]\left(g_J^{3}+g_J\right),-\frac{\lambda_{sJ}^{(k)}}{\mu^{(k)}}\right\} \\
h^{v}(\overline{\boldsymbol{\rho}}, \boldsymbol{u})& =\max \left[\frac{\sum_{J=1}^{N} \bar{\rho}_{J} V_{J}}{\mathrm{V}_{f}},-\frac{\lambda_{v}^{(k)}}{\mu^{(k)}}\right] \\
h^{iso}(\overline{\boldsymbol{\rho}}, \boldsymbol{u}) &=\max \left[\sum_{i, j=1}^{d} \frac{\left(C_{i j}^{H}-C_{i j}^{i s o}\right)^{2}}{\left(C_{i j}^{i s o}+\epsilon\right)^{2}},-\frac{\lambda_{iso}^{(k)}}{\mu^{(k)}}\right]
\end{aligned}
\end{equation}

The terms $P_{s}$, $P_{v}$, and $P_{iso}$ represent the penalty terms associated with stress, volume, and isotropic constraints, each embedding its corresponding equality constraint $h^{s}$, $h^{v}$, and $h^{iso}$. The number of the stress constraint $N_s$ equals the product of the number of independent load cases and the number of elements $N$. The polynomial vanishing constraint is adopted here to impose a more severe penalty on regions that violate the local stress limits, thereby accelerating the convergence and facilitating the minimization of the AL function (\cite{chen_stress-constrained_2024}). In the volume constraint, $V_f$ denotes the prescribed volume fraction. The isotropic constraint is activated only for the minimization of the Poisson’s ratio problem, where it ensures that the resulting Poisson’s ratio remains directionally uniform. The $ij$-th entry of the three-dimensional isotropic constitutive matrix $C_{ij}^{i s o}$ is constructed from the homogenized constitutive matrix $C_{i j}^{H}$ as follows:

\begin{equation}\label{Eq. (12)}
\left\{\begin{array}{l}
C_{i i}^{i s o}=\left(C_{11}^{H}+C_{22}^{H}+C_{33}^{H}\right) / 3, \quad i=1,2,3 \\[8pt]
C_{i j}^{i s o}=\left(C_{12}^{H}+C_{13}^{H}+C_{23}^{H}\right) / 3, \quad i, j=1,2,3, \quad i \neq j \\[8pt]
C_{i i}^{i s o}=\left(C_{11}^{iso}-C_{12}^{iso}\right) / 2, \quad i=4,5,6 \\[8pt]
C_{i j}^{i s o}=0, \quad \text { else }
\end{array}\right.
\end{equation}

The Lagrange multiplier estimator uses the same initial guess across all constraints, and is updated together with the penalty at each iteration as follows: 

\begin{equation}\label{Eq. (13)}
\begin{array}{c}
\mu^{(k+1)}=\min \left(\alpha \mu^{(k)}, \mu_{max}\right) \\ \\
\boldsymbol{\lambda}^{(k+1)}=\boldsymbol{\lambda}^{(k)}+\mu^{(k)}  \boldsymbol{h}\left(\overline{\boldsymbol{\rho}}^{(k)}, \boldsymbol{u}\right)
\end{array}
\end{equation}
where $\alpha>1$ is a constant parameter and $\mu_{max}$ is an upper limit to prevent numerical instability.

In the AL algorithm, the topology optimization problem is solved using a nested iterative scheme consisting of inner and outer loops. In the outer loop, the Lagrange multipliers $\boldsymbol{\lambda}$ and the penalty parameter $\mu$ are updated. Within each outer iteration, $\boldsymbol{\lambda}$ and $\mu$ are fixed, and the minimization of the AL function in Eq. \eqref{Eq. (10)} is approximated by solving a series of convex sub-problems using the method of moving asymptotes through several inner iterations (\cite{giraldo-londono_polystress_2021}). The algorithm terminates when the maximum change in the design variables between two successive iterations is below a specified tolerance $\delta$, and all constraints are satisfied within a separate tolerance $\delta_{\mathrm{s}}$. In the event that the solution does not converge, a predefined maximum number of iterations is imposed. The topology optimization parameters applied in this study are listed in Table \eqref{tab1}.

\begin{table}[!ht]
\centering
\caption{Topology optimization parameters}\label{tab1}
\newsavebox{\tablebox}
\sbox{\tablebox}{%
  \begin{tabular}{@{}lll@{}}
    \toprule
    Parameter & Description & Value \\
    \midrule
   $p$    & Penalty factor   & 5 \\
   $\epsilon$    & Ersatz parameter   & 1e-9 \\
   $\beta^{(0)}{}^{\ast}$  & Initial Heaviside projection penalization factor & 1 \\
   $\beta_{\max}$    & Maximum Heaviside projection penalty factor   & 10\\
   $\eta$    & Heaviside projection threshold   & 0.5\\
   $\mu^{(0)}{}^{\dagger}$    & Initial penalty coefficient   & 10\\
   $\boldsymbol{\lambda}^{(0)}{}^{\ddagger}$  & Initial Lagrange multiplier vector   & $\boldsymbol{0}$\\
   $\mu_{\max}$    & Maximum penalty coefficient   & 10000\\
   $\alpha$    & Penalty factor updating parameter   & 1.1\\
   $q$    & Nonlinear filter index   & 3.5\\
   $\delta$    & Convergence tolerance of design variables for AL   & 0.005\\
   $\delta_{s}$    & Convergence tolerance of constraints for AL   & 0.005\\
   $m$    & Move limit   & 0.15\\
   $\text{MaxIter}$    & Maximum number of external loops   & 100\\
   $\text{MaxIn}$    & Maximum number of internal loops per AL step   & 15\\
    \bottomrule
  \end{tabular}%
}
\usebox{\tablebox}
\vspace{10pt} 
\begin{minipage}{\wd\tablebox}
  \footnotesize\raggedright
  $^{*}$ The parameter $\beta$ is incremented by 1 every 5 AL steps until it reaches $\beta_{\max}$. \\
  $^{\dagger}$ The value of $\mu^{(0)}$ is adjusted to 100 for 3D problems. \\
  $^{\ddagger}$ The vector $\boldsymbol{\lambda}^{(0)}$ comprises $\boldsymbol{\lambda}_s$ (stress constraint), $\lambda_v$ (volume constraint), and $\lambda_{iso}$ (isotropic constraint).
\end{minipage}
\end{table}

\section{Material failure criteria} \label{sec3}
The theory of material failure provides the fundamental framework for assessing structural integrity under external load. The core idea is the reduction of complex multi-axial stress states to an equivalent uni-axial measure that facilitates direct comparison with experimental data from standard tests, such as uni-axial tension or fatigue experiments. Failure is assumed to occur when this equivalent stress reaches a critical threshold, typically defined by the yield strength or fatigue limit of the material. This section provides a review of several classical yield and fatigue criteria, which are used to predict the failure of metallic materials under static and cyclic load conditions, respectively.

\subsection{Yield criteria}\label{subsec3.1}
\begin{figure}[ht]%
\centering
\includegraphics[width=0.6\textwidth]{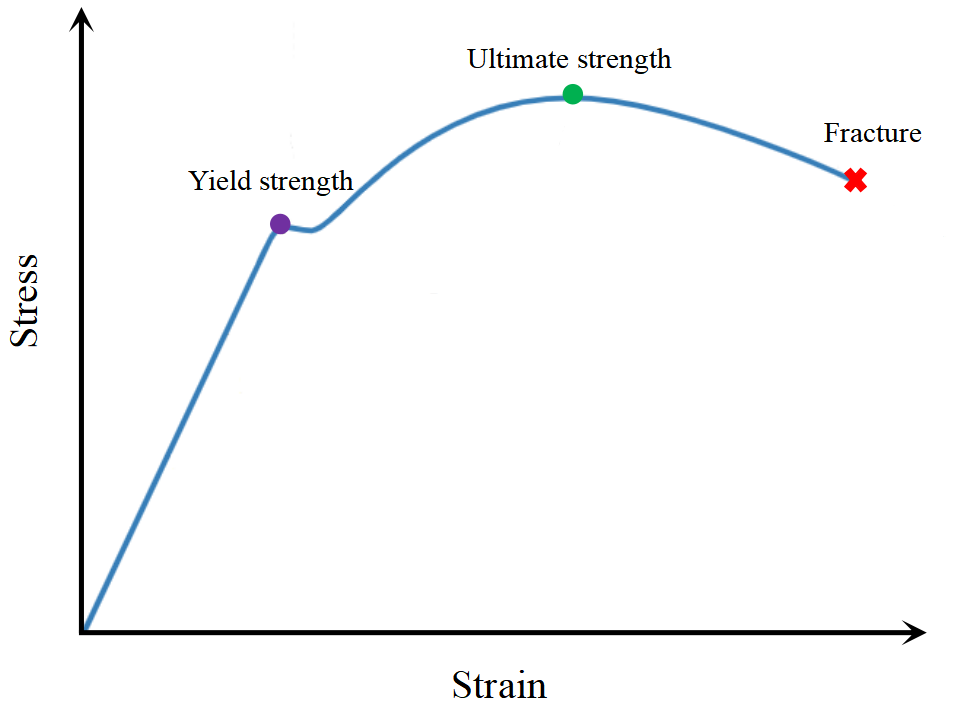}
\caption{Typical stress–strain curve for a metallic material}
\label{fig2}
\end{figure}

A representative stress–strain curve for metallic materials under uni-axial tensile loading is illustrated in Fig. \eqref{fig2}. Before the yield strength, the material deforms elastically, with stress and strain maintaining a linear relationship, and the deformation is fully recoverable upon unloading. Once the applied stress exceeds the yield strength, the material enters the plastic region, where deformation becomes permanent and irreversible. In engineering structural design, the stress level is usually strictly controlled below the yield strength of the material to ensure that the structure remains elastic throughout its service life. 

In yield analysis, equivalent stress measures such as the von Mises and Tresca criteria are extensively applied to predict the initiation of plastic deformation in ductile metallic materials. The constraint functions of these two yield criteria at the element level are expressed as follows:

\begin{equation}\label{Eq. (14)}
\left\{\begin{array}{ll}
g =\sqrt{\left(\boldsymbol{\sigma}\right)^{T} \boldsymbol{M} \boldsymbol{\sigma}} \leq \bar{\sigma} & \text { (von Mises criterion) } \\
\\
g =\sigma^{1}-\sigma^{3} \leq \bar{\sigma} & \text { (Tresca criterion) }
\end{array}\right.
\end{equation}
where $\boldsymbol{\sigma}=\left[\begin{array}{llllll}
\sigma_{xx} & \sigma_{yy} & \sigma_{zz} & \tau_{xy}  & \tau_{yz}  & \tau_{xz}
\end{array}\right]^{T}$ is Cauchy stress vector of the element, $\sigma^{1}$ and $\sigma^{3}$ respectively represent the maximum and minimum principal stresses in the element level, $\bar{\sigma}$ stands for the yield stress of the material and $\boldsymbol{M}$ equals to

\begin{equation}\label{Eq. (15)}
\boldsymbol{M}=\left[\begin{array}{rrcrrr}
1 & -0.5 & -0.5 & \ \ 0 & \ \ 0 & \ \ 0 \\
-0.5 & 1 & -0.5 & 0 & 0 & 0 \\
-0.5 & -0.5 & \ \ \ \ 1 & 0 & 0 & 0 \\
0 & 0 & \ \ \ \ 0 & 3 & 0 & 0 \\
0 & 0 & \ \ \ \ 0 & 0 & 3 & 0 \\
0 & 0 & \ \ \ \ 0 & 0 & 0 & 3
\end{array}\right]
\end{equation}
in three-dimensional case. 

The von Mises criterion is derived from the second deviatoric stress invariant, whereas the Tresca criterion is based on the maximum difference between principal stresses. Experimental evidence has shown that, for most ductile metallic materials, the von Mises criterion provides a more accurate description of yielding behavior under complex stress states (\cite{aleksandrova_efficiency_2019, jin_yielding_2023}). In contrast, the Tresca criterion tends to predict yielding at lower stress levels, leading to overly conservative designs and inefficient utilization of material strength (\cite{shi_equivalent_2025}). Furthermore, the Tresca yield criterion contains non-differentiable regions, which may cause challenges for topology optimization that rely on gradient-based algorithm. Although it can be smoothed into a differentiable form, such approximations do not fully preserve the physical significance of the original Tresca criterion (\cite{giraldo-londono_unified_2020}). Therefore, this study considers the von Mises criterion for the design of metamaterial microstructures under static load conditions.

\subsection{Fatigue criteria}\label{subsec3.2}
\begin{figure}[ht]%
\centering
\includegraphics[width=0.6\textwidth]{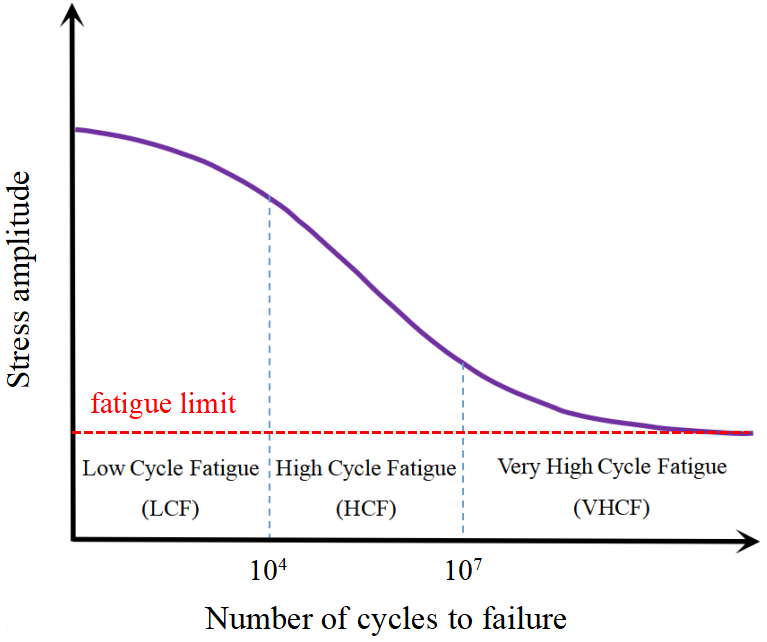}
\caption{General representation of the S-N curve}
\label{fig4}
\end{figure}

Multi-axial fatigue is a prevalent issue in many engineering structures and components. Accurate prediction of fatigue life is essential to ensure their long-term structural integrity and reliable performance. The fatigue behavior of ductile metals is typically characterized by the $S-N$ curve, which illustrates the relationship between the applied stress amplitude and the number of cycles to failure, as shown in Fig \eqref{fig4}. The fatigue behavior can be generally categorized into low-cycle fatigue (LCF), high-cycle fatigue (HCF) and very high-cycle fatigue (VHCF), based on the number of loading cycles to failure and the associated strain levels.

The LCF typically occurs under high stress or strain conditions that induce significant plastic deformation, with fatigue life generally below $10^4$ cycles. The HCF takes place under stress levels well below the yield strength, where the material deformation remains predominantly elastic. The fatigue life typically ranges from $10^4$ to $10^7$ cycles. The VHCF refers to fatigue failure that occurs beyond $10^7$ loading cycles under extremely low stress amplitudes. In such cases, failure often initiates from internal defects rather than surface cracks. This study focuses on the topology optimization of metamaterial microstructures under HCF conditions, as HCF is the predominant failure mode for metallic components under typical service loading scenarios (\cite{liu_review_2024}).

Among the widely studied multiaxial fatigue criteria, the critical plane approach represents an important category. This approach is formulated based on a function defined over a series of planes, which quantifies the fatigue loading induced by the stress history on that plane. The plane that experiences the maximum fatigue loading, referred to as the critical plane, is assumed to control fatigue crack initiation. The corresponding effective stress provides a scalar measure of this loading, and fatigue failure is expected to occur when it exceeds a critical threshold. Several representative critical plane-based criteria, namely the Findley \cite{findley_theory_1959}, Matake \cite{matake_explanation_1977} and Dang Van \cite{Dang_Van_pdf_1989} criteria, are summarized in Table \eqref{tab2}. For a comprehensive overview of various HCF models, please refer to the cited literature (\cite{socie_2000_multiaxial}).

\begin{table}[!ht]
\begin{center}
\renewcommand{\arraystretch}{1.5}
\caption{Stress-based HCF criteria}\label{tab2}
\begin{tabular}{@{}lll@{}}  
  \toprule
  Model & Formula & Parameter \\
  \midrule
  Findley criterion 
  & $g=\max\limits_{\theta}\left(\tau_{a}+\alpha_{f} \sigma_{n,\max}\right) \leq \beta_{f}$    
  & $\begin{aligned}
      \alpha_{f}&=\frac{2-\frac{f_{-1}}{t_{-1}}}{2 \sqrt{\frac{f_{-1}}{t_{-1}}-1}}\\
      \beta_{f}&=\frac{f_{-1}}{2 \sqrt{\frac{f_{-1}}{t_{-1}}-1}}
     \end{aligned}$ \\

  \midrule
  Matake criterion
  & $g=\max\limits_{\theta}\left(\tau_{a}\right)+\alpha_{m} \sigma_{n,\max} \leq \beta_{m}$    
  & $\begin{aligned}
      \alpha_{m}&=\frac{2t_{-1}}{f_{-1}}-1\\
      \beta_{m}&=t_{-1}
     \end{aligned}$ \\

  \midrule
  Dang Van criterion
  & $g=\max\limits_{\theta}\left(\tau_{a}\right)+\alpha_{dv} \sigma_{H,\max} \leq \beta_{dv}$   
  & $\begin{aligned}
      \alpha_{dv}&=\frac{3t_{-1}}{f_{-1}}-\frac{3}{2}\\
      \beta_{dv}&=t_{-1}
     \end{aligned}$ \\

  \bottomrule
\end{tabular}
\end{center}
\end{table}

The material parameters $\alpha$ and $\beta$ in Table \eqref{tab2} can be determined through uni-axial fatigue tests, i.e., the fully reversed torsional fatigue limit $t_{-1}$ and the fully reversed bending fatigue limit $f_{-1}$, $\tau_{a}$ is the amplitude of shear stress, $\sigma_{n,\max}$ and $\sigma_{H,max}$ represent the maximum normal stress and the maximum hydrostatic stress, respectively. During the cyclic loading $0\leq t\leq T$, $\sigma_{H,max}$ can be calculated as follows:

\begin{equation}\label{Eq. (16)}
\sigma_{H, max}= \max\limits_{t} \left[\sigma_{x x}(t)+\sigma_{y y}(t)+\sigma_{z z}(t)\right]/3
\end{equation}

\begin{figure}[ht]%
\centering
\includegraphics[width=0.9\textwidth]{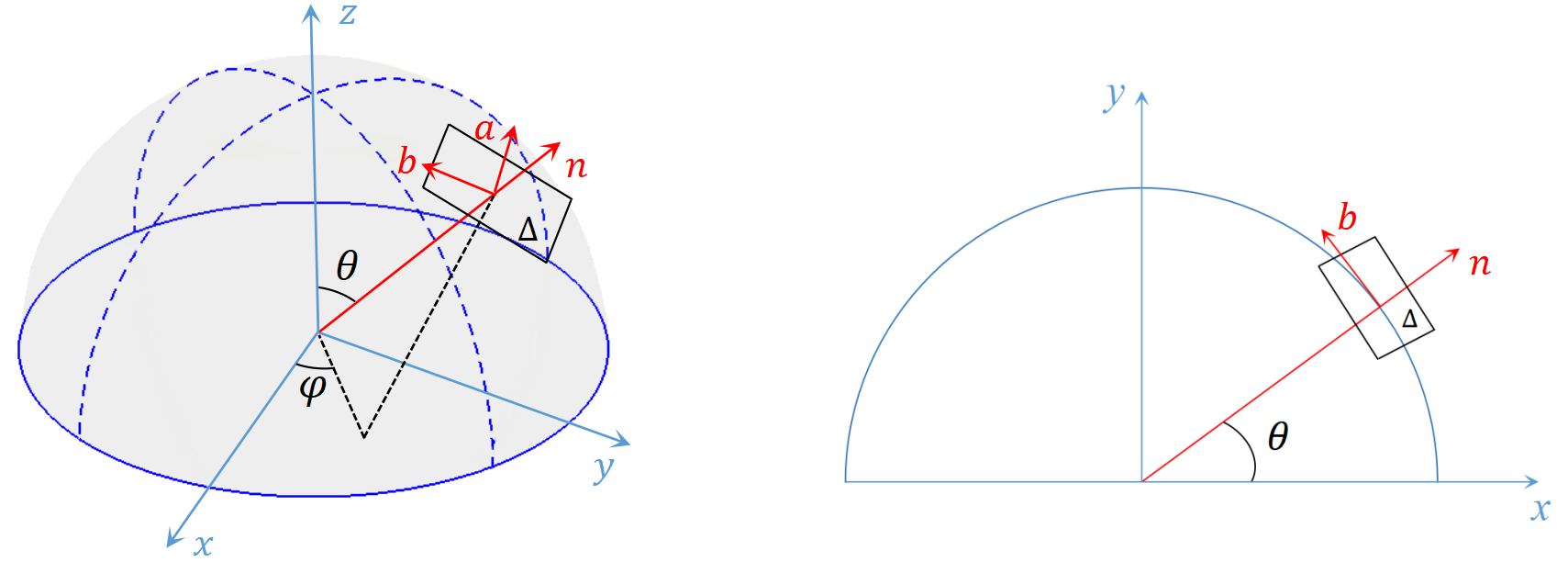}
\caption{Critical plane search space in 3D (left) and 2D (right) } 
\label{fig5}
\end{figure}

The critical plane search is plotted in Fig \eqref{fig5}. At each position of interest, a general plane $\Delta$ can be defined by its unit normal vector $\boldsymbol{n}$, which is characterized by the spherical angles $\theta$ and $\varphi$ in three-dimensional state. Here, $\theta$ is the angle between $\boldsymbol{n}$ and the $z$ axis, while $\varphi$ is the angle between the projection of $\boldsymbol{n}$ onto the $a-b$ plane and the $x$ axis. In the case of plane stress, the orientation of the plane is determined only by $\theta$, which corresponds to the angle between $\boldsymbol{n}$ and the $x$ axis. Since the negative normal vector describes the same plane, the critical plane search space is further reduced to a hemisphere in 3D and a semicircle in 2D.

Given the orientation of a generic material plane $\Delta$, the instantaneous stress components $\boldsymbol{\sigma}(t) = \left[\begin{array}{llllll}
\sigma_{n}(t) & \tau_{na}(t) & \tau_{nb}(t) \end{array}\right]^{T}$ acting on this plane can be calculated by projecting the stress tensor from the global coordinate system $(x,y,z)$ to the local coordinate system $(n,a,b)$:

\begin{equation}\label{Eq. (18)}
\left[\begin{array}{c}
\sigma_{n}(t) \\
\tau_{na}(t) \\
\tau_{nb}(t)
\end{array}\right]=\left[\begin{array}{ccc}
n_x & n_y & n_z \\
a_x & a_y & a_z \\
b_x & b_y & b_z
\end{array}\right]\left[\begin{array}{ccc}
\sigma_{xx}(t) & \tau_{xy}(t) & \tau_{xz}(t) \\
\tau_{xy}(t) & \sigma_{yy}(t) & \tau_{yz}(t) \\
\tau_{xz}(t) & \tau_{yz}(t) & \sigma_{zz}(t)
\end{array}\right]\left[\begin{array}{c}
n_x \\
n_y  \\
n_z
\end{array}\right]
\end{equation}
where

\begin{equation}\label{Eq. (19)}
\left[\begin{array}{ccc}
n_x & n_y & n_z \\
a_x & a_y & a_z \\
b_x & b_y & b_z
\end{array}\right] = \left[\begin{array}{ccc}
\sin (\theta) \cos (\varphi) & \sin (\theta) \sin (\varphi) & \cos (\theta) \\
\sin (\varphi) & -\cos (\varphi) & 0 \\
\cos (\theta) \cos (\varphi) & \cos (\theta) \sin (\varphi) & -\sin (\theta)
\end{array}\right]
\end{equation}
are the direction cosines of the local basis vectors with respect to the global coordinate.

In plane stress conditions, the transformation of the stress vector becomes:

\begin{equation}\label{Eq. (20)}
\left[\begin{array}{c}
\sigma_{n}(t) \\
\sigma_{b}(t) \\
\tau_{nb}(t)
\end{array}\right]=
\left[\begin{array}{ccc}
\cos (\theta)^{2} & \sin (\theta)^{2} & \sin (2\theta) \\
\sin (\theta)^{2} & \cos (\theta)^{2} & -\sin (2\theta) \\
-\frac{1}{2}\sin (2\theta) & \frac{1}{2}\sin (2\theta) & \cos (2\theta)
\end{array}\right]\left[\begin{array}{c}
\sigma_{xx}(t)\\
\sigma_{yy}(t)\\
\tau_{xy}(t)
\end{array}\right]
\end{equation}

\section{Numerical results}
\label{sec4}
\begin{figure}[ht]%
\centering
\includegraphics[width=0.6\textwidth]{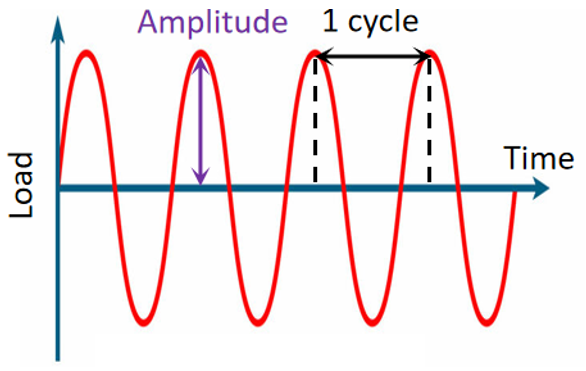}
\caption{Sinusoidal cyclic loading} 
\label{fig6}
\end{figure}

This section presents several numerical examples to demonstrate the effectiveness of the proposed method. The material considered is additive manufactured Ti-6Al-4V (\cite{mower_mechanical_2016,fatemi_torsional_2017}), with the mechanical properties listed in Table \eqref{tab3}. This titanium alloy is known for its high strength, stiffness, and excellent corrosion resistance, making it suitable for manufacturing complex geometries used in the aerospace and biomedical industries. The design domain is a cube with dimensions of 10 $mm$ × 10 $mm$ × 10 $mm$, which degenerates to a square of 10 $mm$ × 10 $mm$ in 2D. During fatigue analysis, the structure is subjected to a sinusoidal waveform with constant amplitude (Fig \eqref{fig6} and is designed to endure $10^{5}$ cycles without the initiation of fatigue cracks. Under such cyclic loading, the critical plane search can be carried out by discretizing the hemisphere using predefined angular increments, such as $\delta \theta = \delta \varphi = 5^\circ$. In the 2D case, the search can be simplified to a smaller angular increment, e.g., $\delta \theta = 1^\circ$. A right-handed Cartesian coordinate system is employed throughout this study, where the $x$-axis denotes the horizontal direction, the $y$-axis the vertical direction, and the $z$-axis is normal to the $xy$-plane, pointing outward.

\begin{table}[!ht]
\centering
\caption{Mechanical parameters of Ti-6Al-4V alloy}
\label{tab3}
\sbox{\tablebox}{%
  \begin{tabular}{@{}lll@{}}
    \toprule
    Parameter & Description & Value \\
    \midrule
    $E_0$    & Young’s modulus & 108.8 GPa \\
    $\nu$    & Poisson’s ratio & 0.29 \\
    $\overline{\sigma}^{\dagger}$ & Yield stress & 972 MPa \\
    $f_{-1}$ & Fully reversed bending fatigue limit & 454 MPa \\
    $t_{-1}$ & Fully reversed torsional fatigue limit & 300 MPa \\
    \bottomrule
  \end{tabular}%
}
\usebox{\tablebox}
\vspace{3pt} 
\begin{minipage}{\wd\tablebox}
  \footnotesize\raggedright
  $^{\dagger}$The yield stress under pure shear is 561.18 MPa.
\end{minipage}
\end{table}

\subsection{Maximization of the bulk modulus}\label{subsec4.1}
The first example involves maximizing the material bulk modulus under hydrostatic loading. A uniform triaxial strain state $\boldsymbol{\varepsilon} = [-0.5\% \ \ -0.5\% \ \ -0.5\% \ \ 0 \ \ 0 \ \ 0]^{T}$ is imposed equally along all three principal axes in 3D case. Under the plane stress assumption, the load can be simplified to equi-biaxial strain $\boldsymbol{\varepsilon} = [-0.5\% \ \ -0.5\% \ \ 0]^{T}$ along the $x$ and $y$ directions. The bulk modulus quantifies the strength of the material in resisting elastic deformation under hydrostatic pressure. The objective function $c$ in the AL function in Eq. \eqref{Eq. (10)} can be expressed as follows:

\begin{equation}\label{Eq. (21)}
c(\overline{\boldsymbol{\rho}})=-\sum_{i, j=1}^{D} C_{i j}^{H}(\overline{\boldsymbol{\rho}})
\end{equation}
where $D$ represents the spatial dimension ($D=3$ in 3D and $D=2$ in 2D).

To demonstrate the effectiveness of the proposed method in mitigating stress concentrations for metamaterial microstructure design, Table \eqref{tab4} presents a comparison of two-dimensional topologies obtained without and with stress constraints at a volume fraction of $V_f = 0.6$. The table displays the resulting RUC (left), their corresponding normalized von Mises stress distributions (center), and the homogenized constitutive matrices $\boldsymbol{C}^H$ (right). It can be observed that the design without stress considerations results in a nearly circular central void, whereas the stress-constrained topology yields a square-like cavity with rounded corners. This trend is consistent with the findings reported by Collet et al.(\cite{collet_topology_2018}). Compared with the compliance-driven topology, the introduction of stress constraints leads to a decrease in the objective function value from $1.27$ to $1.23$ ($\times 10^{5}$), while a more significant effect lies in the reduction of the maximum von Mises stress from $1139.96 \ \text{MPa}$ to $971.58 \ \text{MPa}$ (a $14.8\%$ decrease). This considerable mitigation in stress concentration contributes directly to a notable enhancement in structural reliability.

\begin{table}[ht]
\centering
\caption{2D metamaterial microstructures with maximum bulk modulus ($V_f = 0.6$)}
\renewcommand{\arraystretch}{0.5}
\begin{tabular}{c m{2.5cm} m{3cm} m{3cm} c}
\toprule
Case & \multicolumn{1}{c}{RUC} & \multicolumn{1}{c}{$\frac{\boldsymbol{\sigma}^{vm}}{\bar{\sigma}}$} & \multicolumn{1}{c}{$\boldsymbol{C}^H$} \\
\midrule
\begin{tabular}{@{}c@{}}compliance-\\driven\end{tabular} &
\includegraphics[width=2.5cm]{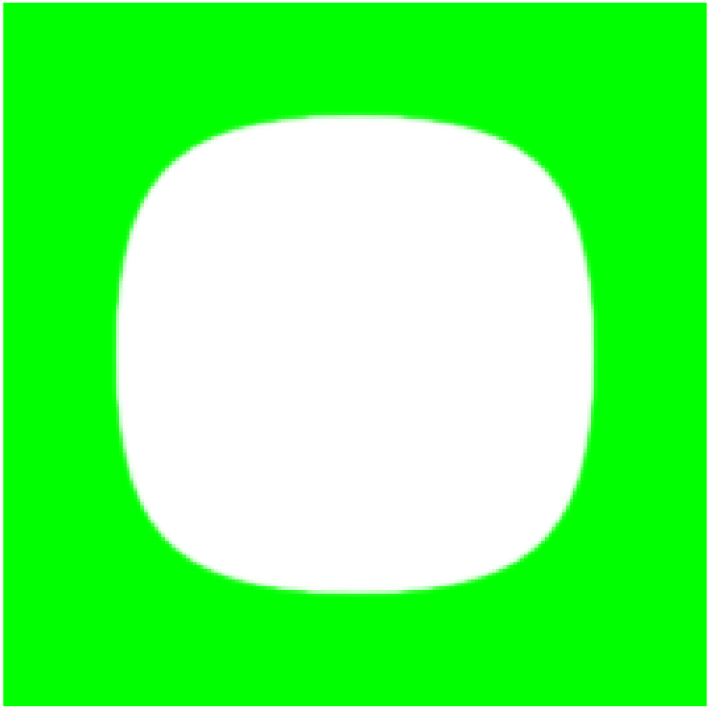} &
\includegraphics[width=3cm]{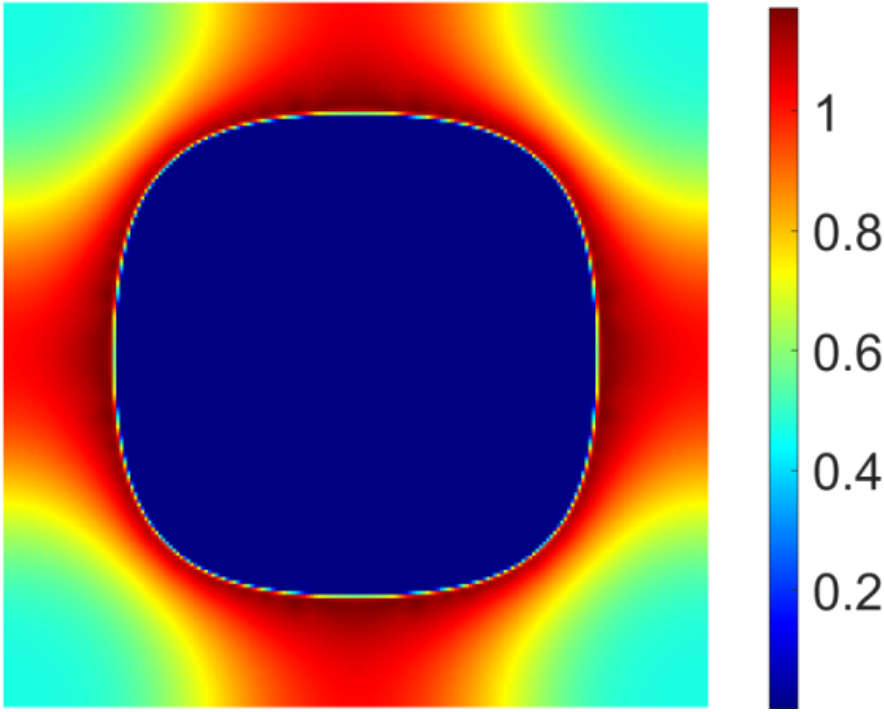} &
\includegraphics[width=3cm]{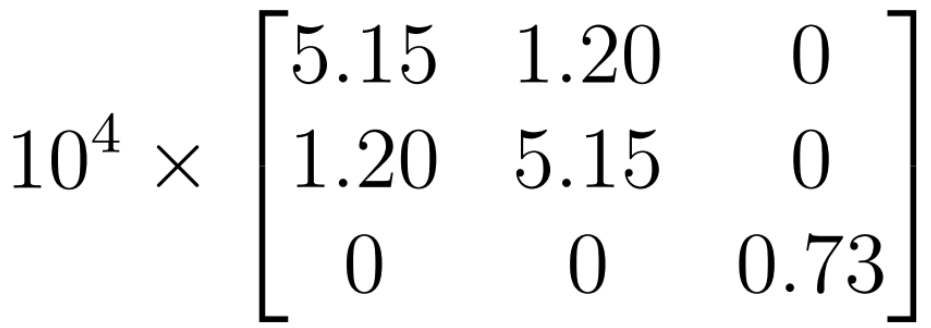} & \\
\midrule
\begin{tabular}{@{}c@{}}stress-\\constrained\end{tabular} &
\includegraphics[width=2.5cm]{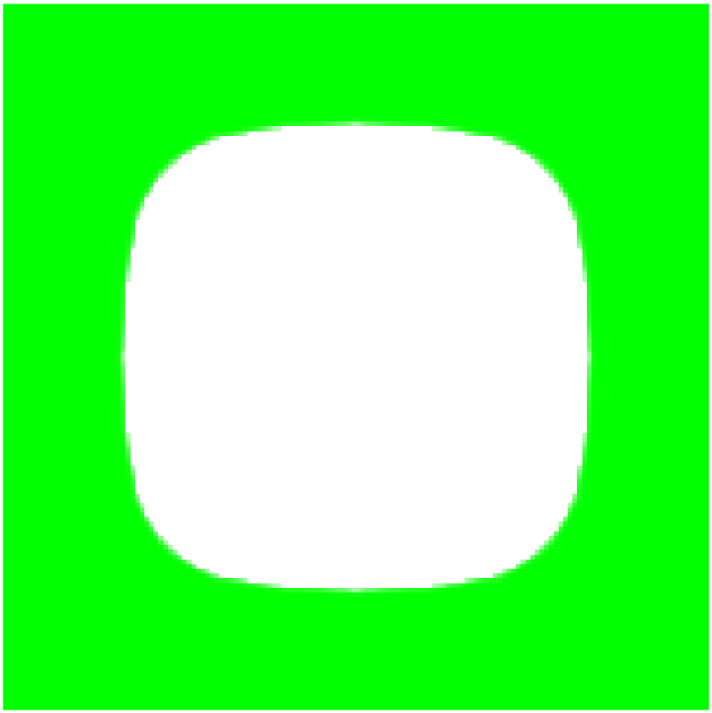} &
\includegraphics[width=3cm]{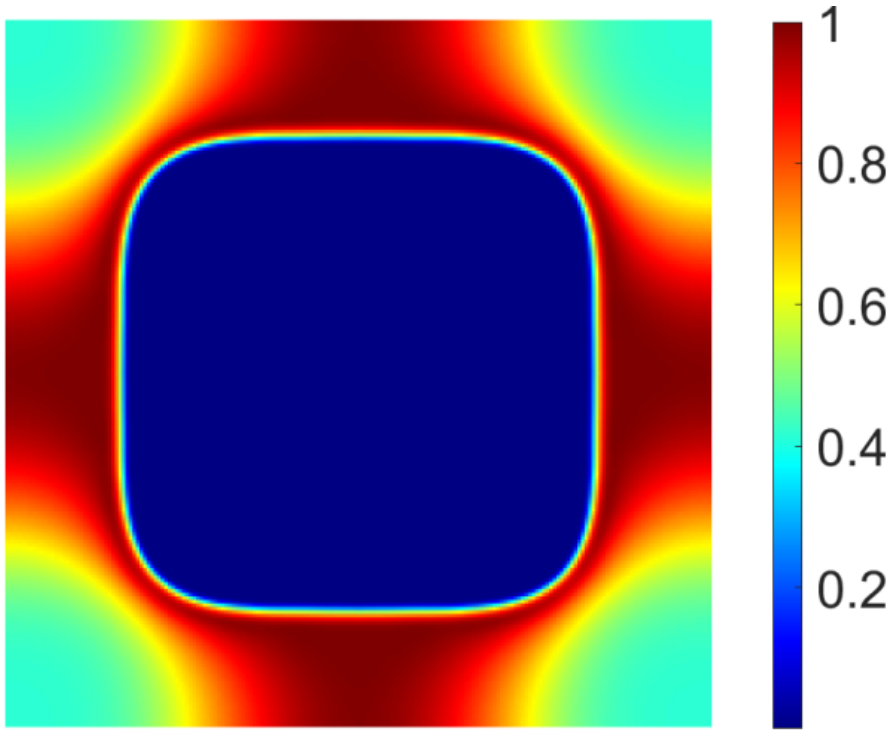} &
\includegraphics[width=3cm]{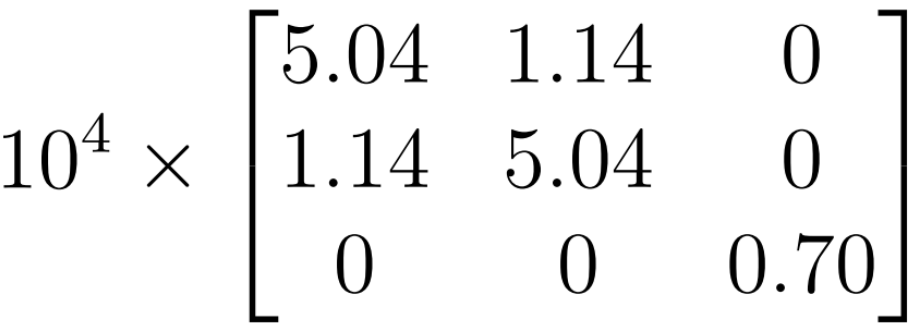} & \\
\bottomrule
\end{tabular}
\label{tab4}
\end{table}

The three-dimensional topologies with von Mises stress constraints at different volume fractions are depicted in Table \eqref{tab5}. To reveal the internal geometric features, a half RUC is plotted in the middle column. A comparison of the topologies clearly demonstrates the structural evolution with increasing volume fraction. At a volume fraction of $0.4$,  the architecture forms a highly porous, star-shaped framework that efficiently distribute loads in multiple directions while maintaining material efficiency. As the volume fraction increases to $0.6$, the structure becomes significantly denser with smaller internal voids. Furthermore, the cross-section at high volume fraction exhibits geometric characteristics similar to those in the 2D case. The topology evolution directly enhances the macroscopic elastic response, as evidenced by the homogenized stiffness matrices: the normal stiffness terms rise from approximately $2.57$ to $4.54$ ($\times 10^{4}$) and the shear components increase from $0.85$ to $1.28$ ($\times 10^{4}$). This uniform enhancement confirms that the denser topology provides superior stiffness in both normal and shear modes.

\begin{table}[!ht]
\centering
\caption{3D stress-constrained metamaterial microstructures with maximum bulk modulus}
\renewcommand{\arraystretch}{0.5}
\begin{tabular}{c m{3cm} m{3cm} m{4cm} c}
\toprule
Case & \multicolumn{1}{c}{RUC} & \multicolumn{1}{c}{Half RUC} & \multicolumn{1}{c}{$\boldsymbol{C}^H$} \\
\midrule
$V_f = 0.4$ &
\includegraphics[width=3cm]{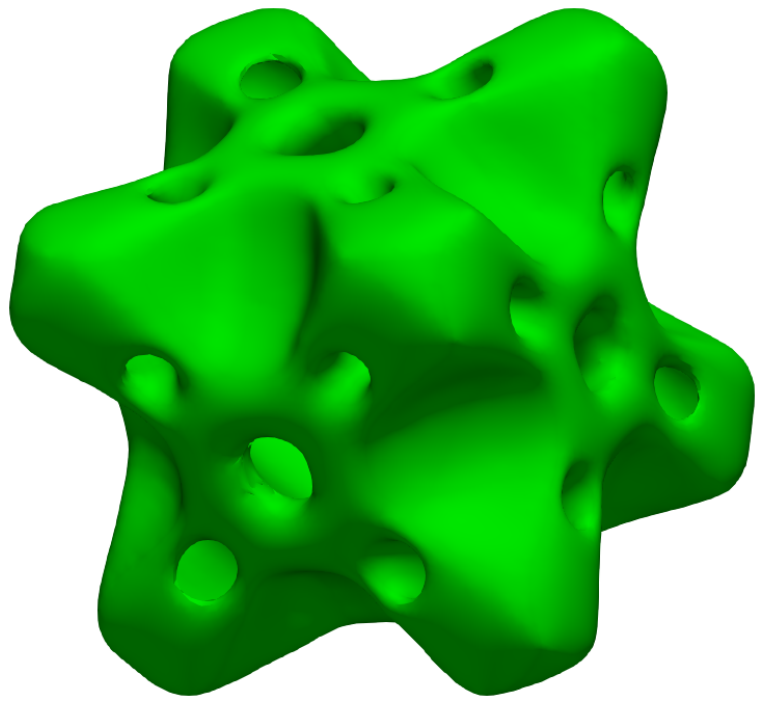} &
\includegraphics[width=3cm]{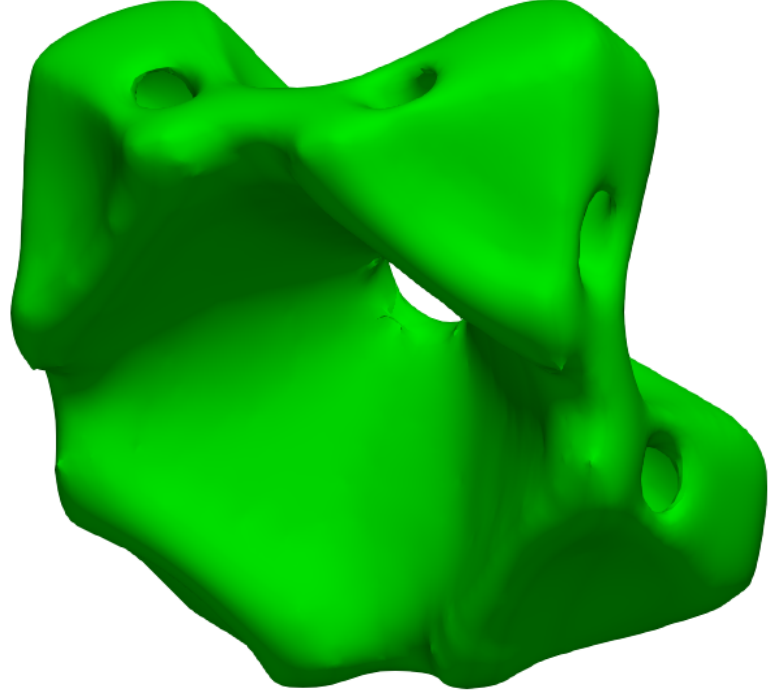} &
\includegraphics[width=4cm]{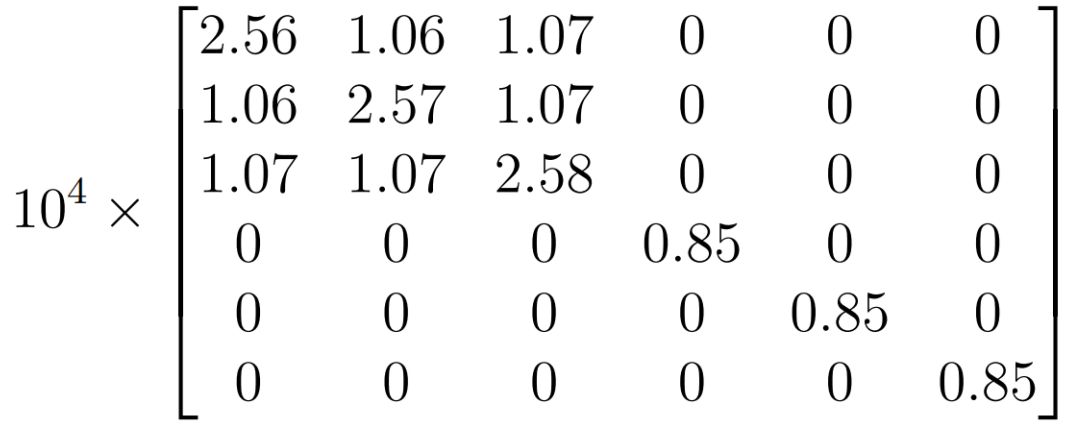} & \\
\midrule
$V_f = 0.6$  &
\includegraphics[width=3cm]{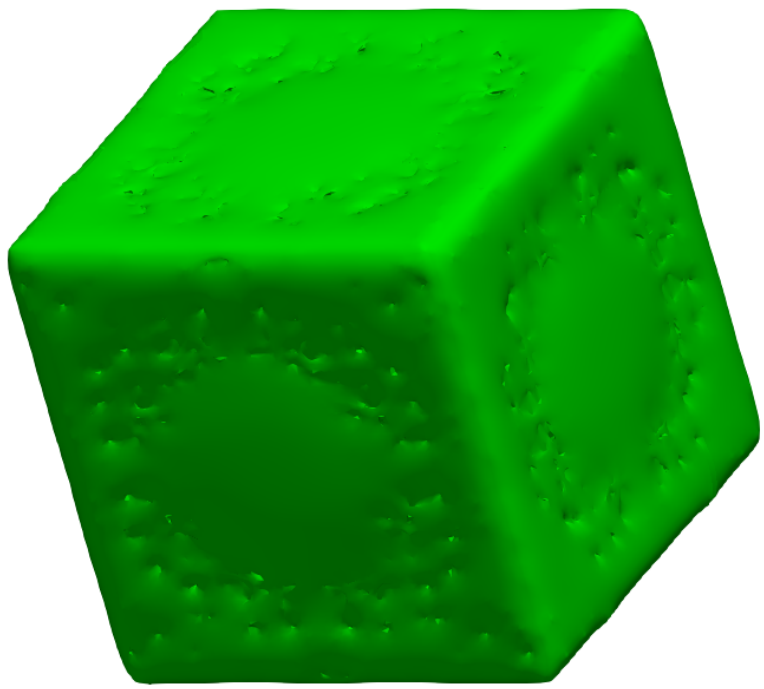} &
\includegraphics[width=3cm]{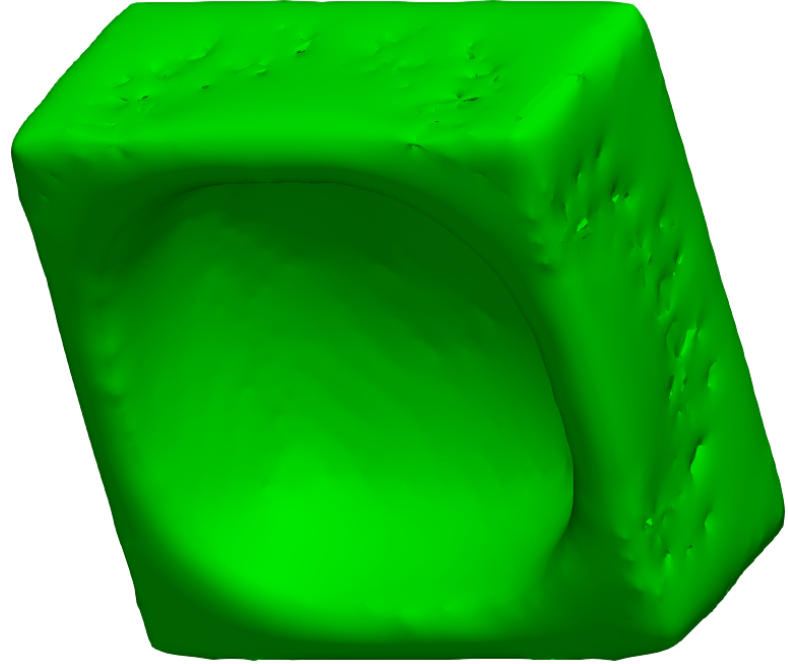} &
\includegraphics[width=4cm]{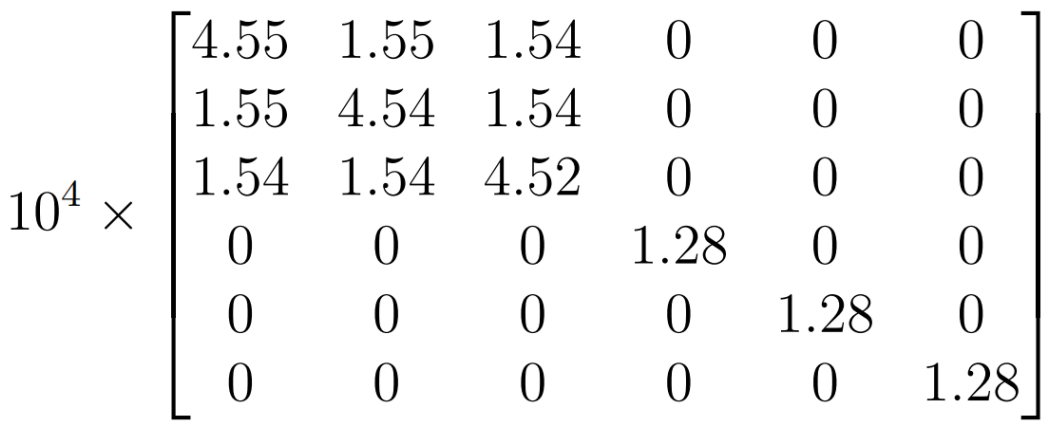} & \\
\bottomrule
\end{tabular}
\label{tab5}
\end{table}

Table \eqref{tab6} presents the three-dimensional metamaterial microstructures optimized for achieving maximum bulk modulus under different fatigue criteria constraints. The applied hydrostatic strain has an amplitude of $0.26\%$ and a zero mean value. Although all topologies are optimized under identical cyclic normal load conditions, their resulting geometries exhibit distinct morphological characteristics associated with each fatigue criterion. The topology optimized under the Findley criterion exhibits a cross-sectional evolution from a quasi-octagonal profile with slightly curved edges at a lower volume fraction $V_f=0.4$ to a more square-like contour as the volume fraction increases to $V_f=0.6$. In contrast, the structure constrained by the Dang Van criterion transforms from a rhombic to an octagonal cross-section. The design obtained under the Matake criterion maintains a consistent nearly square cross-sections across both volume fractions.

\begin{table}[!htbp]
\centering
\caption{3D fatigue-constrained metamaterial microstructures with maximum bulk modulus under cyclic hydrostatic strain loading}
\renewcommand{\arraystretch}{0.1}
\begin{tabular}{c m{2.8cm} m{2.8cm} m{3.8cm} c}
\toprule
Case & \multicolumn{1}{c}{RUC} & \multicolumn{1}{c}{Half RUC} & \multicolumn{1}{c}{$\boldsymbol{C}^H$} \\
\midrule
\multirow{2}{*}{\centering\begin{tabular}{@{}c@{}} Findley \\ $V_f = 0.4$ \end{tabular}} 
& \includegraphics[width=2.8cm]{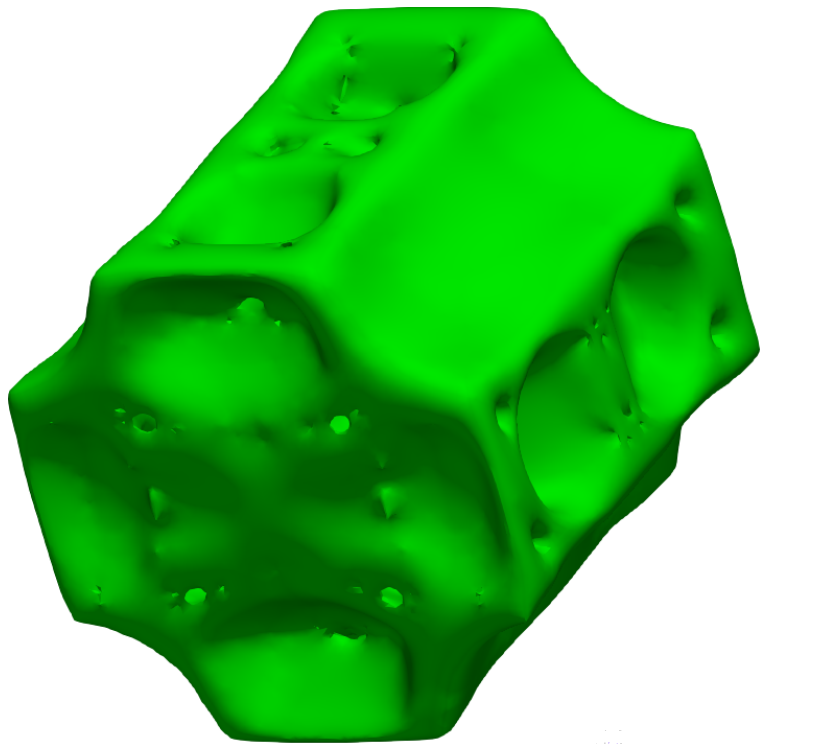} 
& \includegraphics[width=2.8cm]{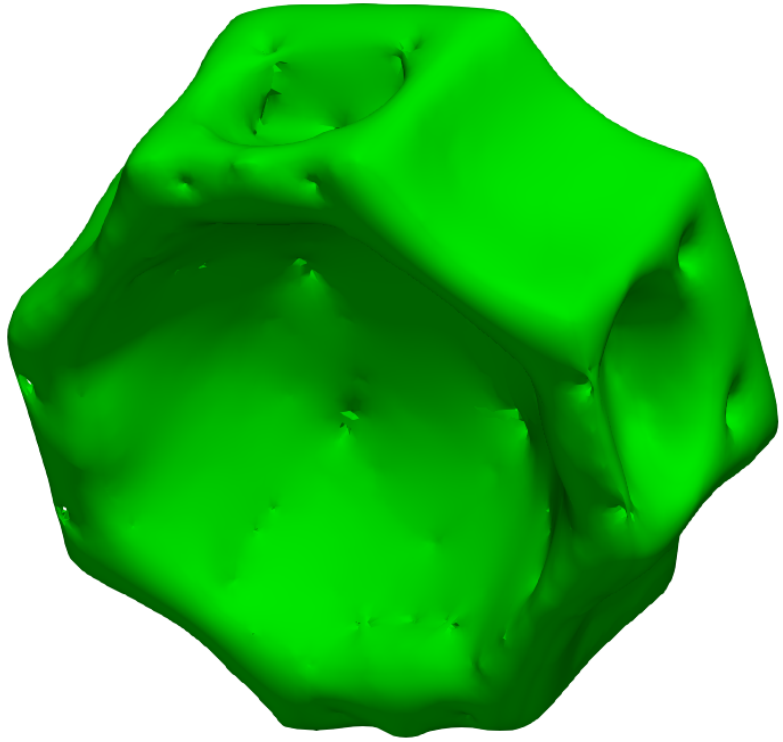} 
& \includegraphics[width=3.8cm]{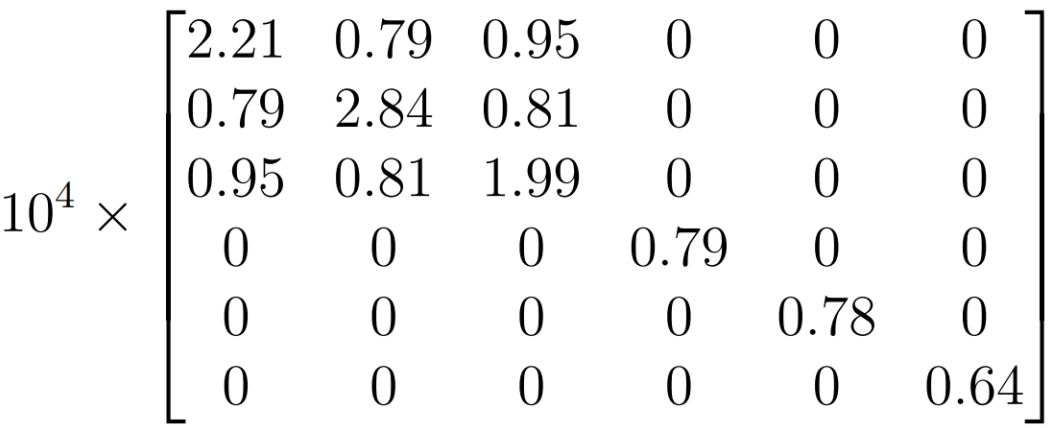} 
& \\
& & & & \\
\midrule
\multirow{2}{*}{\centering\begin{tabular}{@{}c@{}} Findley \\ $V_f = 0.6$ \end{tabular}} 
& \includegraphics[width=2.8cm]{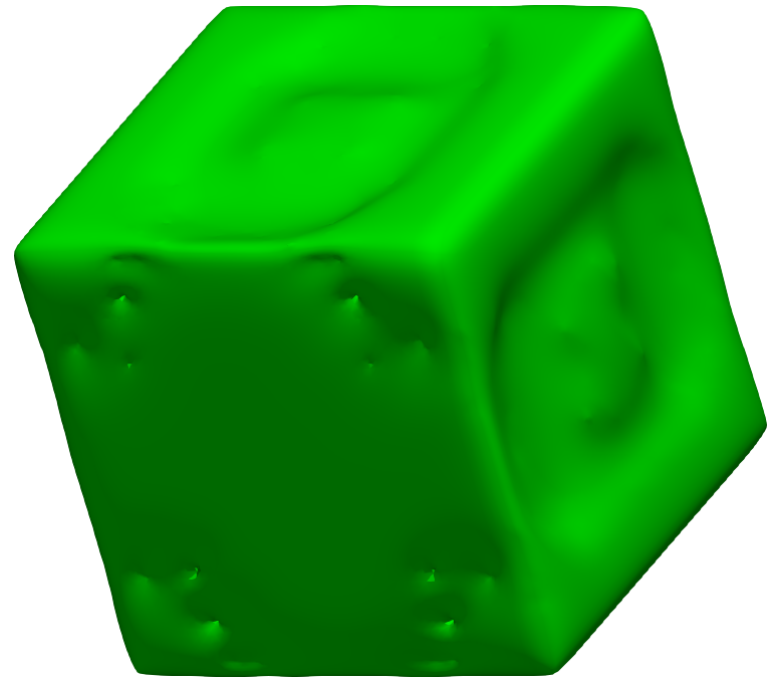} 
& \includegraphics[width=2.8cm]{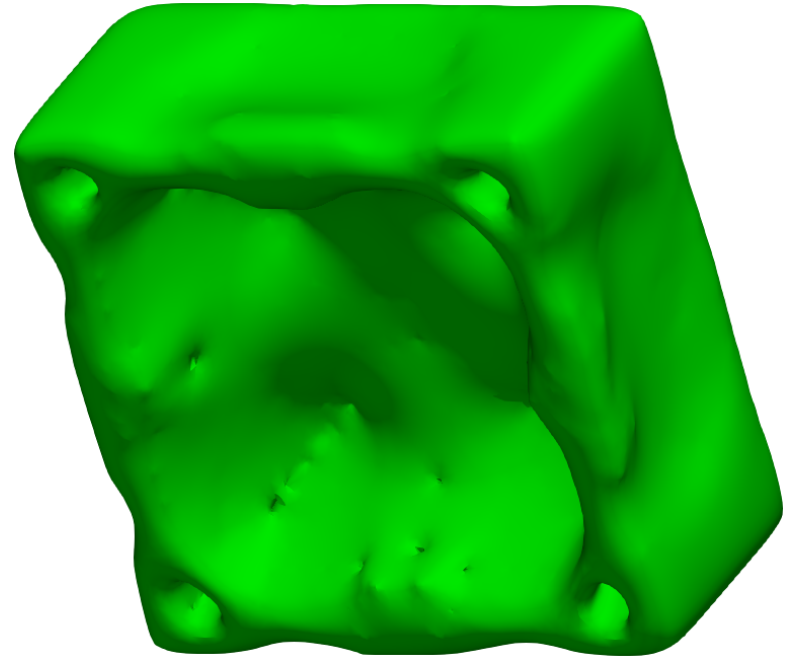} 
& \includegraphics[width=3.8cm]{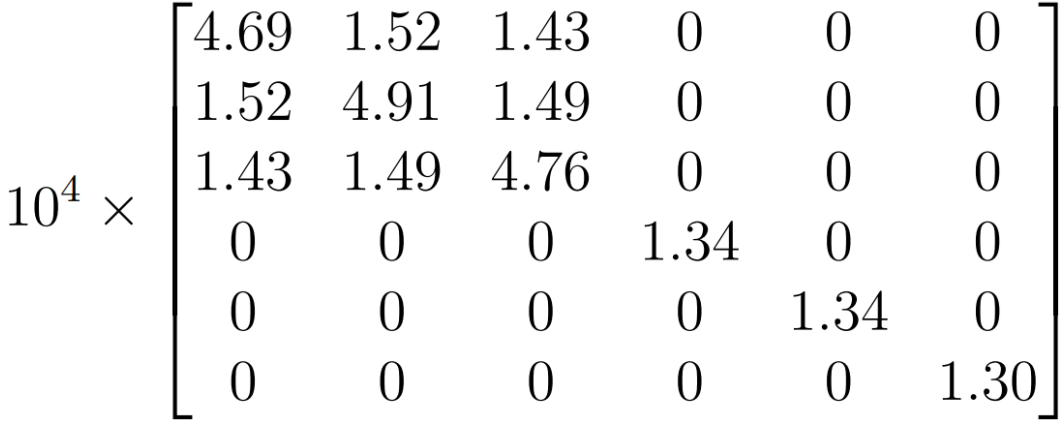} 
& \\
& & & & \\
\midrule
\multirow{2}{*}{\centering\begin{tabular}{@{}c@{}} Matake \\ $V_f = 0.4$ \end{tabular}} 
& \includegraphics[width=2.8cm]{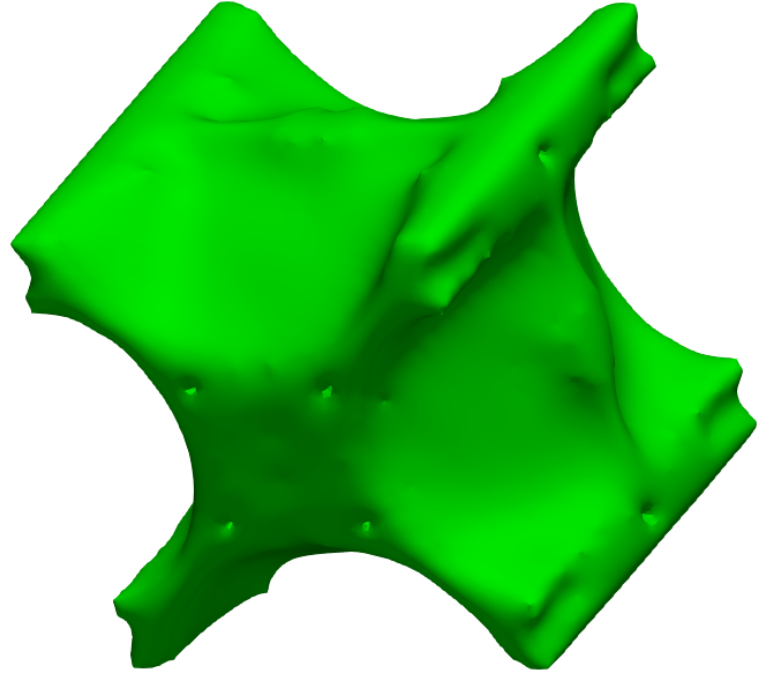} 
& \includegraphics[width=2.8cm]{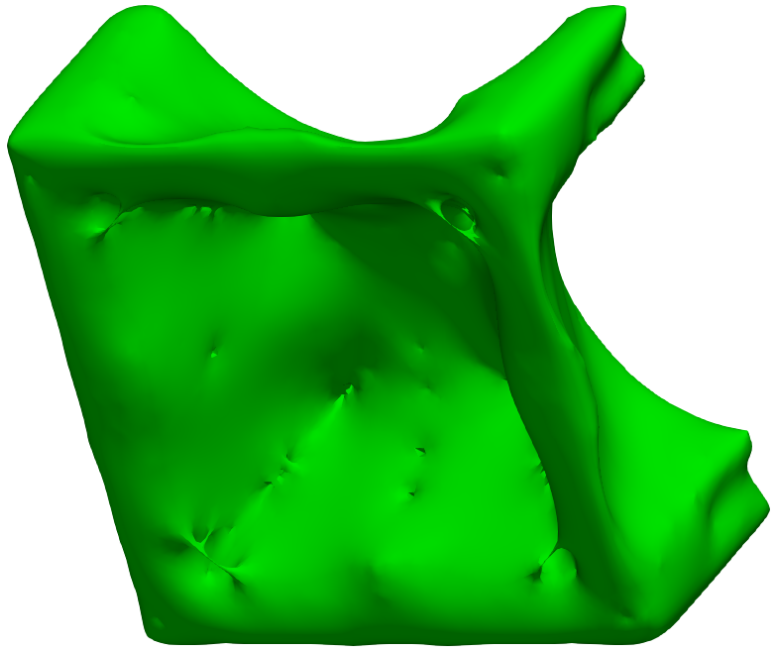} 
& \includegraphics[width=3.8cm]{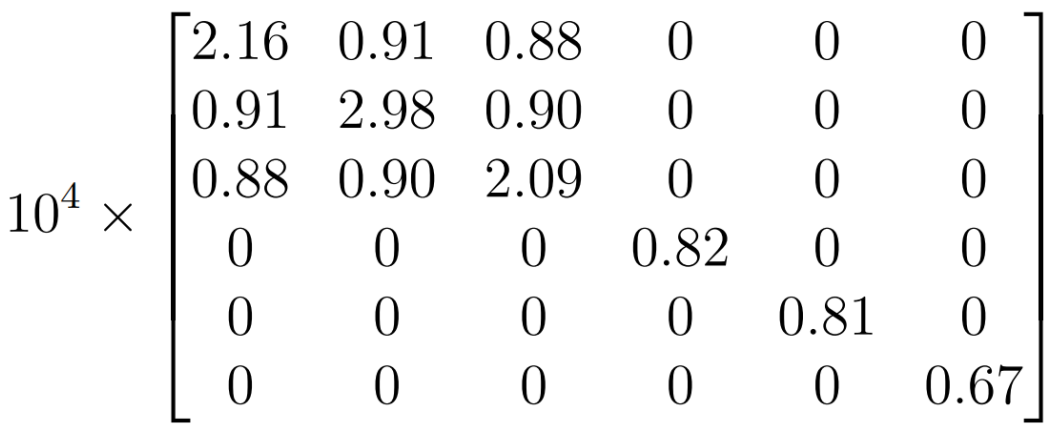} 
& \\
& & & & \\
\midrule
\multirow{2}{*}{\centering\begin{tabular}{@{}c@{}} Matake \\ $V_f = 0.6$ \end{tabular}} 
& \includegraphics[width=2.8cm]{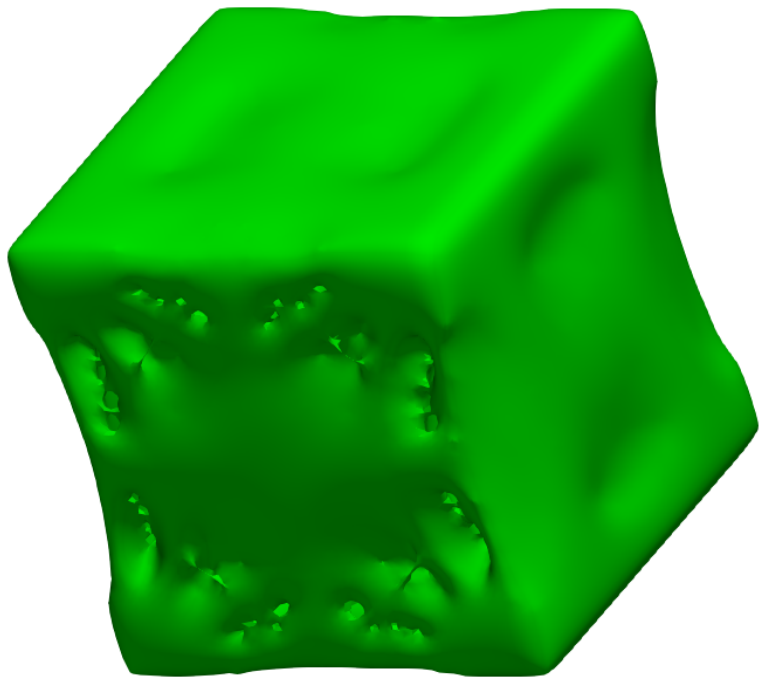} 
& \includegraphics[width=2.8cm]{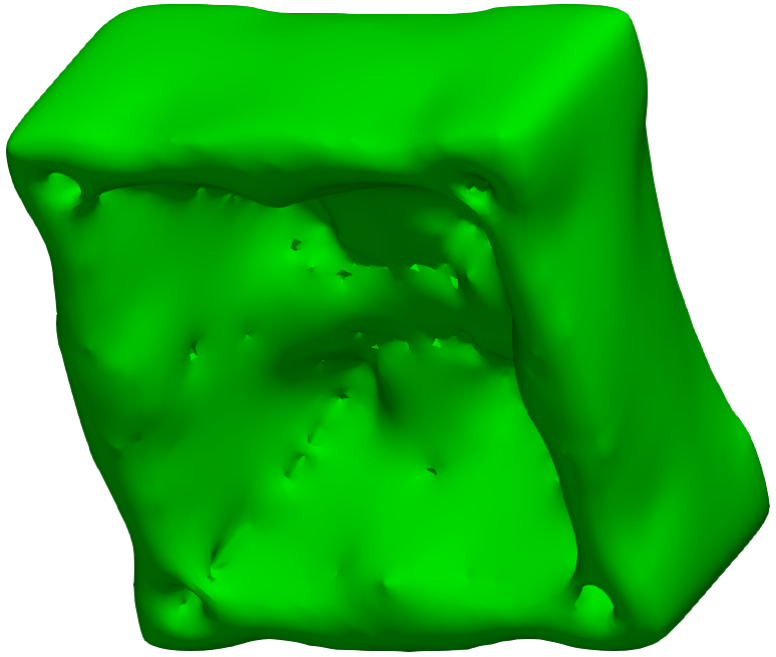} 
& \includegraphics[width=3.8cm]{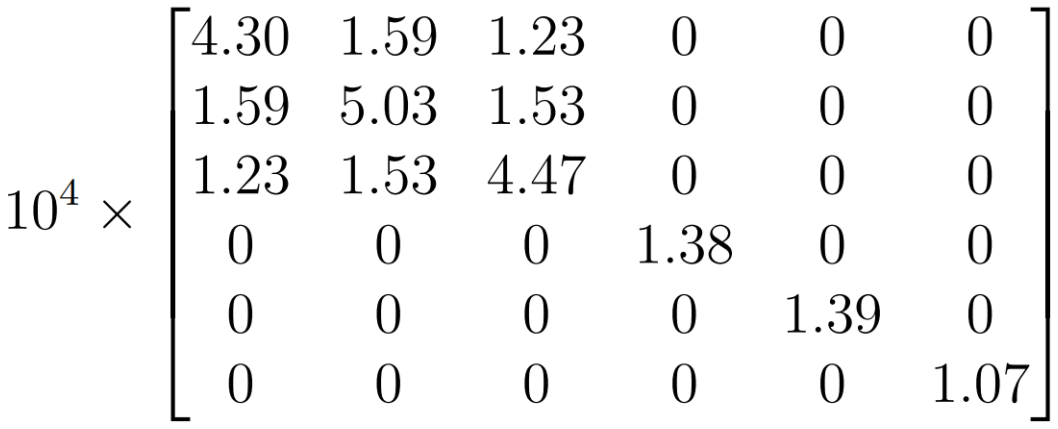} 
& \\
& & & & \\
\midrule
\multirow{2}{*}{\centering\begin{tabular}{@{}c@{}} Dang van \\ $V_f = 0.4$ \end{tabular}} 
& \includegraphics[width=2.8cm]{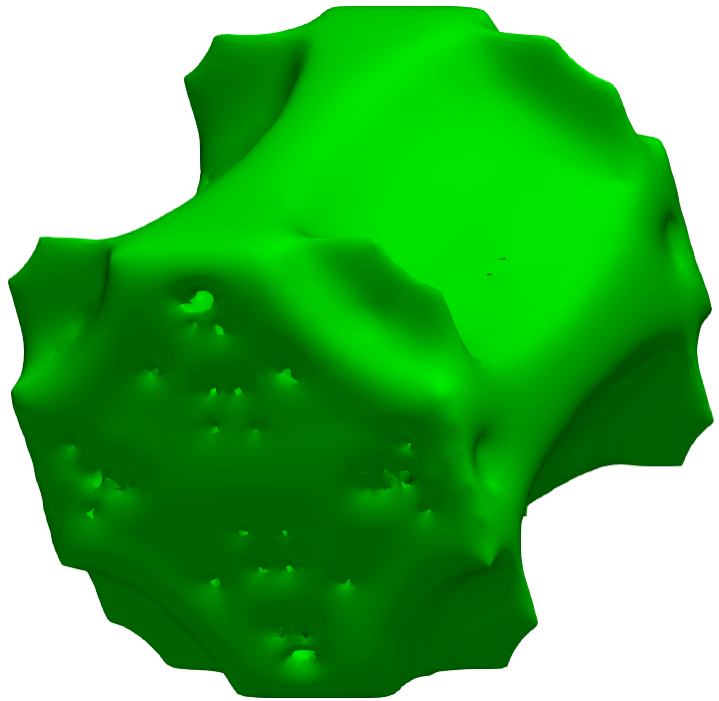} 
& \includegraphics[width=2.8cm]{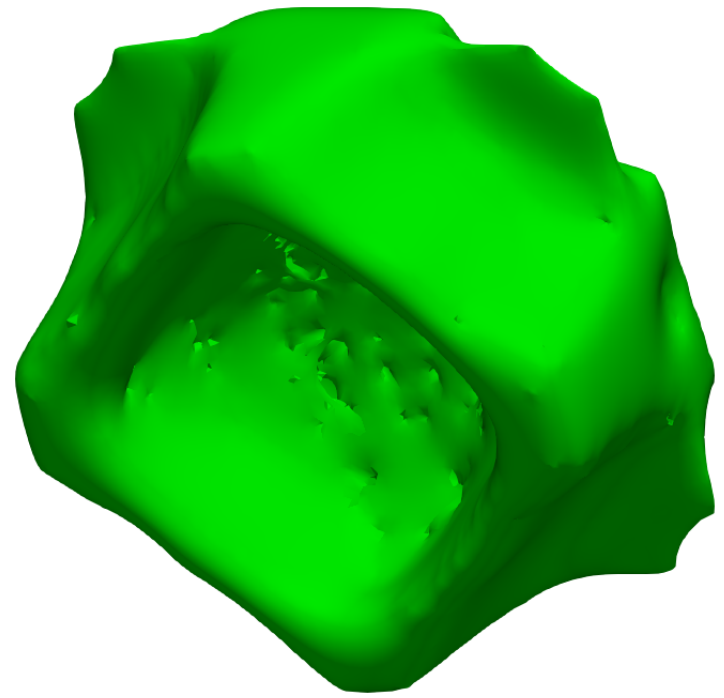} 
& \includegraphics[width=3.8cm]{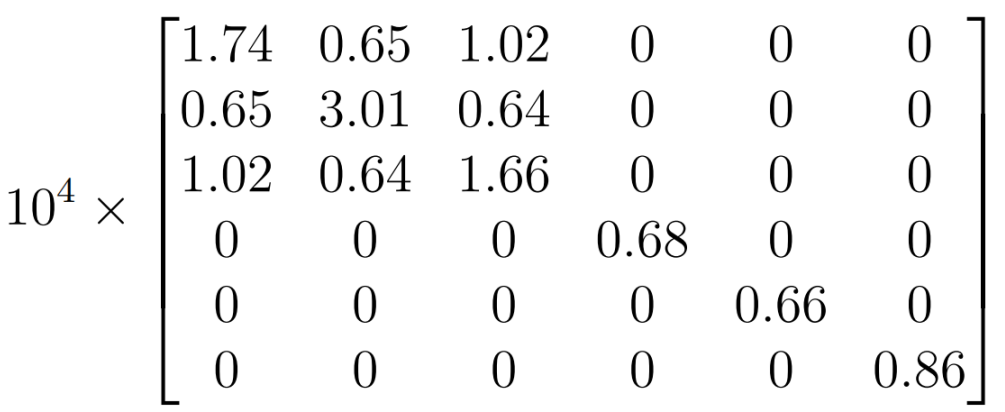} 
& \\
& & & & \\
\midrule
\multirow{2}{*}{\centering\begin{tabular}{@{}c@{}} Dang van \\ $V_f = 0.6$ \end{tabular}} 
& \includegraphics[width=2.8cm]{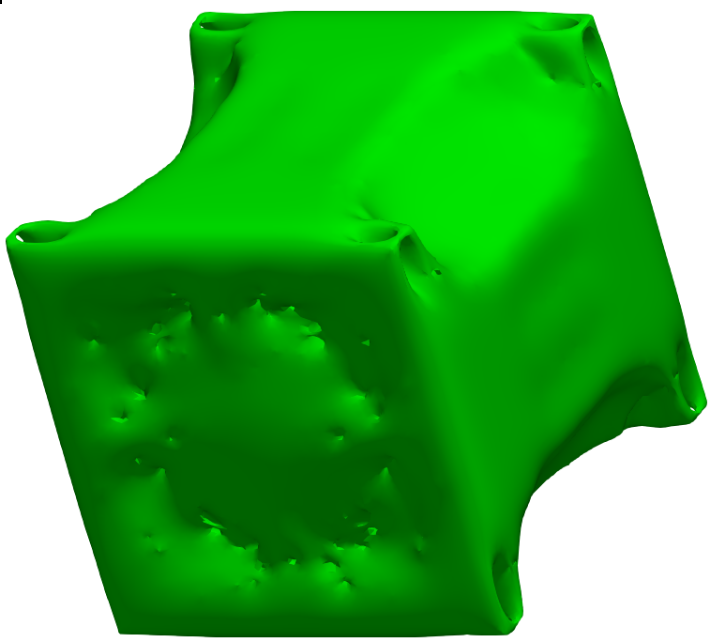} 
& \includegraphics[width=2.8cm]{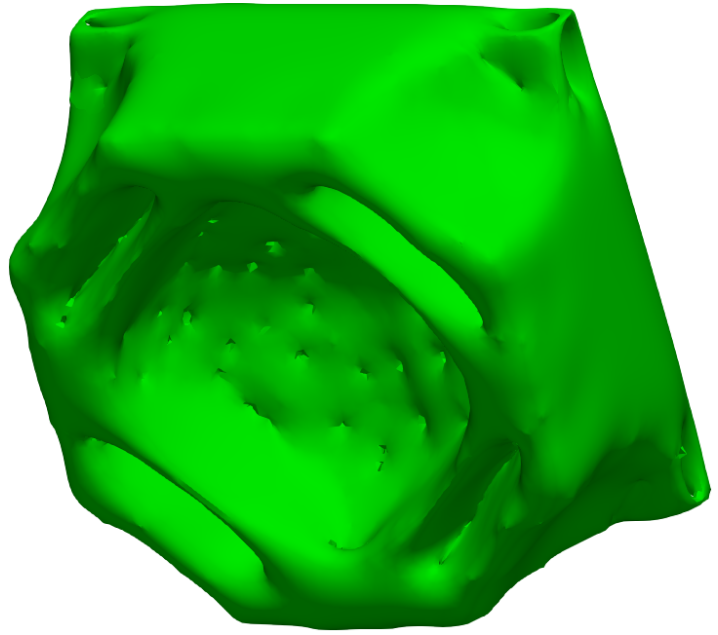} 
& \includegraphics[width=3.8cm]{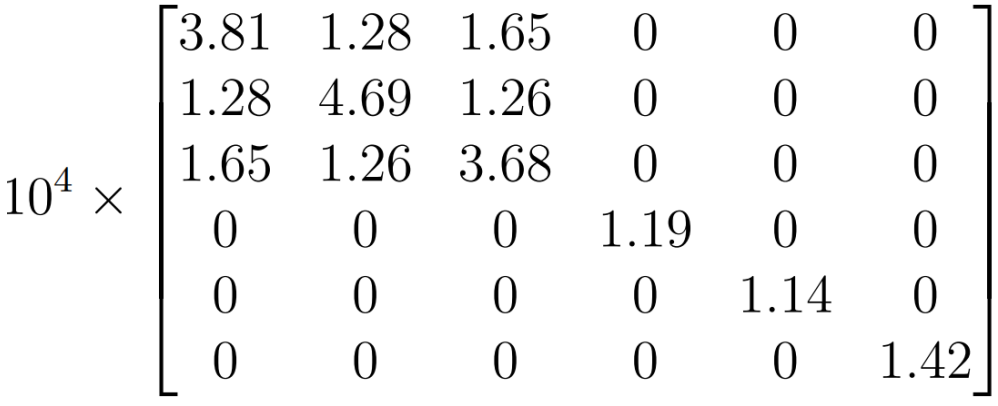} 
& \\
& & & & \\
\bottomrule
\end{tabular}
\label{tab6}
\end{table}

\subsection{Maximization of the shear modulus}\label{subsec4.2}
The second example aims to maximize the shear modulus of the material under shear loading. In the three-dimensional case, a shear strain of $\boldsymbol{\varepsilon} = [0 \ \ 0 \ \ 0 \ \ 1.4\% \ \ 0 \ \ 0]^{T}$is applied in the $xy$-plane, while in the two-dimensional case, the load condition is simplified to $\boldsymbol{\varepsilon} = [0\% \ \ 0\% \ \ 1.4\%]^{T}$. The shear modulus represents the resistance of the material to elastic deformation induced by applied shear stresses. The AL formulation employed here follows the same framework as that presented in Section \eqref{subsec4.1}, except that the objective function is redefined to maximize the effective shear modulus:

\begin{equation}\label{Eq. (22)}
c(\overline{\boldsymbol{\rho}})=-\sum_{i=D+1}^{d} C_{i i}^{H}(\overline{\boldsymbol{\rho}}) 
\end{equation}

Table \eqref{tab7} presents the two-dimensional metamaterial topologies optimized with and without stress constraints at a volume fraction of 0.6. Overall, both designs exhibit a diamond-shaped geometry that is favorable for enhancing shear resistance. In the compliance-driven case, both the inner and outer contours are characterized by smoothly curved boundaries with continuous transitions, suggesting that the material distribution follows global stiffness optimization rather than directional stress alignment. With the introduction of the stress constraint, these curved profiles transform into nearly straight contours aligned with the principal stress directions. Although the stress-constrained topology appears more angular at the corners, the redistribution and alignment of material along the dominant stress paths strengthen the load-bearing framework, contributing to enhanced mechanical efficiency. Consistent with the results obtained for the bulk modulus maximization problem, the introduction of the stress constraint reduces the peak von Mises stress from $610.22 \ \text{MPa}$ to $558.92 \ \text{MPa}$ (a $8.41\%$ decrease), while maintaining a comparably high shear modulus performance.

\begin{table}[ht]
\centering
\caption{2D metamaterial microstructures with maximum shear modulus ($V_f = 0.6$)}
\renewcommand{\arraystretch}{0.5}
\begin{tabular}{c m{2.5cm} m{3cm} m{3cm} c}
\toprule
Case & \multicolumn{1}{c}{RUC} & \multicolumn{1}{c}{$\frac{\boldsymbol{\sigma}^{vm}}{\bar{\sigma}}$} & \multicolumn{1}{c}{$\boldsymbol{C}^H$} \\
\midrule
\begin{tabular}{@{}c@{}}compliance-\\driven\end{tabular} &
\includegraphics[width=2.5cm]{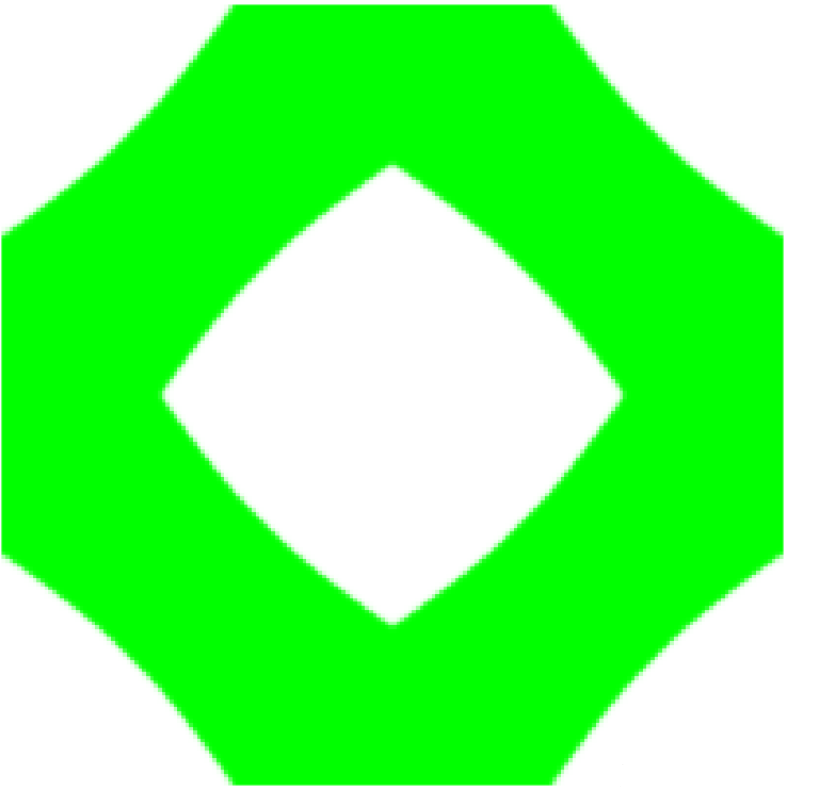} &
\includegraphics[width=3cm]{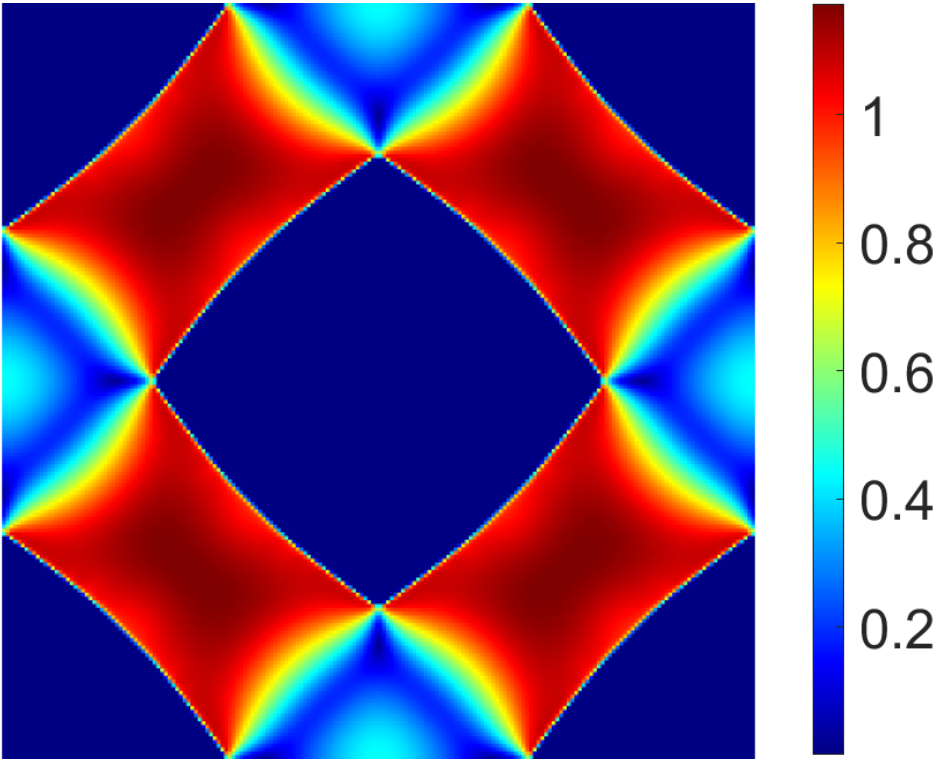} &
\includegraphics[width=3cm]{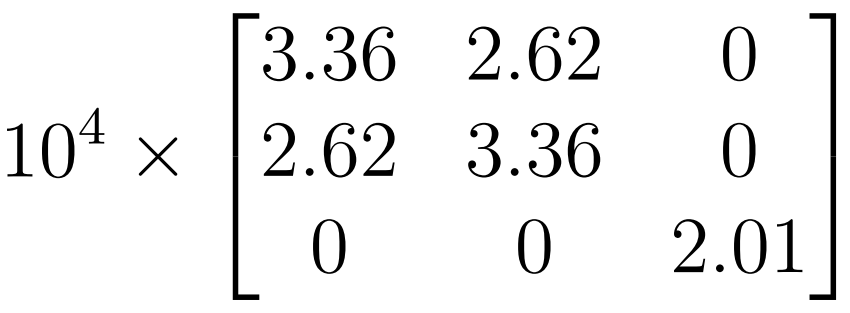} & \\
\midrule
\begin{tabular}{@{}c@{}}stress-\\constrained\end{tabular} &
\includegraphics[width=2.5cm]{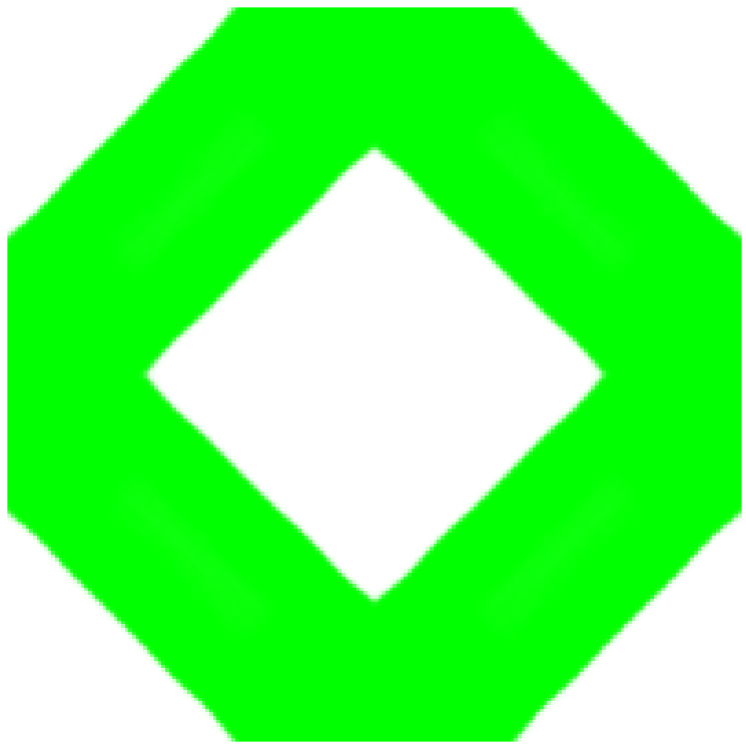} &
\includegraphics[width=3cm]{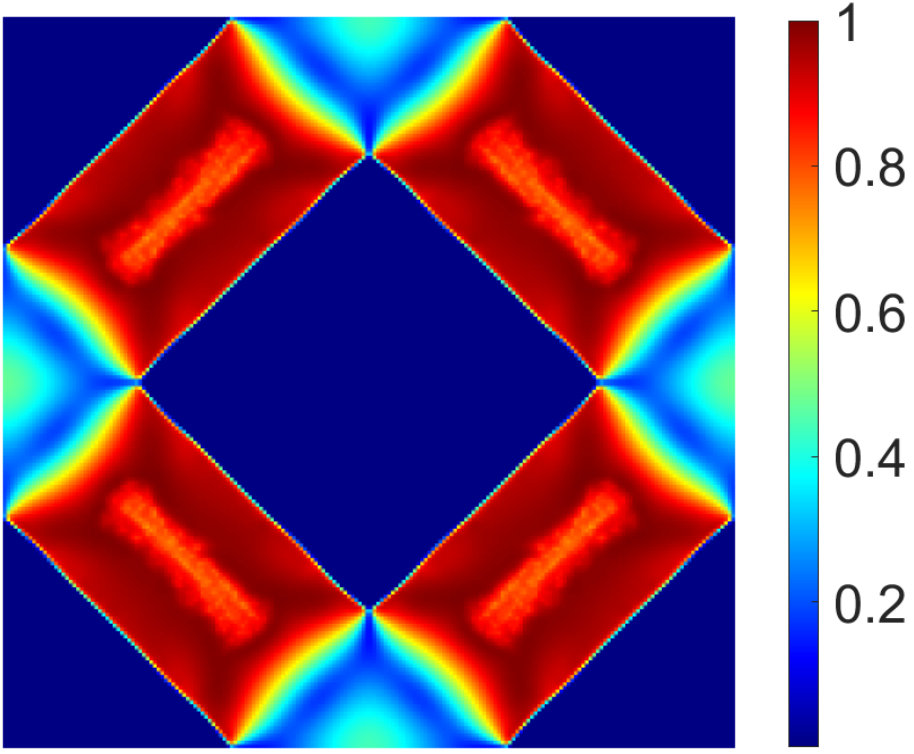} &
\includegraphics[width=3cm]{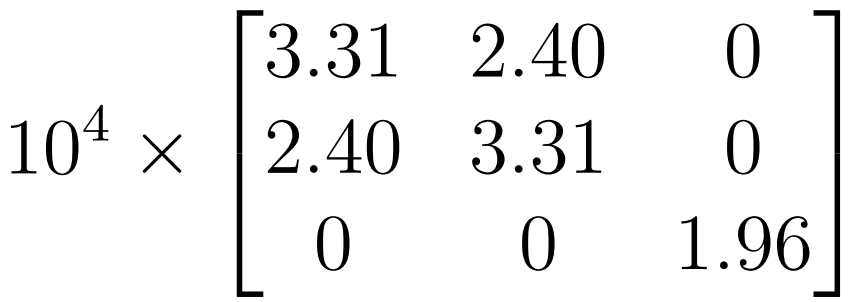} & \\
\bottomrule
\end{tabular}
\label{tab7}
\end{table}

The optimized three-dimensional topologies under von Mises stress constraints at varying volume fractions are summarized in Table \eqref{tab8}. At $V_f = 0.4$, the topology manifests a star-like open framework composed of slender diagonal members that align with the principal shear directions, thereby effectively accommodating the imposed deformation. Increasing the volume fraction to  $V_f = 0.6$ yields a denser topology characterized by thickened load-bearing struts and reduced porosity. Due to the applied shear strain in the $xy$-plane, rectangular regions appear at the midsections of the lateral surfaces, while cross-like connections are preserved at the top and bottom boundaries. The homogenized constitutive matrices further reveal anisotropic features induced by this shear-dominated loading mode: the in-plane shear modulus $C_{44}$ exceeds $C_{55}$ and $C_{66}$. This directional stiffness enhancement indicates that the optimized framework preferentially reinforces the $xy$-plane, consistent with the shear-driven deformation pattern. Moreover, the cross-sectional morphologies evolve into octagonal contours at both volume fractions, reflecting a geometric adaptation that mitigates local shear concentration while ensuring continuous load transfer. Similar to the observations in Table \eqref{tab5}, the denser topology strengthens the framework against both shear and normal deformation modes and achieves superior mechanical efficiency.

\begin{table}[!htbp]
\centering
\caption{3D stress-constrained metamaterial microstructures with maximum shear modulus}
\renewcommand{\arraystretch}{0.5}
\begin{tabular}{c m{3cm} m{2.8cm} m{4cm} c}
\toprule
Case & \multicolumn{1}{c}{RUC} & \multicolumn{1}{c}{Half RUC} & \multicolumn{1}{c}{$\boldsymbol{C}^H$} \\
\midrule
$V_f = 0.4$ &
\includegraphics[width=3cm]{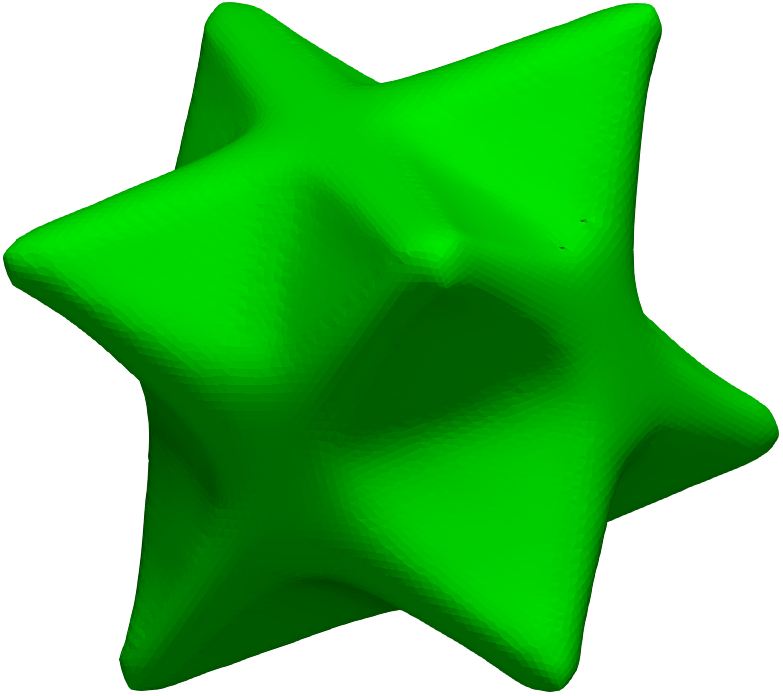} &
\includegraphics[width=2.8cm]{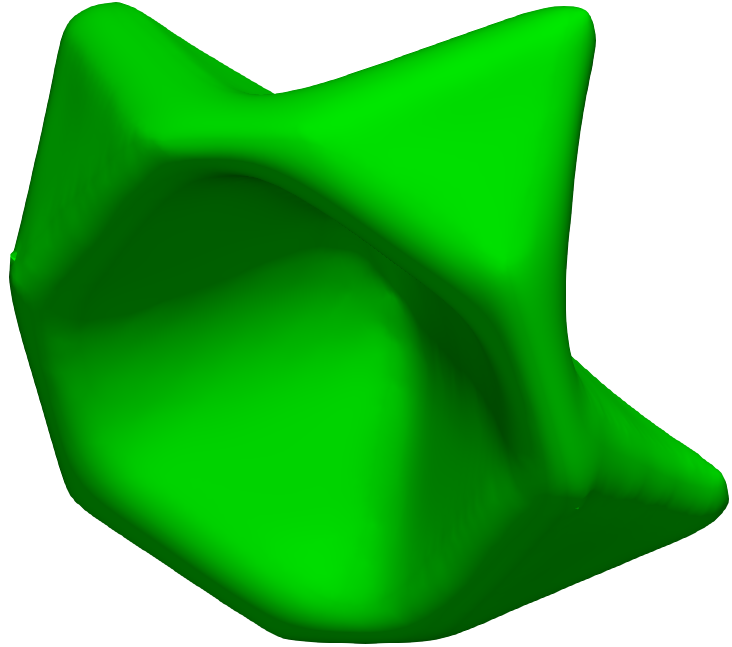} &
\includegraphics[width=4cm]{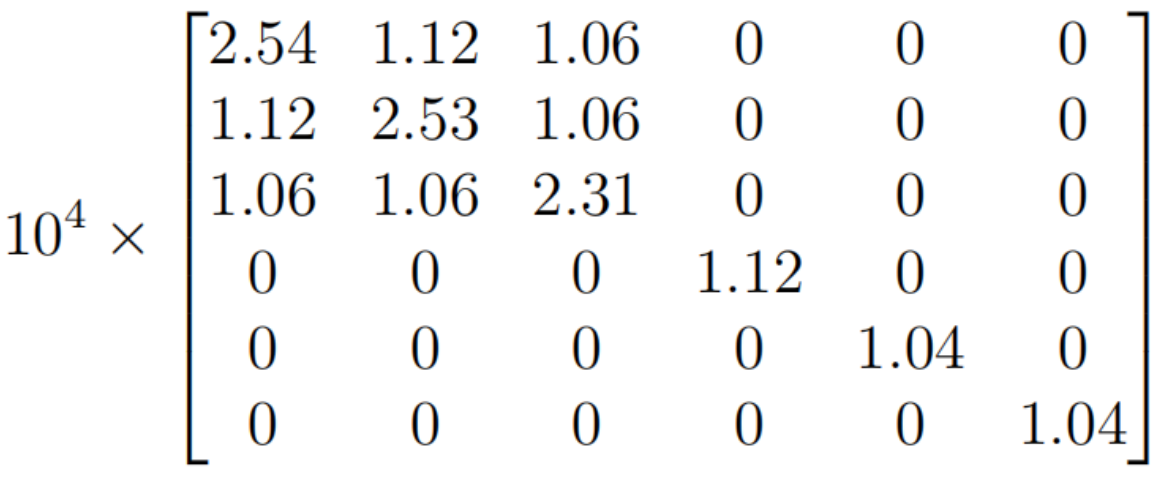} & \\
\midrule
$V_f = 0.6$  &
\includegraphics[width=3cm]{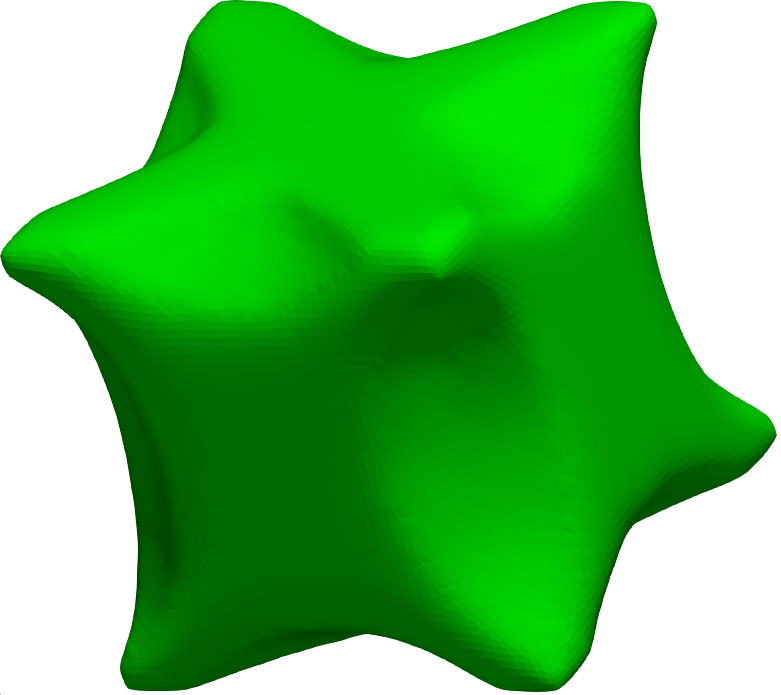} &
\includegraphics[width=2.8cm]{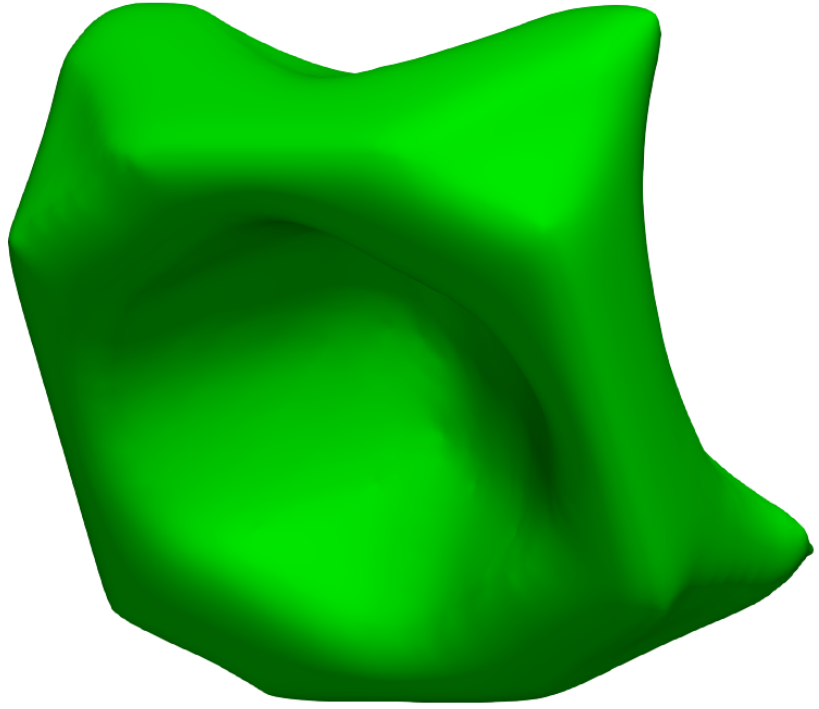} &
\includegraphics[width=4cm]{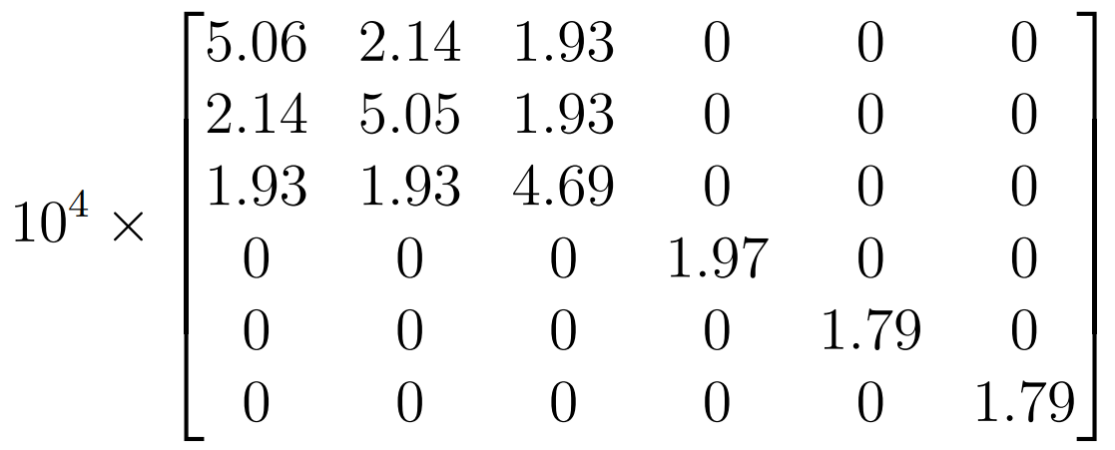} & \\
\bottomrule
\end{tabular}
\label{tab8}
\end{table}

Table \eqref{tab9} summarizes the three-dimensional metamaterial microstructures optimized for maximum shear modulus under different fatigue constraints. The structures are subjected to cyclic shear strain loading in the xy-plane with an amplitude of $0.8\%$ and a zero mean value. In contrast to the discernible divergence observed in the topologies optimized for maximum bulk modulus under normal strain, the present designs exhibit nearly identical geometries at the same volume fraction regardless of the fatigue criterion applied. All resulting architectures feature external surfaces composed of triangular or diamond-like facets, a morphology that effectively accommodates the imposed shear deformation and highlights the shear-dominated nature of the mechanical response. Furthermore, the cross-sectional views reveal that the internal void boundaries consistently exhibit rhombic contours across different volume fractions. This geometric convergence stems from the fact that under such load conditions, the Matake, Findley, and Dang Van criteria are all predominantly governed by the shear stress amplitude. The contributions of the normal stress components are almost negligible. Therefore, the optimizations constrained by different fatigue criteria converge to structurally similar configurations with nearly identical homogenized constitutive constitutive matrices.

\begin{table}[!htbp]
\centering
\caption{3D fatigue-constrained metamaterial microstructures with maximum shear modulus under cyclic shear strain loading}
\renewcommand{\arraystretch}{0}
\setlength{\tabcolsep}{3pt} 
\begin{tabular}{c m{2.6cm} m{2.6cm} m{3.8cm} c}
\toprule
Case & \multicolumn{1}{c}{RUC} & \multicolumn{1}{c}{Half RUC} & \multicolumn{1}{c}{$\boldsymbol{C}^H$} \\
\midrule
\multirow{2}{*}{\centering\begin{tabular}{@{}c@{}} Findley \\ $V_f = 0.4$ \end{tabular}} 
& \includegraphics[width=2.6cm]{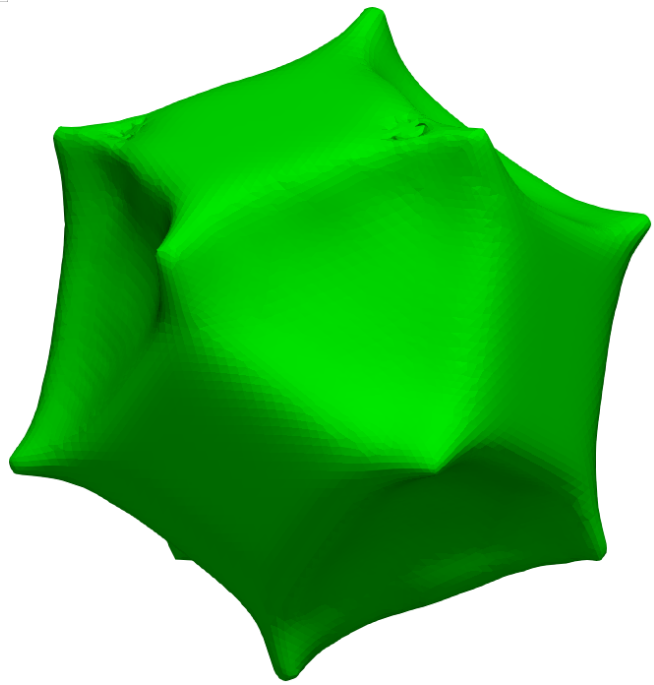} 
& \includegraphics[width=2.5cm]{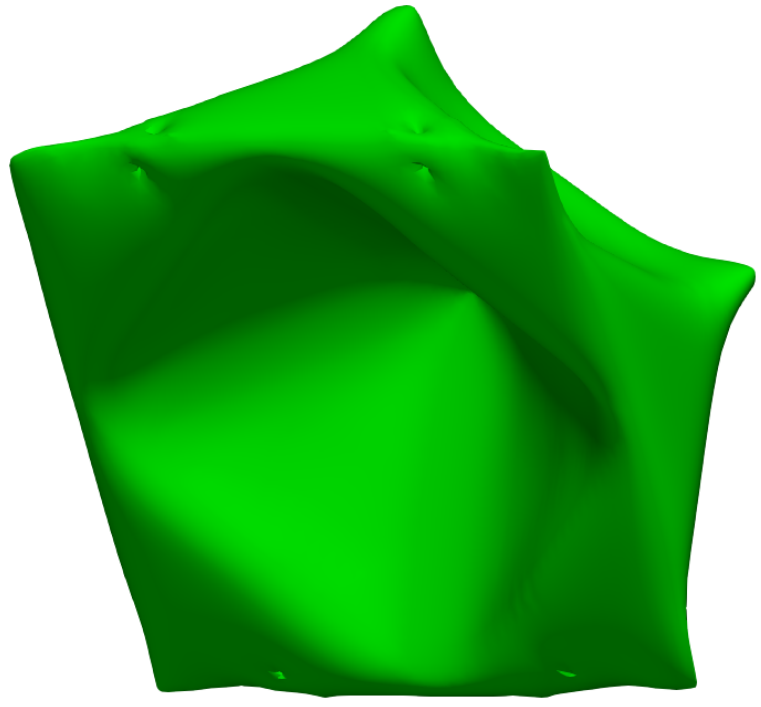} 
& \includegraphics[width=3.8cm]{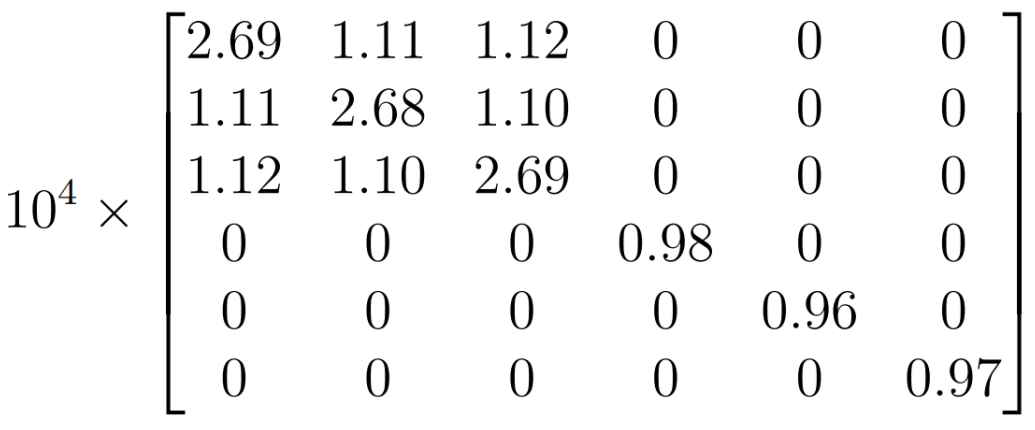} 
& \\
& & & & \\
\midrule
\multirow{2}{*}{\centering\begin{tabular}{@{}c@{}} Findley \\ $V_f = 0.6$ \end{tabular}} 
& \includegraphics[width=2.6cm]{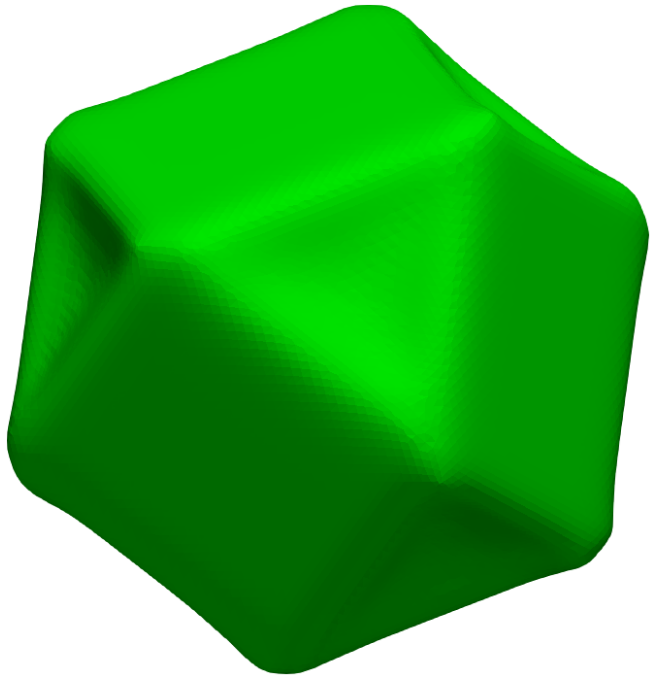} 
& \includegraphics[width=2.5cm]{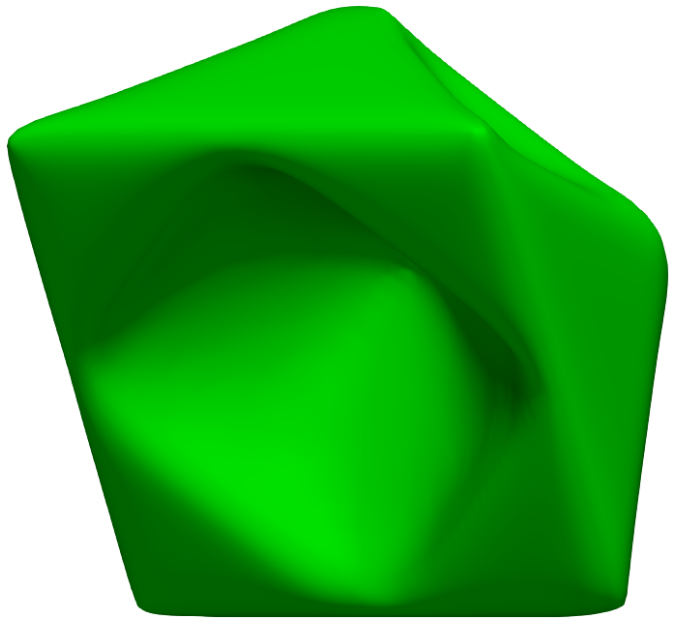} 
& \includegraphics[width=3.8cm]{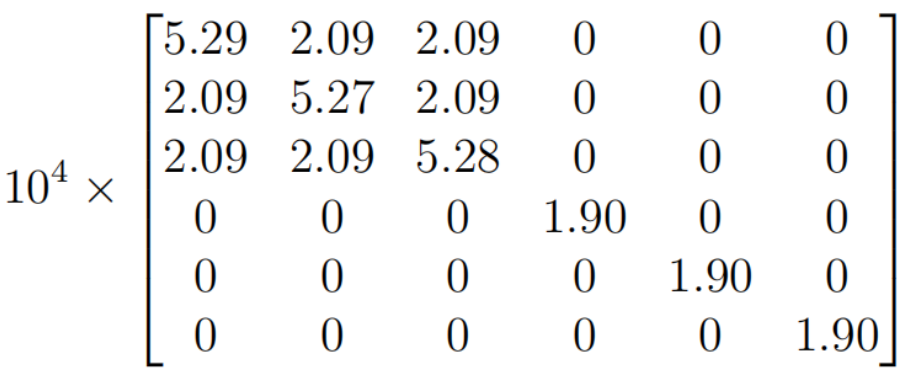} 
& \\
& & & & \\
\midrule
\multirow{2}{*}{\centering\begin{tabular}{@{}c@{}} Matake \\ $V_f = 0.4$ \end{tabular}} 
& \includegraphics[width=2.6cm]{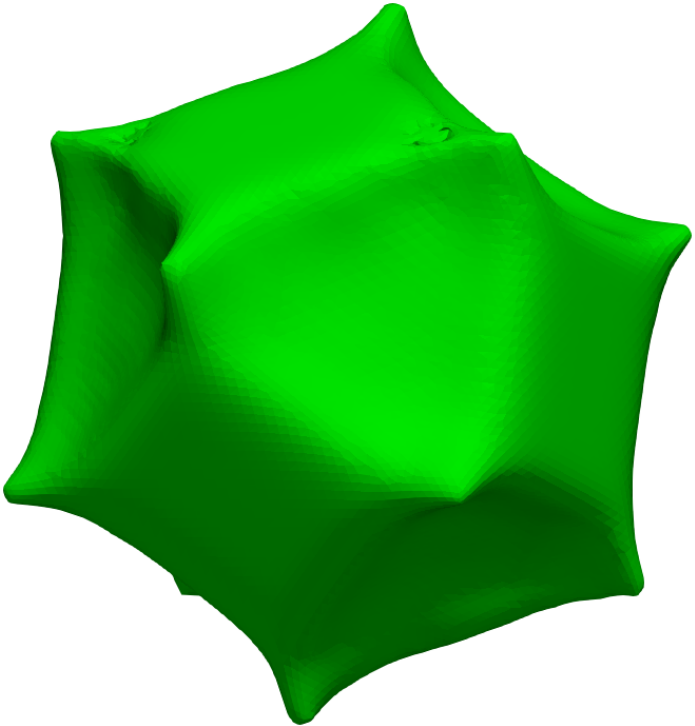} 
& \includegraphics[width=2.5cm]{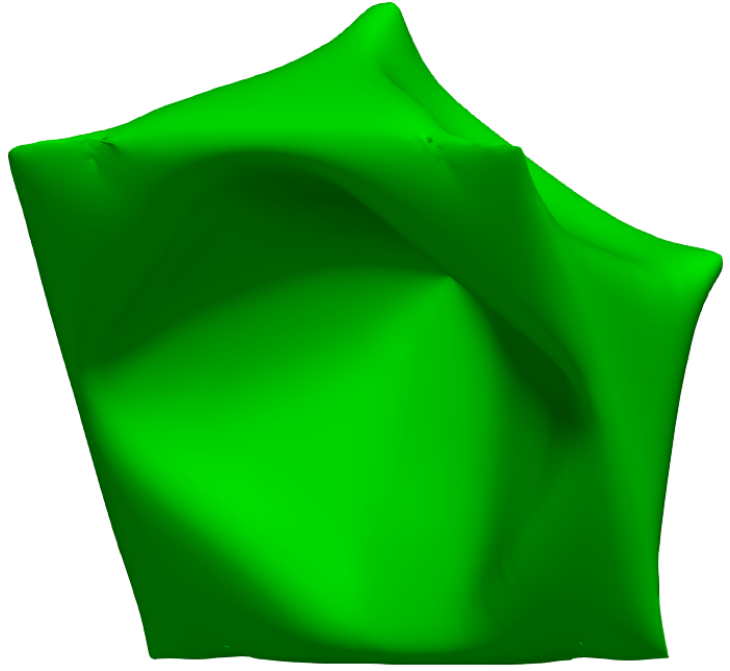} 
& \includegraphics[width=3.8cm]{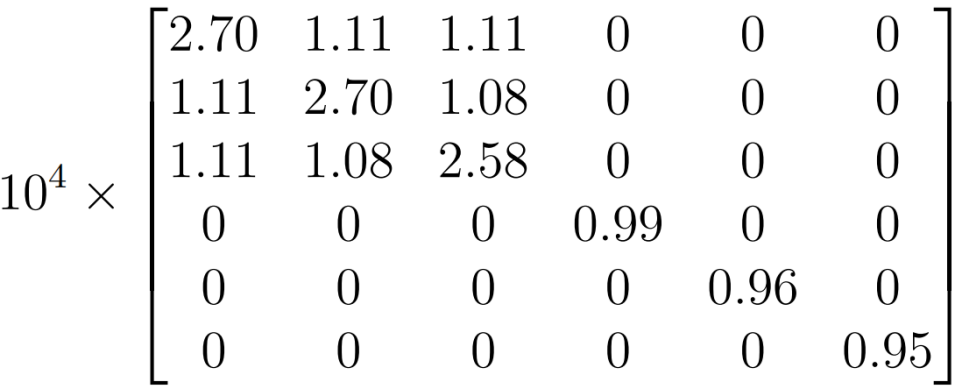} 
& \\
& & & & \\
\midrule
\multirow{2}{*}{\centering\begin{tabular}{@{}c@{}} Matake \\ $V_f = 0.6$ \end{tabular}} 
& \includegraphics[width=2.6cm]{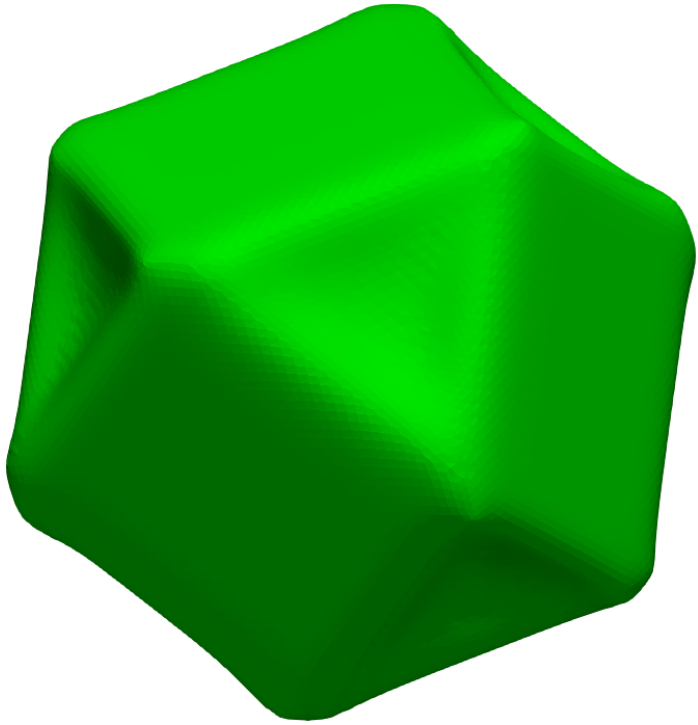} 
& \includegraphics[width=2.5cm]{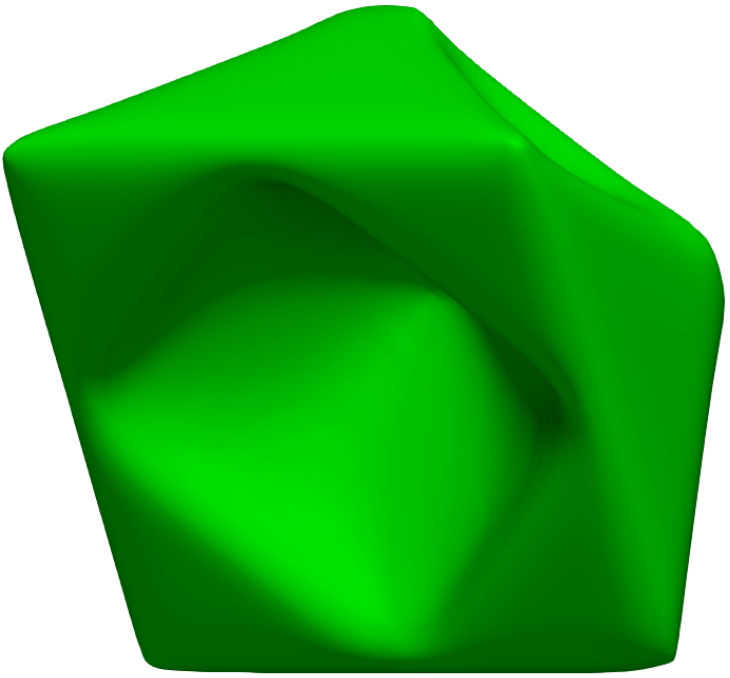} 
& \includegraphics[width=3.8cm]{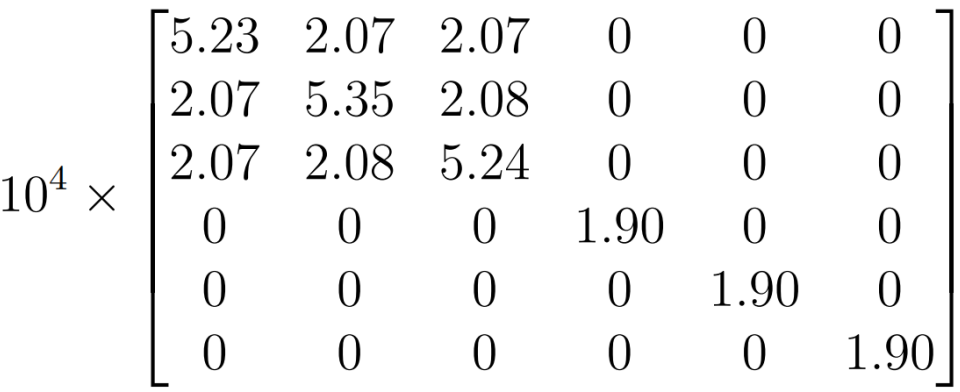} 
& \\
& & & & \\
\midrule
\multirow{2}{*}{\centering\begin{tabular}{@{}c@{}} Dang van \\ $V_f = 0.4$ \end{tabular}} 
& \includegraphics[width=2.6cm]{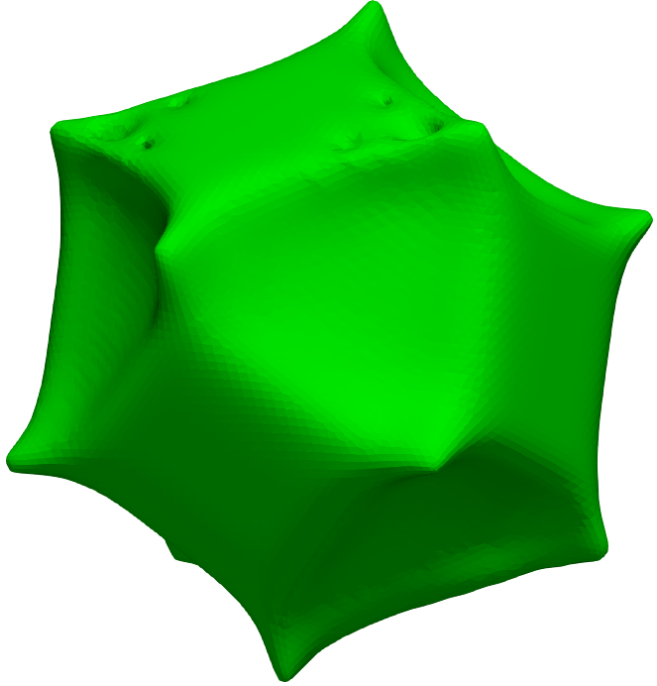} 
& \includegraphics[width=2.5cm]{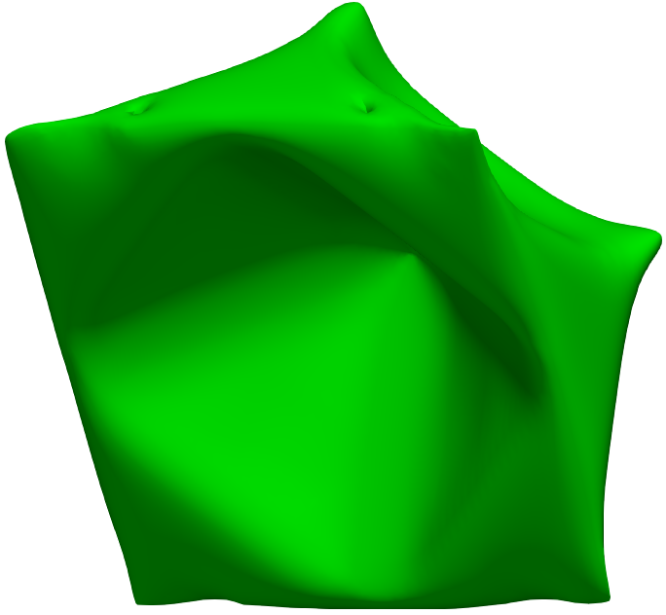} 
& \includegraphics[width=3.8cm]{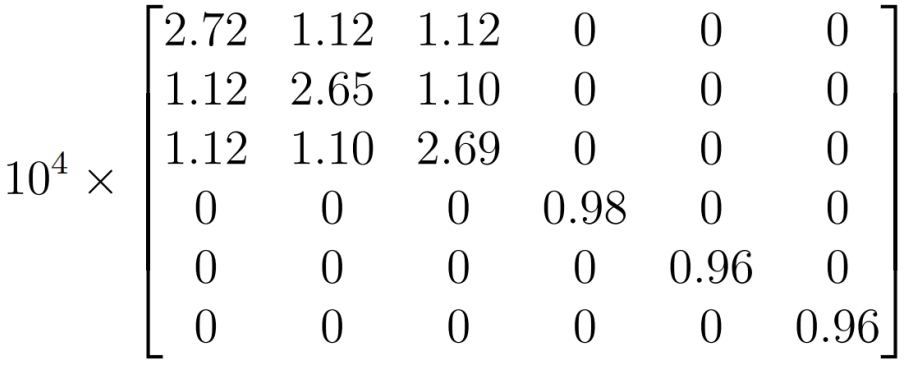} 
& \\
& & & & \\
\midrule
\multirow{2}{*}{\centering\begin{tabular}{@{}c@{}} Dang van \\ $V_f = 0.6$ \end{tabular}} 
& \includegraphics[width=2.6cm]{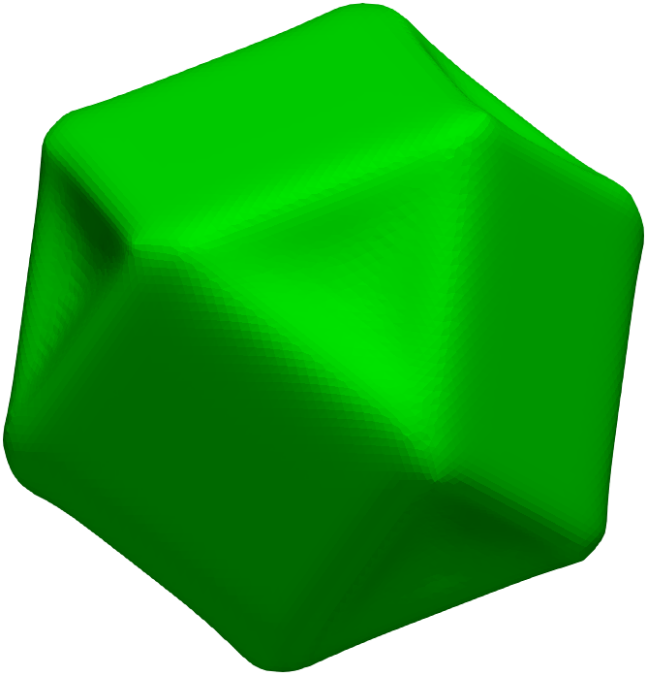} 
& \includegraphics[width=2.5cm]{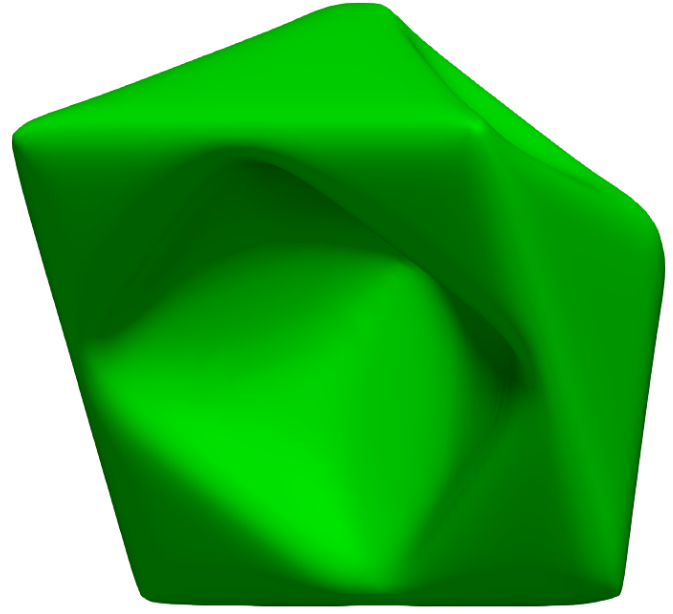} 
& \includegraphics[width=3.8cm]{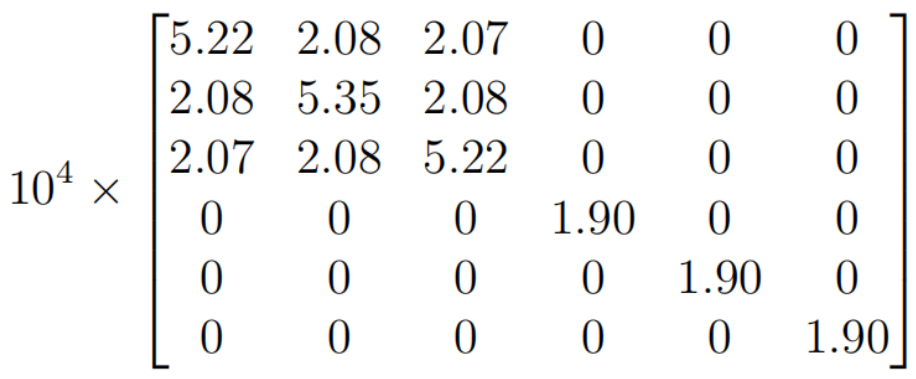} 
& \\
& & & & \\
\bottomrule
\end{tabular}
\label{tab9}
\end{table}

\subsection{Minimization of the Poisson’s ratio}\label{subsec4.3}
The third example focuses on the minimization of the Poisson's ratio of the material under axial strain, and such materials are also referred to as auxetics. In both 2D and 3D cases, a uniform axial strain of $-0.7\%$ is imposed along the $x$-axis. The metamaterial microstructures with a negative Poisson's ratio exhibit lateral expansion under tensile loading and lateral contraction under compressive loading. The topology optimization in this subsection is conducted to design isotropic metamaterials, and the objective function is shown below:

\begin{equation}\label{Eq. (23)}
c(\overline{\boldsymbol{\rho}})=- \frac{C_{1 2}^{H}(\overline{\boldsymbol{\rho}})}{C_{1 1}^{H}(\overline{\boldsymbol{\rho}})} 
\end{equation}

Table \eqref{tab10} illustrates the topologies with and without stress constraints at $V_f = 0.4$. It can be found that the omission of stress constraints in topology optimization generally leads to slender frameworks connected by compliant hinges, which promote stress concentrations at joints and geometric discontinuities. Such layouts often require additional post-processing to satisfy manufacturability requirements. In contrast, the incorporation of stress constraints produces fundamentally different structural configurations that are not only structurally robust but also more resistant to localized stress concentrations. The thin members and weak connections are effectively eliminated, and geometric transitions become smoother. Consequently, the manufacturability of the resulting designs is enhanced and the need for post-design smoothing is substantially reduced. Incorporating the stress constraint results in a increase in the Poisson’s ratio, rising from $-0.90$ to $-0.85$. However, this adjustment contributes to a $6.85\%$ reduction in the maximum von Mises stress, decreasing from $1039.67$ MPa to $968.41$ MPa.

\begin{table}[!htbp]
\centering
\caption{2D metamaterial microstructures with minimum Poisson’s ratio ($V_f = 0.4$)}
\renewcommand{\arraystretch}{0.5}
\begin{tabular}{c m{2.5cm} m{3cm} m{3cm} c}
\toprule
Case & \multicolumn{1}{c}{RUC} & \multicolumn{1}{c}{$\frac{\boldsymbol{\sigma}^{vm}}{\bar{\sigma}}$} & \multicolumn{1}{c}{$\boldsymbol{C}^H$} \\
\midrule
\begin{tabular}{@{}c@{}}compliance-\\driven\end{tabular} &
\includegraphics[width=2.5cm]{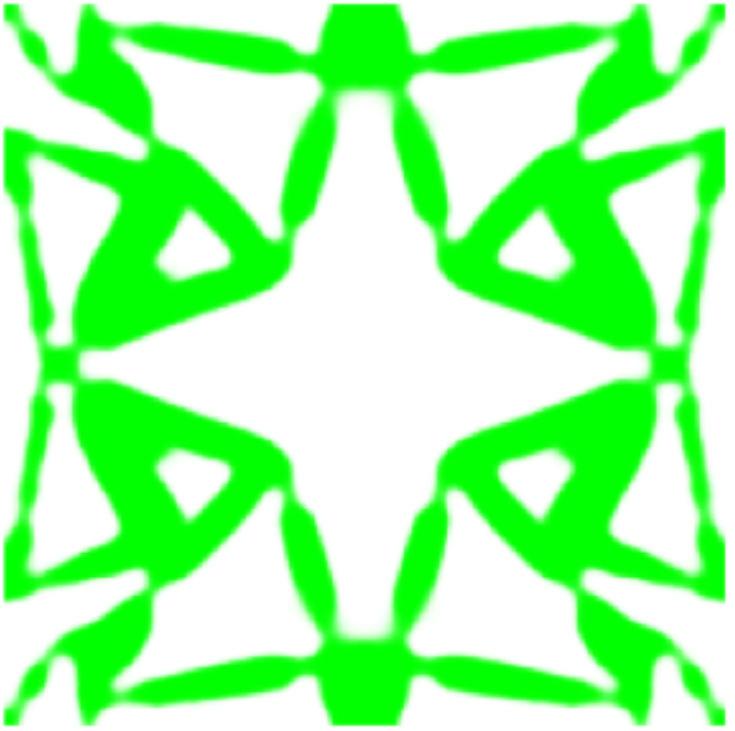} &
\includegraphics[width=3cm]{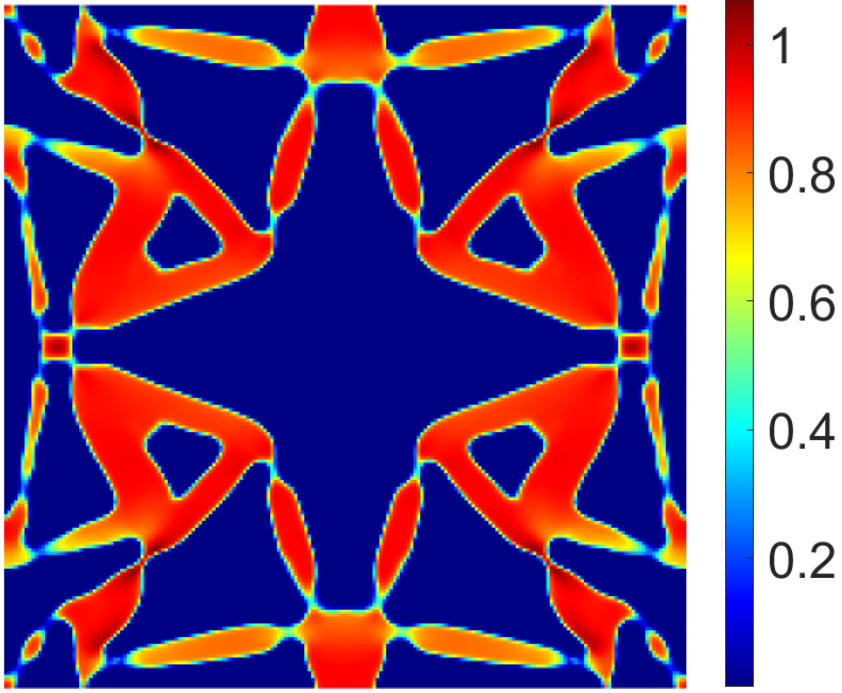} &
\includegraphics[width=3cm]{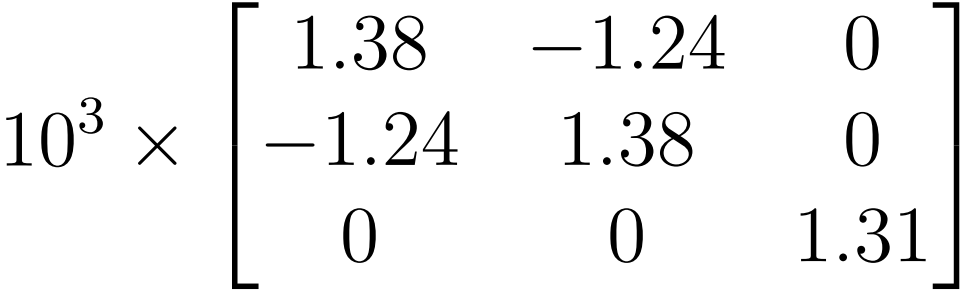} & \\
\midrule
\begin{tabular}{@{}c@{}}stress-\\constrained\end{tabular} &
\includegraphics[width=2.5cm]{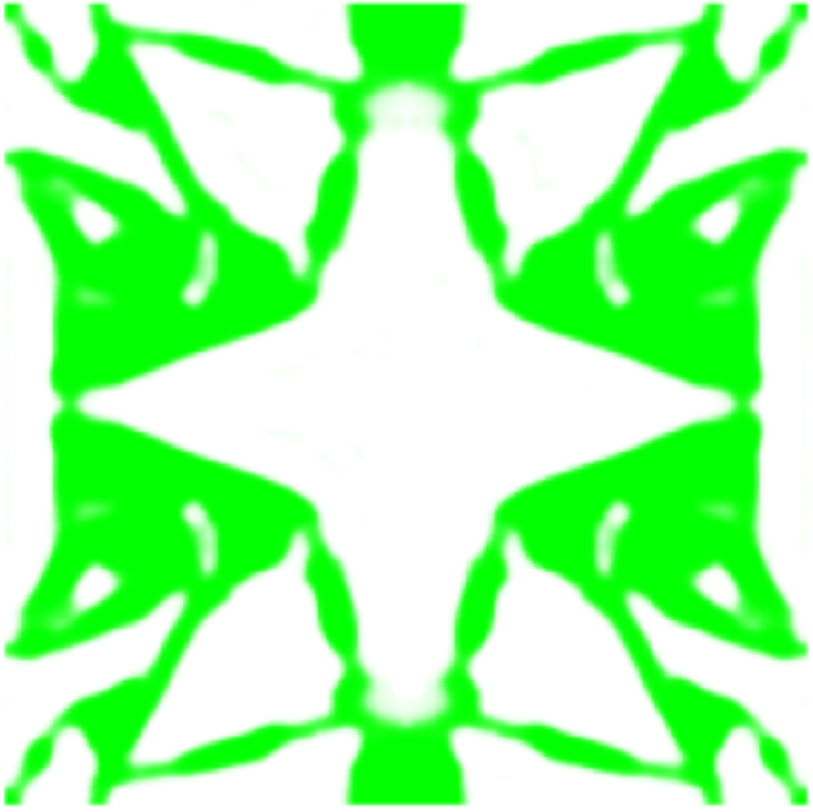} &
\includegraphics[width=3cm]{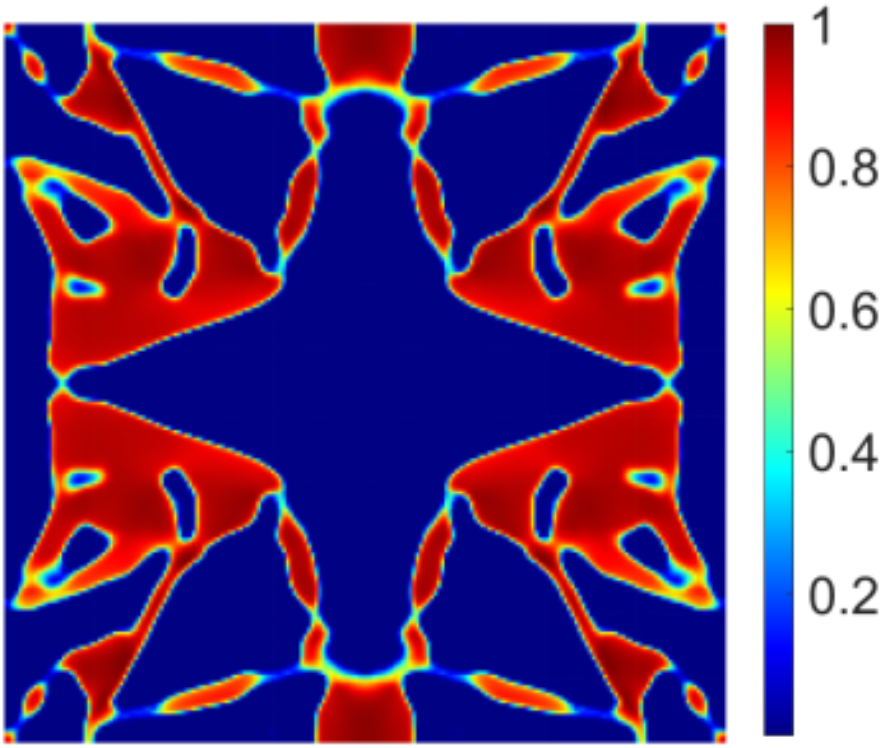} &
\includegraphics[width=3cm]{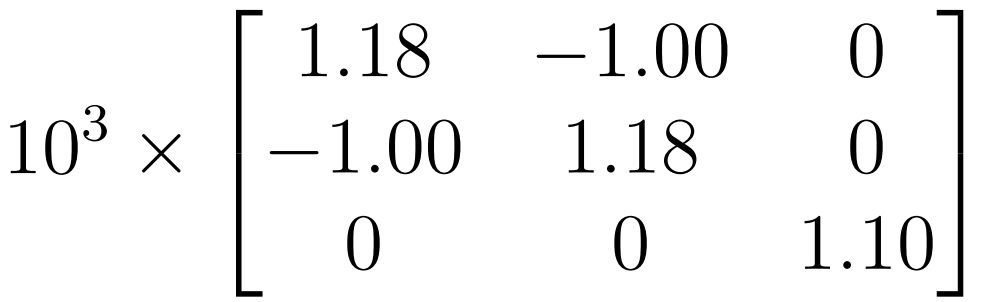} & \\
\bottomrule
\end{tabular}
\label{tab10}
\end{table}

Table \eqref{tab11} summarizes the optimized three-dimensional topologies under von Mises stress constraints for different volume fractions. The framework proposed can be directly extended to multi-load conditions. Table \eqref{tab12} presents the resulting topologies for different volume fractions under the simultaneous application of two load cases, corresponding to strains of $\boldsymbol{\varepsilon} = [-0.5\% \ \ 0 \ \ 0 \ \ 0\ \ 0 \ \ 0]^{T}$ and $\boldsymbol{\varepsilon} = [0\ \ 0 \ \ 0 \ \ 1\%\ \ 0 \ \ 0]^{T}$, respectively. Overall, the optimized topologies exhibit highly open and geometrically intricate network-like frameworks. At lower volume fractions, the structures are primarily composed of slender load-bearing struts forming continuous lattice networks, where the negative Poisson’s ratio originates predominantly from a rotational deformation mechanism. As the volume fraction increases, these thin members gradually vanish within the interior, giving rise to densely packed and interlocking solid blocks that dominate the overall framework. The auxetic behavior mainly stems from three-dimensional rotational interactions among adjacent solid domains rather than simple geometric expansion. Under uniaxial compression along the  $x$ -axis, the adjacent blocks undergo coordinated rotation and reorientation, resulting in simultaneous lateral contraction along both the  $y$- and $z$-directions. is cooperative deformation leads to an overall negative Poisson’s ratio effect.

\begin{table}[!htbp]
\centering
\caption{3D stress-constrained metamaterial microstructures with minimum Poisson’s ratio}
\renewcommand{\arraystretch}{0.5}
\begin{tabular}{c m{3cm} m{3cm} m{3.8cm} c}
\toprule
Case & \multicolumn{1}{c}{RUC} & \multicolumn{1}{c}{Half RUC} & \multicolumn{1}{c}{$\boldsymbol{C}^H$} \\
\midrule
$V_f = 0.4$ &
\includegraphics[width=3cm]{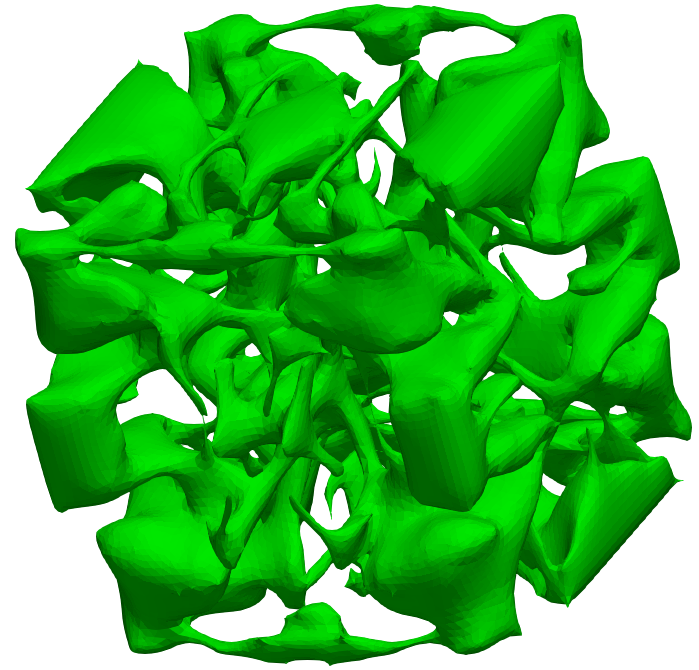} &
\includegraphics[width=3cm]{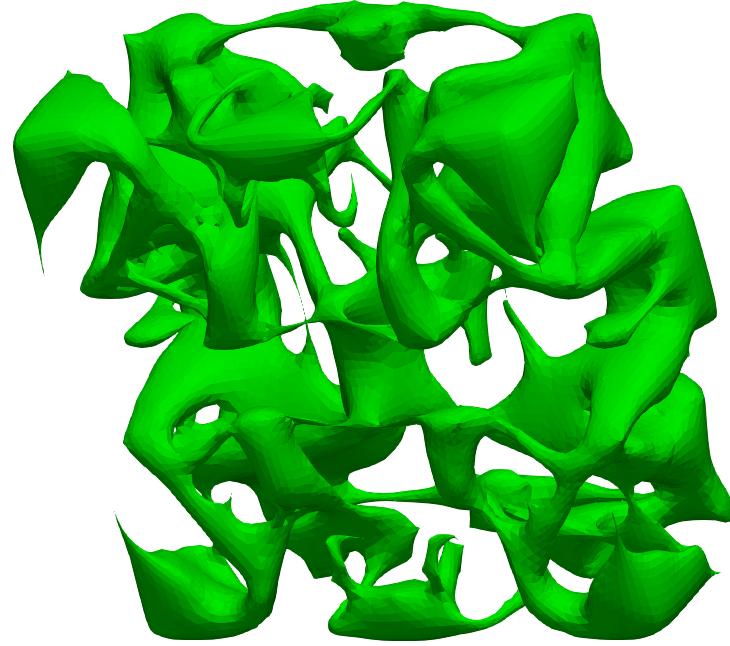} &
\includegraphics[width=3.8cm]{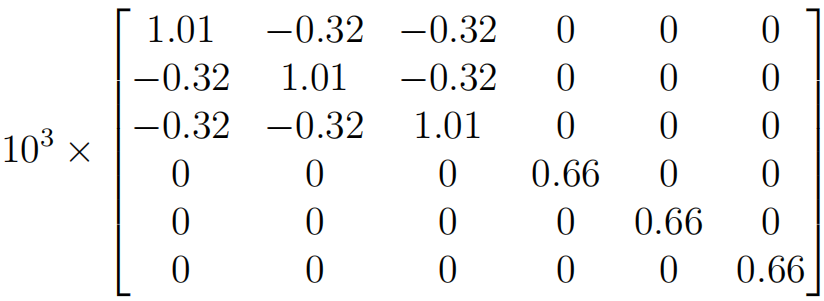} & \\
\midrule
$V_f = 0.6$  &
\includegraphics[width=3cm]{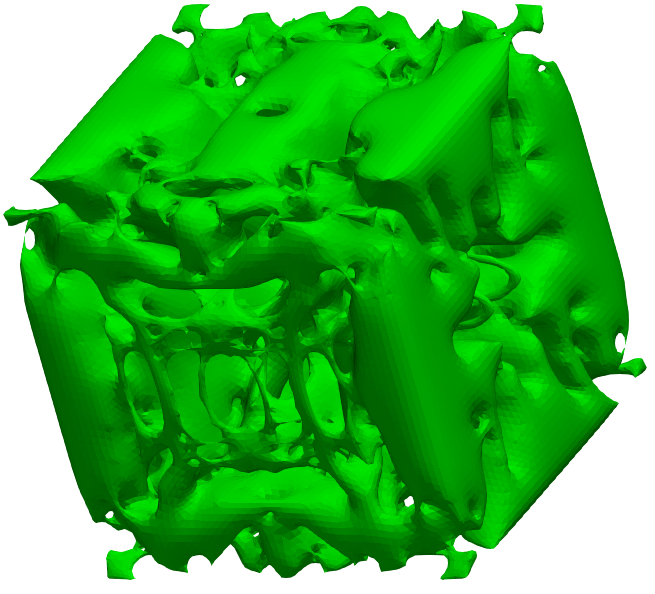} &
\includegraphics[width=3cm]{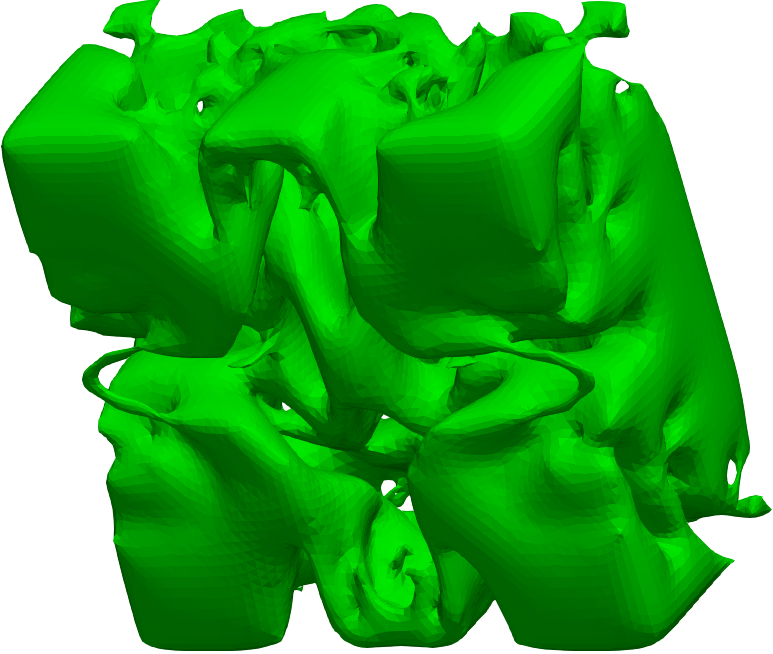} &
\includegraphics[width=3.8cm]{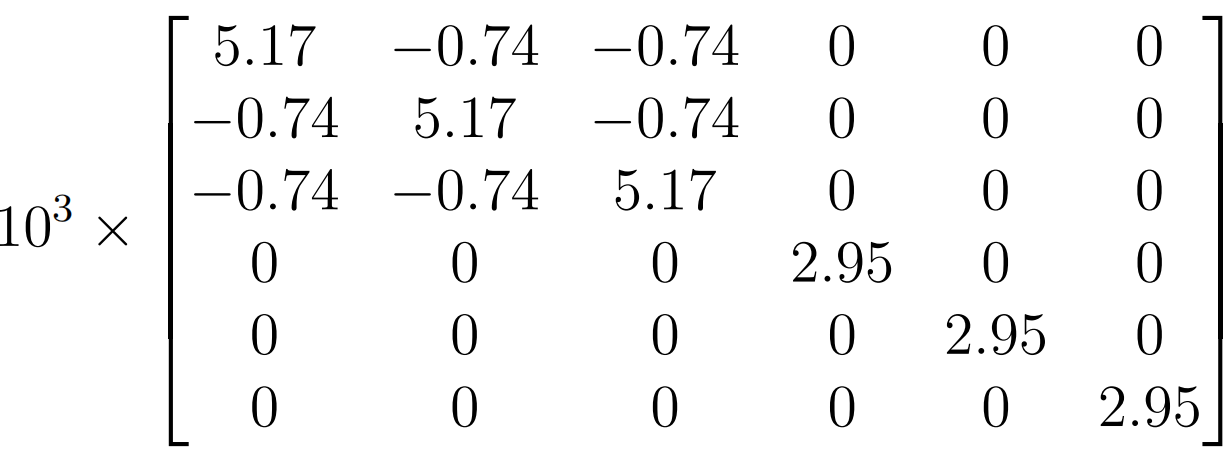} & \\
\bottomrule
\end{tabular}
\label{tab11}
\end{table}

\begin{table}[!htbp]
\centering
\caption{3D stress-constrained metamaterial microstructures with minimum Poisson’s ratio}
\renewcommand{\arraystretch}{0.5}
\begin{tabular}{c m{3cm} m{3cm} m{3.8cm} c}
\toprule
Case & \multicolumn{1}{c}{RUC} & \multicolumn{1}{c}{Half RUC} & \multicolumn{1}{c}{$\boldsymbol{C}^H$} \\
\midrule
$V_f = 0.4$ &
\includegraphics[width=3cm]{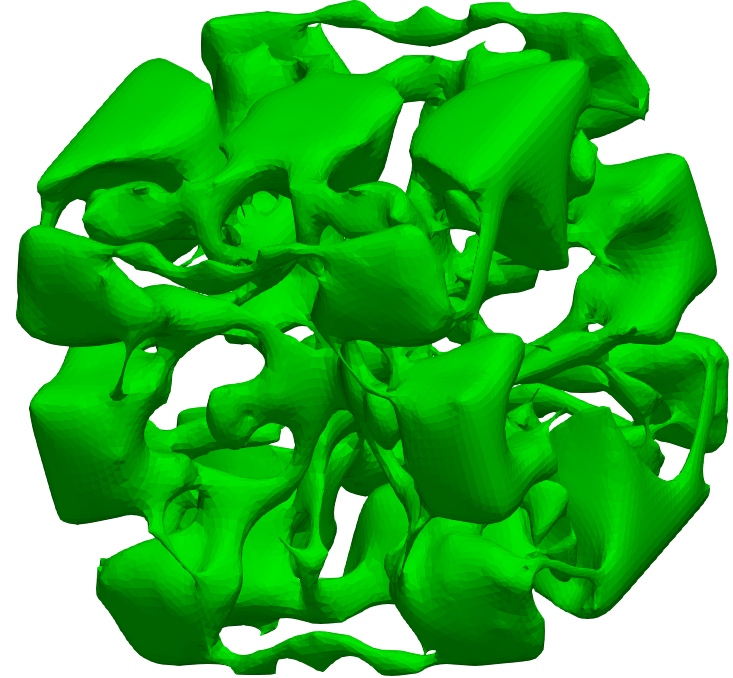} &
\includegraphics[width=3cm]{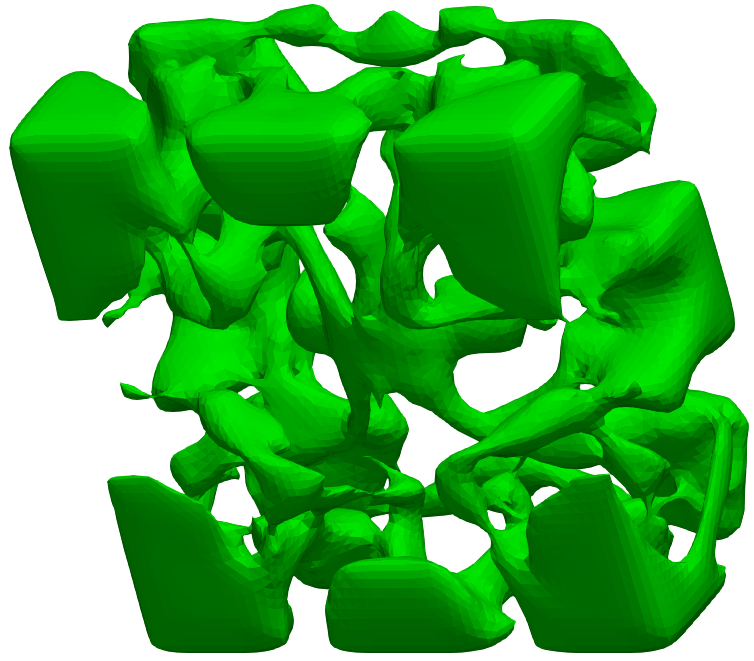} &
\includegraphics[width=3.8cm]{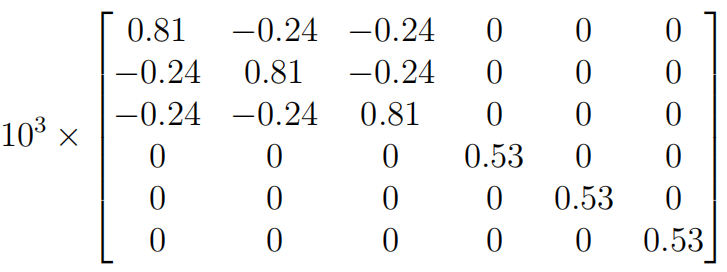} & \\
\midrule
$V_f = 0.6$  &
\includegraphics[width=3cm]{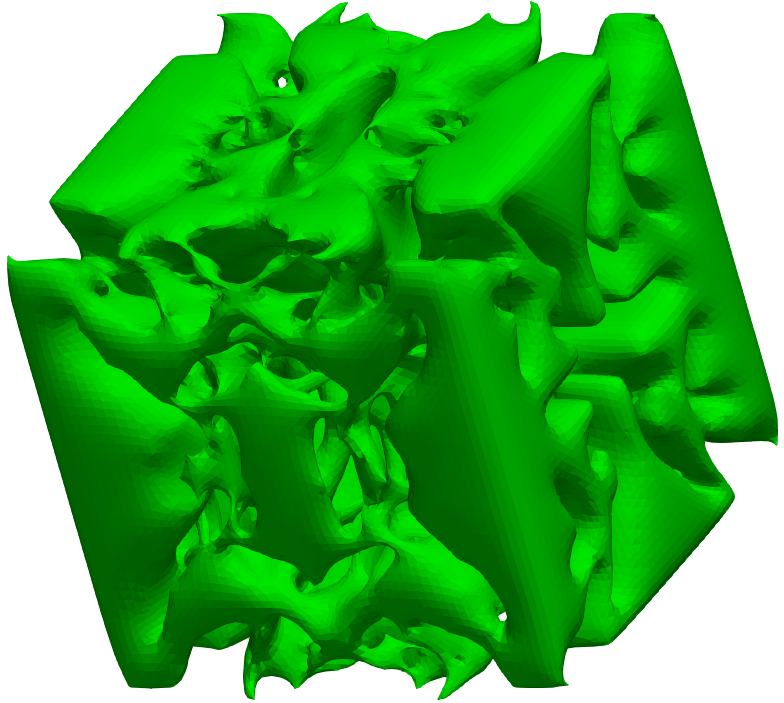} &
\includegraphics[width=3cm]{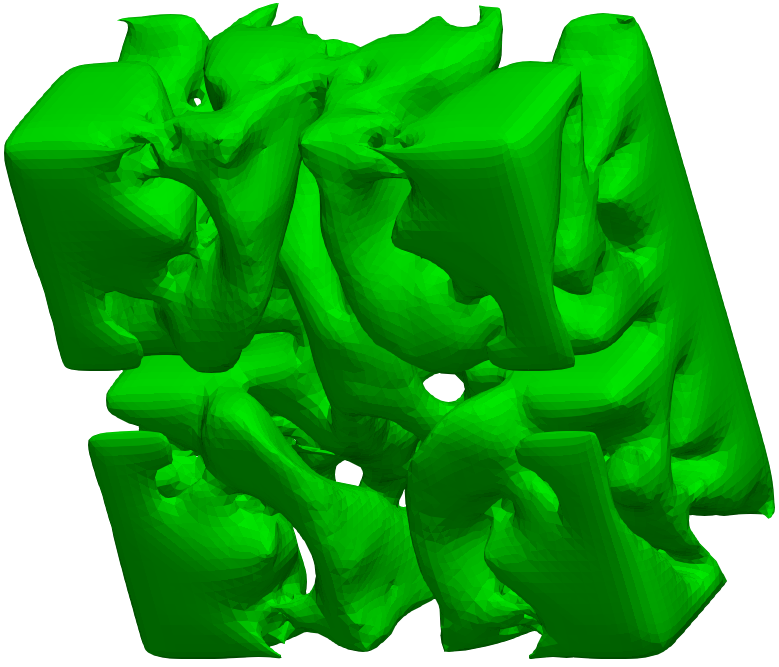} &
\includegraphics[width=3.8cm]{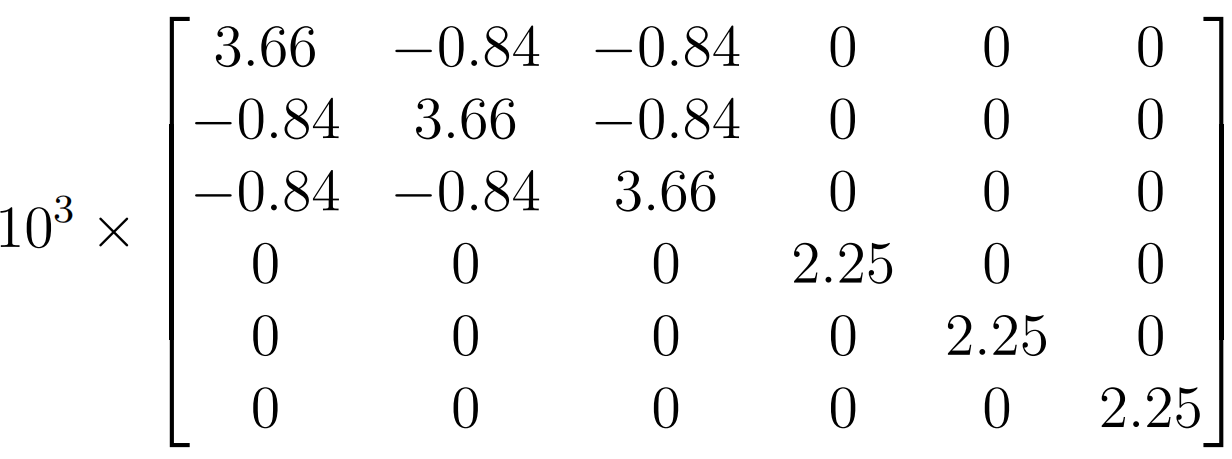} & \\
\bottomrule
\end{tabular}
\label{tab12}
\end{table}

\section{Discussion}
\label{sec5}
The introduction of local stress constraints within the same volume fraction framework leads to a substantial improvement in strength performance with only minor geometric modifications, while the stiffness degradation remains limited. These small yet effective structural adjustments are unlikely to be achieved through intuitive or experience-based design, demonstrate the complexity between topology and material behavior and confirm the effectiveness of the proposed approach. Particularly in stress-informed multiscale structural topology optimization, employing metamaterial microstructures designed with stress constraints holds great potential for substantially enhancing mechanical performance and further reducing material usage.

In the fatigue-constrained topology optimization of metamaterials, several classical critical plane-based criteria are considered. These criteria exhibit comparable predictive behavior under pure shear loading. However, under normal loading conditions, they produce distinct topologies, reflecting the intrinsic fatigue mechanisms inherent of each criterion. Although the Findley and Matake criteria involve the same stress components, their differing definitions of the critical plane lead to entirely dissimilar topological features at a lower volume fraction ($V_f = 0.4$). Owing to the sensitivity to the normal stress component in determining the critical plane, the Findley criterion typically generates more uniform geometrical configurations. The Matake criterion prioritizes the reduction of shear stress fluctuations, leading to the formation of branched load paths in regions of high shear intensity. These branches redistribute the concentrated shear loads along directions approximately parallel to the maximum shear stress planes. At higher volume fractions ($V_f = 0.6$), the decrease of shear stress amplitude on the critical planes can be achieved through the thickening of structural members, reducing the necessity for branching mechanisms used at lower volume fractions. This transition results in more comparable topologies predicted by the Findley and Matake criteria, with both displaying small circular voids at the four corners in cross-sectional views. 

The Dang Van fatigue damage model accounts for the hydrostatic pressure throughout the entire loading cycle rather than the normal stress on a fixed plane. Consequently, the predicted topologies exhibit distinctive morphological features compared with those derived from the other two criteria across different volume fractions. At a lower volume fraction, the optimized topology develops a diamond shaped cross-section primarily composed of thin walls inclined along the principal shear directions. This configuration effectively alleviates the coupled effects of shear stress amplitude and local hydrostatic pressure while minimizing material usage. As the volume fraction increases, the structure evolves into an octagonal cross-section featuring an additional internal octagonal cavity. This morphological transition reflects the increasing influence of hydrostatic pressure in the optimization process, promoting a more uniform stress distribution across the section.

In contrast to von Mises stress–constrained topologies that produce highly symmetric and nearly isotropic frameworks, fatigue criteria based on the critical plane approach yield predictions with pronounced anisotropy. The von Mises stress is insensitive to loading direction since it depends solely on the second invariant of the deviatoric stress tensor, which represents an equivalent scalar measure of the overall stress state. In comparison, critical plane–based criteria explicitly account for direction-dependent stress components, and the resulting spatial distribution of critical planes is highly discontinuous, thereby introducing abrupt variations in stress orientation. In addition, topologies obtained from Poisson’s ratio minimization differ fundamentally from those derived for maximizing the bulk or shear modulus. In designs that consider stiffness, the resulting geometries are typically compact and highly efficient in transmitting loads, with deformation primarily governed by axial tension and compression. By contrast, structures optimized for a minimal Poisson’s ratio achieve the negative Poisson’s ratio effect through local rotational and bending mechanisms embedded within their microarchitecture (\cite{Huang_negative_2016}). Consequently, the homogenized elastic moduli of these auxetic configurations are considerably lower than those of stiffness-oriented designs. This behavior highlights the inherent trade-off between attaining a large negative Poisson’s ratio, which enhances deformability, and maintaining high overall structural stiffness.

\section{Conclusion and perspective}
\label{sec6}
This study proposes a topology optimization framework that generalizes the AL formulation to simultaneously manage global and local constraints within a unified scheme. The AL formulation streamlines sensitivity analysis by requiring only a single adjoint vector per independent load case, which dramatically reduces the computational cost of managing numerous constraints while ensuring their precise local satisfaction. Two representative classes of engineering problems are investigated in this study: von Mises stress–constrained optimization under static loading and  HCF–constrained optimization under cyclic loading. Given the highly nonlinear nature of topology optimization, the choice of fatigue criterion sometimes can lead to markedly different predictions. However, since the applicability of each criterion depends on the material behavior and loading characteristics, incorporating a range of classical fatigue models within the design framework provides valuable comparative insights and enables diverse solution pathways.

Topology optimization only provides a preliminary local optimal solution. Therefore, extensive post-processing and experimental verification is still needed to verify the reliability of the optimized design. This study does not yet account for the influence of the additive manufacturing process on the mechanical behavior of the material. In laser powder bed fusion, commonly adjusted processing parameters include laser power, scanning speed, layer thickness, and scan strategy. Variations in these parameters can significantly affect the grain size and orientation, as well as the formation and distribution of defects such as pores and cracks, thereby influencing the mechanical properties of the fabricated components. Future work will incorporate the effects of these parameters to improve the accuracy and engineering relevance of the simulation results.

The future work will also focus on two main directions. First, a metamaterial library comprising a series of predefined RUCs with varied mechanical properties will be developed as the foundation for constructing a data-driven multiscale topology optimization framework capable of efficiently generating high-performance metamaterials across multiple design scales. Second, to further address the computational challenges associated with large-scale three-dimensional designs, which remain both memory and computation intensive due to the need to solve equilibrium and adjoint equations, a deep generative design model based on machine learning will be established to enable rapid interpolation and prediction of metamaterial microstructures under different volume fractions.

\section{Conflict of interest statement}

On behalf of all authors, the corresponding author states that there is no conflict of interest.

\appendix 
\section{Sensitivity analysis}\label{secA}
The sensitivity of the AL function can be calculated using the chain rule as follows:

\begin{equation}\label{Eq. (A1)}
\frac{\partial \mathcal{J}}{\partial \rho_{I}}=\sum_{J=1}^{N}(\frac{\partial c}{\partial \tilde{\rho}_{J}} \frac{d \tilde{\rho}_{J}}{d \rho_{I}}+\frac{1}{N_s} \frac{\partial P_s}{\partial \tilde{\rho}_{J}} \frac{d \tilde{\rho}_{J}}{d \rho_{I}}+\frac{\partial P_v}{\partial \tilde{\rho}_{J}} \frac{d \tilde{\rho}_{J}}{d \rho_{I}}+\frac{\partial P_{iso}}{\partial \tilde{\rho}_{J}} \frac{d \tilde{\rho}_{J}}{d \rho_{I}})
\end{equation}

The derivative of the penalization term $P$ to the $I$-th design variable is expressed as:

\begin{equation}\label{Eq. (A2)}
\frac{\partial P_s^{(k)}}{\partial \tilde{\rho}_{I}}=\sum_{J=1}^{N_s}\left[\left(\lambda_{sJ}^{(k)}+\mu^{(k)} h_{J}^s\right)\left(\frac{\partial h_{J}^s}{\partial \tilde{\rho}_{I}}+\frac{\partial h_{J}^s}{\partial \boldsymbol{u}} \frac{\partial \boldsymbol{u}}{\partial \tilde{\rho}_{I}}\right) \right]
\end{equation}

\begin{equation}\label{Eq. (A3)}
\frac{\partial P_v^{(k)}}{\partial \tilde{\rho}_{I}}=\left(\lambda_{v}^{(k)}+\mu^{(k)} h^v\right) \frac{\partial h^v}{\partial \tilde{\rho}_{I}}
\end{equation}

\begin{equation}\label{Eq. (A4)}
\frac{\partial P_{iso}^{(k)}}{\partial \tilde{\rho}_{I}}=\left(\lambda_{iso}^{(k)}+\mu^{(k)} h^{iso}\right) \frac{\partial h^{iso}}{\partial \tilde{\rho}_{I}} 
\end{equation}

The derivatives of the volume and isotropic constraints with respect to density are very straightforward:

\begin{equation}\label{Eq. (A5)}
\frac{\partial h^v}{\partial \tilde{\rho}_{I}} = \frac{V_{I}}{\lvert \Omega \rvert} \frac{d \bar{\rho}_{I}}{d \tilde{\rho}_{I}}
\end{equation}

\begin{equation}\label{Eq. (A6)}
\begin{aligned}
\frac{\partial h^{i s o}}{\partial \tilde{\rho}_I} = & \frac{1}{(C_{i j}^{i s o}+\epsilon)^{4}} 
\bigg[ \sum_{i, j=1}^{d} 2(C_{i j}^{H}-C_{i j}^{i s o})(C_{i j}^{i s o}+\epsilon) 
\bigg[\bigg(\frac{\partial c_{i j}^{H}}{\partial \rho_{I}}-\frac{\partial c_{i j}^{i s o}}{\partial \rho_{I}}\bigg) \\
 & \times (C_{i j}^{i s o}+\epsilon)-(C_{i j}^{H}-C_{i j}^{i s o}) \frac{\partial c_{i j}^{i s o}}{\partial \rho_{I}}\bigg] \bigg] 
\frac{d \bar{\rho}_I}{d \tilde{\rho}_I}
\end{aligned}
\end{equation}

\noindent when $\frac{\sum_{J=1}^{N} \bar{\rho}_{J} V_{J}}{\mathrm{V}_{f}} \geq-\lambda_{v}^{(k)} / \mu^{(k)} $ and $\frac{\left(C_{i j}^{H}-C_{i j}^{i s o}\right)^{2}}{\left(C_{i j}^{i s o}+\epsilon\right)^{2}} \geq-\lambda_{iso}^{(k)} / \mu^{(k)}$, otherwise $\partial h^v / \partial \tilde{\rho}_I =\partial h^{iso} / \partial \tilde{\rho}_I = 0$.

In order to avoid calculating the derivatives of the displacements with respect to the design variables in Eq. \eqref{Eq. (A2)}, the adjoint method is used here to simplify the calculation:

\begin{equation}\label{Eq. (A7}
\widehat{P}_s^{(k)}=P_s^{(k)}+\sum_{l=1}^{N_{l}} \left[ \boldsymbol{\eta}_{l}^{T}\left(\boldsymbol{K} \boldsymbol{u}_{l, r e f}-\boldsymbol{f}_{l, r e f}\right) \right]
\end{equation}
where $N_l$ is the number of load cases, $\boldsymbol{f}_{l, r e f}$ and $\boldsymbol{u}_{l, r e f}$ represent the load and displacement vectors of the $l$-th load case at a chosen reference time point during cyclic loading, respectively. For static loading, the subscript “ref” can be omitted.

Then Eq. \eqref{Eq. (A2)} can be replaced by: 

\begin{equation}\label{Eq. (A8)}
\frac{\partial \widehat{P}_s^{(k)}}{\partial \tilde{\rho}_{I}}=\frac{\partial P_s^{(k)}}{\partial \tilde{\rho}_{I}}+ \sum_{l=1}^{N_l} \boldsymbol{\eta}_{l}^{T} \left(\frac{\partial \boldsymbol{K}}{\partial \tilde{\rho}_{I}} \boldsymbol{u}_{l, r e f} - \frac{\partial \boldsymbol{f}_{l, r e f}}{\partial \tilde{\rho}_{I}}\right)
\end{equation}
with the adjoint equation: 

\begin{equation}\label{Eq. (A9)}
\boldsymbol{\eta}_{l}=-\boldsymbol{K}^{-T}\left(\frac{\partial P_s^{(k)}}{\partial \boldsymbol{u}_{l,ref}}\right)^{T}=-\boldsymbol{K}^{-T}\left\{\sum_{J=1}^{N}\left[\left(\lambda_{sJ}^{(k)}+\mu^{(k)} h^s_{J}\right)\left(\frac{\partial h^s_{J}}{\partial \boldsymbol{u}_{l,ref}}\right)\right]\right\}^{T}
\end{equation}
where

\begin{equation}\label{Eq. (A10)}
\frac{\partial \boldsymbol{f}_{l, r e f}}{\partial \tilde{\rho}_{I}} =p(1-\epsilon) \bar{\rho}_{I}^{(p-1)} \boldsymbol{K} \boldsymbol{B} \boldsymbol{\varepsilon}^{0}_l
\end{equation}

The first item $\partial \widehat{P}_s^{(k)} / \partial \tilde{\rho}_{J}$ in Eq. \eqref{Eq. (A8)} is determined as:

\begin{equation}\label{Eq. (A11)}
\frac{\partial P_s^{(k)}}{\partial \widetilde{\rho}_{I}}=\left(\lambda_{sI}^{(k)}+\mu^{(k)} h^s_{I}\right) p(1-\epsilon) \bar{\rho}_{I}^{(p-1)} \left(g_{I}^{3}+g_{I}\right)
\end{equation}
and the item $\partial h^s_{J} / \partial \boldsymbol{u}_{l,ref}$ for different stress criteria can be calculated as:

\begin{equation}\label{Eq. (A12)}
\begin{aligned}
\text{von Mises: } \frac{\partial h^s_{J}}{\partial \boldsymbol{u}_{l}} & =\frac{\partial h^s_{J}}{\partial \sigma_{vm}^{J}} \frac{\partial  \sigma_{vm}^{J}}{\partial \boldsymbol{\sigma }_{l}} \frac{\partial \boldsymbol{\sigma}_{l}}{\partial \boldsymbol{u}_{l}} 
\end{aligned}
\end{equation}

\begin{equation}\label{Eq. (A13)}
\begin{aligned}
\text{Findley / Matake: } \frac{\partial h^s_{J}(\theta)}{\partial \boldsymbol{u}_{l, r e f}} & =\frac{\partial h^s_{J}(\theta)}{\partial \tau_{a}^{J}(\theta)} \frac{\partial \tau_{a}^{J}(\theta)}{\partial \boldsymbol{\tau }_{l, r e f}(\theta)} \frac{\partial \boldsymbol{\tau}_{l, r e f}(\theta)}{\partial \boldsymbol{\sigma}_{l, r e f}} \frac{\partial \boldsymbol{\sigma}_{l, r e f}}{\partial \boldsymbol{u}_{l, r e f}} \\
& +\frac{\partial h^s_{J}(\theta)}{\partial \sigma_{n, \max }^{J}(\theta)} \frac{\partial \sigma_{n, \max }^{J}(\theta)}{\partial \boldsymbol{\sigma}_{n l, r e f}(\theta)} \frac{\partial \boldsymbol{\sigma}_{n l, r e f}(\theta)}{\partial \boldsymbol{\sigma}_{l, r e f}} \frac{\partial \boldsymbol{\sigma}_{l, r e f}}{\partial \boldsymbol{u}_{l, r e f}}
\end{aligned}
\end{equation}

\begin{equation}\label{Eq. (A14)}
\begin{aligned}
\text{Dang Van: \quad} \frac{\partial h^s_{J}(\theta)}{\partial \boldsymbol{u}_{l, r e f}} & =\frac{\partial h^s_{J}(\theta)}{\partial \tau_{a}^{J}(\theta)} \frac{\partial \tau_{a}^{J}(\theta)}{\partial \boldsymbol{\tau}_{l, r e f}(\theta)} \frac{\partial \boldsymbol{\tau}_{l, r e f}(\theta)}{\partial \boldsymbol{\sigma}_{l, r e f}} \frac{\partial \boldsymbol{\sigma}_{l, r e f}}{\partial \boldsymbol{u}_{l, r e f}} \\
& +\frac{\partial h^s_{J}}{\partial \sigma_{H,max}^{J}} \frac{\partial \sigma_{H,max}^{J}}{\partial \sigma_{H l, r e f}^{J}} \frac{\partial \sigma_{H l, r e f}^{J}}{\partial \boldsymbol{\sigma}_{l, r e f}} \frac{\partial \boldsymbol{\sigma}_{l, r e f}}{\partial \boldsymbol{u}_{l, r e f}}
\end{aligned}
\end{equation}

The non-zero sensitivities of the $\boldsymbol{h}$ function to the stress items of $J$-th element are:

\begin{equation}\label{Eq. (A15)}
\frac{\partial h_{J}^s}{\partial \sigma_{vm}^{J}}=p(1-\epsilon) \bar{\rho}_{J}^{(p-1)}\left(3 g_{J}^{2}+1\right)
\end{equation}

\begin{equation}\label{Eq. (A16)}
\frac{\partial h_{J}^s}{\partial \tau_a^{J}} = \frac{\partial h_{J}^s}{\partial \sigma_{H, \max }^{J}}=\frac{\partial h_{J}^s}{\partial \sigma_{n, \max }^{J}}=\frac{\alpha p(1-\epsilon) \bar{\rho}_{J}^{(p-1)}\left(3 g_{J}^{2}+1\right)}{\beta}
\end{equation}
when $\epsilon+(1-\epsilon) \bar{p}_{J}^{p}\left(g_J^{3}+g_J\right) \geq-\lambda_{sJ}^{(k)} / \mu^{(k)} $, otherwise $\partial h_{J}^s / \partial \boldsymbol{u}_{l,ref}=0$.

The sensitivity of von Mises stress to the Cauchy stress vector is equal to:

\begin{equation}\label{Eq. (A17)}
\frac{\partial \sigma_{vm}^{J}}{\partial \boldsymbol{\sigma}_l}=\frac{\boldsymbol{M} \boldsymbol{\sigma}_l}{\sigma_{vm}^{J}}
\end{equation}

The derivatives of the shear stress amplitude and maximum normal stress with respect to the stress vector over the critical plane under the $l$-th load are:

\begin{equation}\label{Eq. (A18)}
\left\{\begin{array}{l}
\frac{\partial \boldsymbol{\tau}_{a}(\boldsymbol{\theta})}{\partial \boldsymbol{\tau}_{l,ref}(\boldsymbol{\theta})}=\frac{1}{2}\left(\boldsymbol{c}_{\tau l,max}-\boldsymbol{c}_{\tau l,min}\right)
\\ \\
\frac{\partial \boldsymbol{\sigma}_{n, max}(\boldsymbol{\theta})}{\partial \boldsymbol{\sigma}_{nl,ref}(\boldsymbol{\theta})}=\boldsymbol{c}_{nl,max}
\end{array}\right.
\end{equation}
where where $\boldsymbol{c}_{\tau l,max}$, $\boldsymbol{c}_{\tau l,min}$ and $\boldsymbol{c}_{nl,max}$ present the coefficient vectors of stress extremes.

For 3D cases, $\partial \boldsymbol{\sigma}_{Hl, r e f}^{J} / \partial \boldsymbol{\sigma}_{l,r e f}^{J}=\left[\begin{array}{llllll}
1 / 3 & 1 / 3 & 1 / 3 & 0 & 0 & 0
\end{array}\right]^{T}$.

Based on the stress transformation equations in Eq. \eqref{Eq. (18)} for 3D cases, the terms $\partial \boldsymbol{\tau }_{l, r e f}(\theta) / \partial \boldsymbol{\sigma}_{l, r e f}$ and $\partial \boldsymbol{\sigma}_{n l, r e f}(\theta) / \partial \boldsymbol{\sigma}_{l, r e f}$ are expressed as follows:

\begin{equation}\label{Eq. (A19)}
\left\{\begin{array}{l}
\frac{\partial \boldsymbol{\tau }_{nxl, r e f}^{J}(\theta)}{\partial \boldsymbol{\sigma}_{l, r e f}^{J}}=\left[\begin{array}{llllll}
n_{x} a_{x} & n_{y} a_{y} & n_{z} a_{z} & n_{x} a_{y}+a_{x} n_{y} & n_{y} a_{z}+a_{y} n_{z} & n_{x} a_{z}+a_{x} n_{z}
\end{array}\right] \\
\frac{\partial \boldsymbol{\tau }_{nyl, r e f}^{J}(\theta)}{\partial \boldsymbol{\sigma}_{l, r e f}^{J}}=\left[\begin{array}{llllll}
n_{x} b_{x} & n_{y} b_{y} & n_{z} b_{z} & n_{x} b_{y}+b_{x} n_{y} & n_{y} b_{z}+b_{y} n_{z} & n_{x} b_{z}+b_{x} n_{z}
\end{array}\right] \\
\frac{\partial \boldsymbol{\sigma}_{n l, r e f}^{J}(\theta)}{\partial \boldsymbol{\sigma}_{l, r e f}^{J}}=\left[\begin{array}{llllll}
n_{x}^{2} & n_{y}^{2} & n_{z}^{2} & 2n_{x} n_{y} & 2n_{y} n_{z} & 2n_{x} n_{z}
\end{array}\right] \\
\end{array}\right.
\end{equation}

For 2D cases, Eq. \eqref{Eq. (A19)} can be rewritten as follows:

\begin{equation}\label{Eq. (A20)}
\left\{\begin{array}{l}
\frac{\partial \boldsymbol{\tau }_{l, r e f}^{J}(\theta)}{\partial \boldsymbol{\sigma}_{l, r e f}^{J}}=\left[\begin{array}{lll}
-\frac{\sin 2 \theta}{2} & \frac{\sin 2 \theta}{2} & \cos 2 \theta
\end{array}\right] \\
\frac{\partial \boldsymbol{\sigma}_{n l, r e f}^{J}(\theta)}{\partial \boldsymbol{\sigma}_{l, r e f}^{J}}=\left[\begin{array}{lll}
\frac{1+\cos 2 \theta}{2} & \frac{1-\cos 2 \theta}{2} & \sin 2 \theta
\end{array}\right]
\end{array}\right.
\end{equation}

The derivative of the stress to the displacement equals to:

\begin{equation}\label{Eq. (A21)}
\frac{\partial \boldsymbol{\sigma}_{l, r e f}}{\partial \boldsymbol{u}_{l, r e f}}=\boldsymbol{C} \boldsymbol{B}
\end{equation}

The derivative of the homogenized constitutive matrix $\boldsymbol{C}^H$ to the $I$-th design variable is calculated as follows:

\begin{align} \label{Eq. (A22)}
\frac{\partial \boldsymbol{C}^{H}}{\partial \tilde{\rho}_{I}} 
& = \frac{\left(\boldsymbol{u}^{0}-\boldsymbol{u}\right) \frac{\partial \boldsymbol{K}(\overline{\boldsymbol{\rho}})}{\partial \tilde{\rho}_{I}}\left(\boldsymbol{u}^{0}-\boldsymbol{u}\right)^{T}}{|\Omega|} \nonumber \\
& = \frac{\left(\boldsymbol{u}^{0}-\boldsymbol{u}\right) p(1-\epsilon) \bar{\rho}_{I}^{p-1} \boldsymbol{K}_{0}\left(\boldsymbol{u}^{0}-\boldsymbol{u}\right)^{T}}{|\Omega|} \frac{d \bar{\rho}_I}{d \tilde{\rho}_I}
\end{align}

\noindent where $\boldsymbol{u}^{0} = \boldsymbol{B} \boldsymbol{\varepsilon}^{0}$.

Finally, the following equation can be obtained:

\begin{equation}\label{Eq. (A23)} 
\begin{aligned}
\frac{\partial \mathcal{J}}{\partial \rho_{I}} = & \left(\boldsymbol{\mathcal{F}}^{T}\right)_{I} 
\left\{
\frac{\partial c}{\partial \rho_{I}}
+ \frac{1}{N_s} \left[ \left( \lambda_{sI}^{(k)} + \mu^{(k)} h_{I}^s \right) p(1-\epsilon) \bar{\rho}_{I}^{(p-1)} g_{I} (g_{I}^{2}+1) 
\right. \right. \\
& \left. \left. + \; p(1-\epsilon) \bar{\rho}_{I}^{(p-1)} \sum_{l=1}^{N_{l}} \left( \boldsymbol{\eta}_{l}^{I} \right)^{T} \boldsymbol{k}_{0}^{I} \boldsymbol{u}_{l . r e f}^{I} \right] 
+ \left( \lambda_{v}^{(k)} + \mu^{(k)} h^v \right) \frac{\partial h^v}{\partial \tilde{\rho}_{I}} \right. \\
& \left. + \; \left( \lambda_{iso}^{(k)} + \mu^{(k)} h^{iso} \right) \frac{\partial h^{iso}}{\partial \tilde{\rho}_{I}} 
\right\} 
\frac{\beta \left\{ 1 - \tanh \left[ \beta \left( \tilde{\rho}_{j} - \eta \right) \right] \right\}^{2} }{ \tanh \left( \beta \eta + \tanh \left[ \beta (1 - \eta) \right] \right) }
\end{aligned}
\end{equation}
where $\boldsymbol{\mathcal{F}}_{I}$ is the $I$-th row of the filter matrix, $\boldsymbol{k}_{0}^{I}$ is the stiffness matrix of $I$-th solid element.

\bibliographystyle{elsarticle-harv} 
\bibliography{My_Library}

@article{jiao_mechanical_2023,
	title = {Mechanical metamaterials and beyond},
	volume = {14},
	doi = {10.1038/s41467-023-41679-8},
	number = {1},
	journal = {Nature Communications},
	author = {Jiao, Pengcheng and Mueller, Jochen and Raney, Jordan R. and Zheng, Xiaoyu (Rayne) and Alavi, Amir H.},
	year = {2023},
	pages = {6004},
}

@article{zhang_stretchable_2024,
	title = {Stretchable and negative-{Poisson}-ratio porous metamaterials},
	volume = {15},
	doi = {10.1038/s41467-024-44707-3},
	number = {1},
	journal = {Nature Communications},
	author = {Zhang, Xiaoyu and Sun, Qi and Liang, Xing and Gu, Puzhong and Hu, Zhenyu and Yang, Xiao and Liu, Muxiang and Sun, Zejun and Huang, Jia and Wu, Guangming and Zu, Guoqing},
	year = {2024},
	pages = {392},
}

@article{efa_computational_2025,
	title = {Computational design of hybrid metamaterials: {Optimized} topology and lattice infill for complex multi-geometric structures},
	volume = {2},
	doi = {10.1016/j.nexres.2025.100449},
	journal = {Next Research},
	author = {Efa, Dame Alemayehu and Ifa, Dejene Alemayehu and Dejene, Naol Dessalegn},
        year = {2025},
	pages = {100449},
}

@article{zheng_data-driven_2021,
	title = {Data-driven topology optimization of spinodoid metamaterials with seamlessly tunable anisotropy},
	volume = {383},
	doi = {10.1016/j.cma.2021.113894},
	journal = {Computer Methods in Applied Mechanics and Engineering},
	author = {Zheng, Li and Kumar, Siddhant and Kochmann, Dennis M.},
	year = {2021},
	pages = {113894},
}

@article{wang_superior_nodate,
	title = {Superior fracture resistance and topology-induced intrinsic toughening mechanism in {3D} shell-based lattice metamaterials},
	volume = {10},
	doi = {10.1126/sciadv.adq2664},
	number = {35},
	journal = {Science Advances},
	author = {Wang, Yujia and Wu, Kunlin and Zhang, Xuan and Li, Xiaoyan and Wang, Yifan and Gao, Huajian},
        year = {2024},
	pages = {eadq2664},
}

@article{wu_superelastic_2022,
	title = {Superelastic graphene aerogel-based metamaterials},
	volume = {13},
	doi = {10.1038/s41467-022-32200-8},
	journal = {Nature Communications},
	author = {Brandenbourger, Martin and Locsin, Xander and Lerner, Edan and Coulais, Corentin},
	year = {2019},
	pages = {4608},
}

@article{Brandenbourger_reciprocal_2019,
	title = {Non-reciprocal robotic metamaterials {\textbar} {Nature} {Communications}},
	volume = {10},
	doi = {10.1038/s41467-019-12599-3},
	journal = {Nature Communications},
	author = {Wu, Mingmao and Geng, Hongya and Hu, Yajie and Ma, Hongyun and Yang, Ce and Chen, Hongwu and Wen, Yeye and Cheng, Huhu and Li, Chun and Liu, Feng and Jiang, Lan and Qu, Liangti},
	year = {2022},
	pages = {4561},
}

@article{zhao_hydrogel-based_2022,
	title = {A hydrogel-based mechanical metamaterial for the interferometric profiling of extracellular vesicles in patient samples},
	volume = {7},
	doi = {10.1038/s41551-022-00954-7},
	journal = {Nature Biomedical Engineering},
	author = {Zhao, Haitao and Pan, Sijun and Natalia, Auginia and Wu, Xingjie and Ong, Chin-Ann J. and Teo, Melissa C. C. and So, Jimmy B. Y. and Shao, Huilin},
	year = {2022},
	pages = {135--148},
}

@article{Cheng_energy_2023,
	title = {Mechanical metamaterials made of freestanding quasi-{BCC} nanolattices of gold and copper with ultra-high energy absorption capacity {\textbar} {Nature} {Communications}},
	volume = {14},
	doi = {10.1038/s41467-023-36965-4},
	journal = {Nature Communications},
	author = {Cheng, Hongwei and Zhu, Xiaoxia and Cheng, Xiaowei and Cai, Pengzhan and Liu, Jie and Yao, Huijun and Zhang, Ling and Duan, Jinglai},
	year = {2023},
	pages = {1243},
	
}

@article{BendsoeKikuchi1988,
  title        = {Generating optimal topologies in structural design using a homogenization method},
  author       = {Bends\o{}e, Martin P. and Kikuchi, Noboru},
  journal      = {Computer Methods in Applied Mechanics and Engineering},
  volume       = {71},
  pages        = {197--224},
  year         = {1988},
  doi          = {10.1016/0045-7825(88)90086-2},
}

@article{Sigmund1994,
  author    = {Sigmund, Ole},
  title     = {Materials with prescribed constitutive parameters: An inverse homogenization problem},
  journal   = {International Journal of Solids and Structures},
  volume    = {31},
  number    = {17},
  pages     = {2313-2329},
  year      = {1994},
  doi       = {10.1016/0020-7683(94)90154-6},
}

@article{groen_homogenization-based_2018,
	title = {Homogenization-based topology optimization for high-resolution manufacturable microstructures},
	volume = {113},
	doi = {10.1002/nme.5575},
	journal = {International Journal for Numerical Methods in Engineering},
	author = {Groen, Jeroen P. and Sigmund, Ole},
	year = {2018},
	pages = {1148--1163},
}

@article{allaire_homogenization_2019,
	title = {The homogenization method for topology optimization of structures: old and new},
        volume = {25},
	doi = {10.48550/arXiv.1901.09336},
        journal = {Interdisciplinary Information Sciences},
	author = {Allaire, Gr{\'e}goire and Cavallina, Lorenzo and Miyake, Nobuhito and Oka, Tomoyuki and Yachimura, Toshiaki},
	year = {2019},
}

@article{wu_topology_2021,
	title = {Topology optimization of multi-scale structures: a review},
	volume = {63},
	doi = {10.1007/s00158-021-02881-8},
	journal = {Structural and Multidisciplinary Optimization},
	author = {Wu, Jun and Sigmund, Ole and Groen, Jeroen P.} ,
	year = {2021},
	pages = {1455-1480},
}

@article{Xia_topology_2016,
	title = {Recent Advances on Topology Optimization of Multiscale Nonlinear Structures},
	volume = {24},
	doi = {10.1007/s11831-016-9170-7},
	journal = {Archives of Computational Methods in Engineering},
	author = {Xia, Liang and Breitkopf, Piotr} ,
	year = {2016},
	pages = {227-249},
}

@article{Xia_design_2015,
	title = {Design of materials using topology optimization and energy-based homogenization approach in {Matlab}},
        volume = {52},
	doi = {10.1007/s00158-015-1294-0},
	journal = {Structural and Multidisciplinary Optimization},
	author = {Xia, Liang and Breitkopf, Piotr},
	year = {2015},
        pages = {1229–1241},
}

@article{zhang_computational_2019,
	title = {Computational design of finite strain auxetic metamaterials via topology optimization and nonlinear homogenization},
	volume = {356},
	doi = {10.1016/j.cma.2019.07.027},
	journal = {Computer Methods in Applied Mechanics and Engineering},
	author = {Zhang, Guodong and Khandelwal, Kapil},
	year = {2019},
	pages = {490--527},
}

@article{chen_topological_2019,
	title = {Topological design of {3D} chiral metamaterials based on couple-stress homogenization},
	volume = {131},
	doi = {10.1016/j.jmps.2019.07.014},
	journal = {Journal of the Mechanics and Physics of Solids},
	author = {Chen, Wei and Huang, Xiaodong},
	year = {2019},
	pages = {372--386},
}

@article{huo_bi-directional_2025,
	title = {Bi-directional homogenization method for the design of multi-scale mechanical metamaterials},
	volume = {353},
	doi = {10.1016/j.compstruct.2024.118678},
	journal = {Composite Structures},
	author = {Huo, Senlin and Du, Bingxiao and Zhao, Yong and Chen, Xiaoqian},
	year = {2025},
	pages = {118678},
}

@article{hassani_review_1998,
	title = {A review of homogenization and topology optimization {I}{\textemdash}homogenization theory for media with periodic structure},
	volume = {69},
	doi = {10.1016/S0045-7949(98)00131-X},
	journal = {Computers \& Structures},
	author = {Hassani, B. and Hinton, E.},
	year = {1998},
	pages = {707--717},
}

@article{hassani_review_1998_II,
	title = {A review of homogenization and topology optimization {II}{\textemdash}analytical and numerical solution of homogenization equations},
	volume = {69},
	doi = {10.1016/S0045-7949(98)00132-1},
	journal = {Computers \& Structures},
	author = {Hassani, B. and Hinton, E.},
	year = {1998},
	pages = {719-738},
}

@article{andreassen_how_2014,
	title = {How to determine composite material properties using numerical homogenization},
	volume = {83},
	doi = {10.1016/j.commatsci.2013.09.006},
	journal = {Computational Materials Science},
	author = {Andreassen, Erik and Andreasen, Casper Schousboe},
	year = {2014},
	pages = {488--495},
}

@article{dong_149_2018,
	title = {A 149 {Line} {Homogenization} {Code} for {Three}-{Dimensional} {Cellular} {Materials} {Written} in matlab},
	volume = {141},
	doi = {10.1115/1.4040555},
	journal = {Journal of Engineering Materials and Technology},
	author = {Dong, Guoying and Tang, Yunlong and Zhao, Yaoyao Fiona},
	year = {2018},
}

@article{cheng_-relaxed_1997,
	title = {$\varepsilon$-relaxed approach in structural topology optimization},
	volume = {13},
	doi = {10.1007/BF01197454},
	number = {4},
	journal = {Structural optimization},
	author = {Cheng, G. D. and Guo, X.},
	year = {1997},
	pages = {258--266},
}

@article{bruggi_alternative_2008,
	title = {On an alternative approach to stress constraints relaxation in topology optimization},
	volume = {36},
	doi = {10.1007/s00158-007-0203-6},
	number = {2},
	journal = {Structural and Multidisciplinary Optimization},
	author = {Bruggi, Matteo},
	year = {2008},
	pages = {125--141},
}

@article{rozvany_difficulties_1996,
	title = {Difficulties in truss topology optimization with stress, local buckling and system stability constraints},
	volume = {11},
	doi = {10.1007/BF01197036},
	number = {3},
	journal = {Structural optimization},
	author = {Rozvany, G. I. N.},
	year = {1996},
	pages = {213--217},
}

@article{paris_topology_2009,
	title = {Topology optimization of continuum structures with local and global stress constraints},
	volume = {39},
	doi = {10.1007/s00158-008-0336-2},
	number = {4},
	journal = {Structural and Multidisciplinary Optimization},
	author = {Par{\'i}s, J. and Navarrina, F. and Colominas, I. and Casteleiro, M.},
	year = {2009},
	pages = {419--437},
}

@article{duysinx_topology_1998,
	title = {Topology optimization of continuum structures with local stress constraints},
	volume = {43},
	doi = {10.1002/(SICI)1097-0207(19981230)43:8<1453::AID-NME480>3.0.CO;2-2},
	number = {8},
	journal = {International Journal for Numerical Methods in Engineering},
	author = {Duysinx, P. and Bends{\o}e, M. P.},
	year = {1998},
	pages = {1453--1478},
}

@article{kreisselmeier_systematic_1979,
	title = {Systematic control design by optimizing a vector performance Index},
	volume = {12},
	doi = {10.1016/S1474-6670(17)65584-8},
	number = {7},
	journal = {IFAC Proceedings Volumes},
	author = {Kreisselmeier, G. and Steinhauser, R.},
	year = {1979},
	pages = {113--117},
}

@article{le_stress-based_2010,
	title = {Stress-based topology optimization for continua},
	volume = {41},
	doi = {10.1007/s00158-009-0440-y},
	number = {4},
	journal = {Structural and Multidisciplinary Optimization},
	author = {Le, Chau and Norato, Julian and Bruns, Tyler and Ha, Christopher and Tortorelli, Daniel},
	year = {2010},
	pages = {605--620},
}

@article{lee_novel_2016,
	title = {A novel {P}-norm correction method for lightweight topology optimization under maximum stress constraints},
	volume = {171},
	doi = {10.1016/j.compstruc.2016.04.005},
	journal = {Computers \& Structures},
	author = {Lee, Kangwon and Ahn, Kisoo and Yoo, Jeonghoon},
	year = {2016},
	pages = {18--30},
}

@misc{collet_topology_2018,
	title = {Topology optimization for microstructural design under stress constraints},
	volume = {58},
	doi = {10.1007/s00158-018-2045-9},
	journal = {Structural and Multidisciplinary Optimization},
	author = {Collet, Maxime and Noël, Lise and Bruggi, Matteo and Duysinx, Pierre},
	year = {2018},
	pages = {2677–2695},
}

@article{giraldo-londono_polystress_2021,
	title = {PolyStress: a Matlab implementation for local stress-constrained topology optimization using the augmented {Lagrangian} method},
	volume = {63},
	doi = {10.1007/s00158-020-02760-8},
	number = {4},
	journal = {Structural and Multidisciplinary Optimization},
	author = {Giraldo-Londo{\~n}o, Oliver and Paulino, Glaucio H.},
	year = {2021},
	pages = {2065--2097},
}

@article{chen_stress-constrained_2024,
	title = {Stress-constrained topology optimization using the constrained natural element method},
	volume = {67},
	doi = {10.1007/s00158-024-03786-y},
	journal = {Structural and Multidisciplinary Optimization},
	author = {Chen, Yanda and Monteiro, Eric and Koutiri, Imade and Favier, Véronique},
	year = {2024},
	pages = {65},
}

@article{chen_fatigue-constrained_2024,
	title = {Fatigue-constrained topology optimization using the constrained natural element method},
	volume = {422},
	doi = {10.1016/j.cma.2024.116821},
	journal = {Computer Methods in Applied Mechanics and Engineering},
	author = {Chen, Yanda and Monteiro, Eric and Koutiri, Imade and Favier, V{\'e}ronique},
	year = {2024},
	pages = {116821},
}

@article{verbart_damage_2016,
	title = {Damage approach: {A} new method for topology optimization with local stress constraints},
	volume = {53},
	doi = {10.1007/s00158-015-1318-9},
	journal = {Structural and Multidisciplinary Optimization},
	author = {Verbart, Alexander and Langelaar, Matthijs and Keulen, Fred van},
	year = {2016},
	pages = {1081--1098},
}

@misc{Coelho_topology_2019,
	title = {Topology optimization of cellular materials with periodic microstructure under stress constraints},
	volume = {59},
	doi = {10.1007/s00158-018-2089-x},
	journal = {Structural and Multidisciplinary Optimization},
	author = {Coelho, Pedro G. and Guedes, José M. and Cardoso, João B.},
	year = {2019},
	pages = {633–645},
}

@article{gupta_computational_2024,
	title = {Computational design of auxetic microstructures via stress-based topology optimization},
	volume = {319},
	doi = {10.1016/j.engstruct.2024.118807},
	journal = {Engineering Structures},
	author = {Gupta, Anurag and Gupta, Abhinav and Chowdhury, Rajib},
	year = {2024},
	pages = {118807},
}

@misc{Thillaithevan_stress-constrained_2020,
	title = {Stress-constrained optimization using graded lattice microstructures},
	volume = {63},
	doi = {10.1007/s00158-020-02723-z},
	journal = {Structural and Multidisciplinary Optimization},
	author = {Thillaithevan, Dilaksan and Bruce, Paul and Santer, Matthew},
	year = {2021},
	pages = {721–740},
}

@article{alacoque_stress-based_2021,
	title = {Stress-based and robust topology optimization for thermoelastic multi-material periodic microstructures},
	volume = {379},
	doi = {10.1016/j.cma.2021.113749},
	journal = {Computer Methods in Applied Mechanics and Engineering},
	author = {Alacoque, Lee and Watkins, Ryan T. and Tamijani, Ali Y.},
	year = {2021},
	pages = {113749},
}

@article{papadopoulos_comparative_1997,
	title = {A comparative study of multiaxial high-cycle fatigue criteria for metals},
	volume = {19},
	doi = {10.1016/S0142-1123(96)00064-3},
	number = {3},
	journal = {International Journal of Fatigue},
	author = {Papadopoulos, Ioannis V. and Davoli, Piermaria and Gorla, Carlo and Filippini, Mauro and Bernasconi, Andrea},
	year = {1997},
	pages = {219--235},
}

@article{carpinteri_multiaxial_2001,
	title = {Multiaxial high-cycle fatigue criterion for hard metals},
	volume = {23},
	doi = {10.1016/S0142-1123(00)00075-X},
	journal = {International Journal of Fatigue},
	author = {Carpinteri, Andrea and Spagnoli, Andrea},
	year = {2001},
	pages = {135--145},
}

@article{miehe_computational_2002,
	title = {Computational micro-to-macro transitions of discretized microstructures undergoing small strains},
	volume = {72},
	doi = {10.1007/s00419-002-0212-2},
	journal = {Archive of Applied Mechanics (Ingenieur Archiv)},
	author = {Miehe, C. and Koch, A.},
	year = {2002},
	pages = {300--317},
}

@article{van_dijk_formulation_2016,
	title = {Formulation and implementation of stress-driven and/or strain-driven computational homogenization for finite strain},
	volume = {107},
	doi = {10.1002/nme.5198},
	journal = {International Journal for Numerical Methods in Engineering},
	author = {van Dijk, N. P.},
	year = {2016},
	pages = {1009--1028},
}

@article{temizer_computation_2008,
	title = {On the {Computation} of the {Macroscopic} {Tangent} for {Multiscale} {Volumetric} {Homogenization} {Problems}},
	volume = {8},
	doi = {10.1002/pamm.200810577},
	journal = {PAMM},
	author = {Temizer, {\.I}lker and Wriggers, Peter},
	year = {2008},
	pages = {10577--10578},
}

@book{bendsoe_2003_topology,
       title     = {Topology Optimization: Theory, Methods, and Applications},
       author    = {Bends{\o}e, Martin P. and Sigmund, Ole},
       year      = {2003},
       publisher = {Springer},
       address   = {Berlin, Heidelberg},
       doi       = {10.1007/978-3-662-05086-6}
}

@article{sigmund_99_2001,
	title = {A 99 line topology optimization code written in {Matlab}},
	volume = {21},
	doi = {10.1007/s001580050176},
	journal = {Structural and Multidisciplinary Optimization},
	author = {Sigmund, O.},
	year = {2001},
	pages = {120--127},
}

@article{andreassen_efficient_2011,
	title = {Efficient topology optimization in {MATLAB} using 88 lines of code},
	volume = {43},
	doi = {10.1007/s00158-010-0594-7},
	journal = {Structural and Multidisciplinary Optimization},
	author = {Andreassen, Erik and Clausen, Anders and Schevenels, Mattias and Lazarov, Boyan S. and Sigmund, Ole},
	year = {2011},
	pages = {1--16},
}

@article{zegard_bridging_2016,
	title = {Bridging topology optimization and additive manufacturing},
	volume = {53},
	doi = {10.1007/s00158-015-1274-4},
	number = {1},
	journal = {Structural and Multidisciplinary Optimization},
	author = {Zegard, Tom{\'a}s and Paulino, Glaucio H.},
	year = {2016},
	pages = {175--192},
}

@article{guest_achieving_2004,
	title = {Achieving minimum length scale in topology optimization using nodal design variables and projection functions},
	volume = {61},
	doi = {10.1002/nme.1064},
	number = {2},
	journal = {International Journal for Numerical Methods in Engineering},
	author = {Guest, J. K. and Pr{\'e}vost, J. H. and Belytschko, T.},
	year = {2004},
	pages = {238--254},
}

@article{wang_projection_2011,
	title = {On projection methods, convergence and robust formulations in topology optimization},
	volume = {43},
	doi = {10.1007/s00158-010-0602-y},
	number = {6},
	journal = {Structural and Multidisciplinary Optimization},
	author = {Wang, Fengwen and Lazarov, Boyan Stefanov and Sigmund, Ole},
	year = {2011},
	pages = {767--784},
}

@article{senhora_topology_2020,
	title = {Topology optimization with local stress constraints: a stress aggregation-free approach},
	volume = {62},
	doi = {10.1007/s00158-020-02573-9},
	journal = {Structural and Multidisciplinary Optimization},
	author = {Senhora, Fernando V. and Giraldo-Londo{\~n}o, Oliver and Menezes, Ivan F. M. and Paulino, Glaucio H.},
	year = {2020},
	pages = {1639--1668},
}

@article{aleksandrova_efficiency_2019,
	title = {Efficiency of the {Tresca} yield criterion in modeling of annular plates with rigid constraint},
	volume = {75},
	doi = {10.1016/j.apm.2019.05.051},
	journal = {Applied Mathematical Modelling},
	author = {Aleksandrova, Nelli},
	year = {2019},
	pages = {371--384},
}

@article{jin_yielding_2023,
	title = {The yielding behavior and plastic deformation of oxygen-free copper under biaxial quasi-static and dynamic loadings},
	volume = {276},
	doi = {10.1016/j.ijsolstr.2023.112333},
	journal = {International Journal of Solids and Structures},
	author = {Jin, Kanghua and Qi, Lin and Kang, Huaipu and Wang, Weibin and Guo, Yazhou and Li, Yulong},
	year = {2023},
	pages = {112333},
}

@article{shi_equivalent_2025,
	title = {An equivalent average shear stress yield criterion based on {Simpson}'s numerical integration rule and its application in burst pressure analysis of thin-walled pipelines},
	volume = {206},
	doi = {10.1016/j.tws.2024.112613},
	journal = {Thin-Walled Structures},
	author = {Shi, Yongsheng and Qie, Yanhui and Song, Jinhua and Li, Yutong},
	year = {2025},
	pages = {112613},
}

@article{giraldo-londono_unified_2020,
	title = {A unified approach for topology optimization with local stress constraints considering various failure criteria: von {Mises}, {Drucker}-{Prager}, {Tresca}, {Mohr}-{Coulomb}, {Bresler}- {Pister} and {Willam}-{Warnke}},
	volume = {476},
	doi = {10.1098/rspa.2019.0861},
	journal = {Proceedings. Mathematical, Physical, and Engineering Sciences},
	author = {Giraldo-Londo{\~n}o, Oliver and Paulino, Glaucio H.},
	year = {2020},
	pages = {20190861},
}

@article{liu_review_2024,
	title = {Review on {Fatigue} of {Additive} {Manufactured} {Metallic} {Alloys}: {Microstructure}, {Performance}, {Enhancement}, and {Assessment} {Methods}},
	volume = {36},
	doi = {10.1002/adma.202306570},
	journal = {Advanced Materials},
	author = {Liu, Hui and Yu, Hanyang and Guo, Chuan and Chen, Xuliang and Zhong, Shiyu and Zhou, Lin and Osman, Amr and Lu, Jian},
	year = {2024},
	pages = {2306570},
}

@article{matake_explanation_1977,
	title = {An {Explanation} on {Fatigue} {Limit} under {Combined} {Stress}},
	volume = {20},
	doi = {10.1299/jsme1958.20.257},
	journal = {Bulletin of JSME},
	author = {Matake, Tomokazu},
	year = {1977},
	pages = {257--263},
}

@article{findley_theory_1959,
	title = {A {Theory} for the {Effect} of {Mean} {Stress} on {Fatigue} of {Metals} {Under} {Combined} {Torsion} and {Axial} {Load} or {Bending}},
	volume = {81},
	doi = {10.1115/1.4008327},
	journal = {Journal of Engineering for Industry},
	author = {Findley, W. N.},
	year = {1959},
	pages = {301--305},
}

@article{Dang_Van_pdf_1989,
	title = {{On} a {New} {Multiaxial} {Fatigue} {Limit} {Criterion}: {Theory} and {Application}},
	publisher = {Mechanical and Multiaxial Fatigue},
               address = {London},
	journal = {Biaxial and Multiaxial Fatigue (EGF 3)},
	author = {Dang Van, K. and Griveau, B. and Message, O.},
	year = {1989},
	pages = {479--496},
}

@book{socie_2000_multiaxial,
  title     = {Multiaxial Fatigue},
  author    = {Socie, Darrell F. and Marquis, Gary B.},
  year      = {2000},
  publisher = {SAE International},
  address   = {Warrendale, PA},
}

@article{mower_mechanical_2016,
	title = {Mechanical behavior of additive manufactured, powder-bed laser-fused materials},
	volume = {651},
	doi = {10.1016/j.msea.2015.10.068},
	journal = {Materials Science and Engineering: A},
	author = {Mower, Todd M. and Long, Michael J.},
	year = {2016},
	pages = {198--213},
}

@article{fatemi_torsional_2017,
	title = {Torsional fatigue behavior of wrought and additive manufactured {Ti}-{6Al}-{4V} by powder bed fusion including surface finish effect},
	volume = {99},
	doi = {10.1016/j.ijfatigue.2017.03.002},
	journal = {International Journal of Fatigue},
	author = {Fatemi, Ali and Molaei, Reza and Sharifimehr, Shahriar and Shamsaei, Nima and Phan, Nam},
	year = {2017},
	pages = {187--201},
}

@article{Huang_negative_2016,
	title = {Negative Poisson's Ratio in Modern Functional Materials},
	volume = {28},
	doi = {10.1002/adma.201601363},
	journal = {Advanced Materials},
	author = {Huang, Chuanwei and Chen, Lang},
	year = {2016},
	pages = {8079-8096},
}





\end{document}